\documentclass[onecolumn, usenatbib]{mnras}
\usepackage{savesym}
\usepackage{graphicx}
\usepackage{longtable}
\usepackage{changepage}

\usepackage{subfig}
\usepackage{amsmath}
\usepackage{amssymb}
\usepackage{verbatim}
\usepackage[yyyymmdd,hhmmss]{datetime}
\usepackage{array}
\usepackage{times}
\usepackage{xcolor}

\usepackage{xspace}

\title[Radiative feedback from TDEs]{ Galaxy scale consequences of tidal disruption events: extended emission line regions, extreme coronal lines and infrared-to-optical light echoes }

\author [Andrew Mummery, et al.]{Andrew Mummery$^1$\thanks{E-mail:
andrew.mummery@physics.ox.ac.uk}, Muryel Guolo$^2$, James Matthews$^3$, Megan Newsome$^{4, 5}$, \newauthor Chris Lintott$^3$, William Keel$^6$ \\
$^1$Oxford Theoretical Physics, Beecroft Building,  Clarendon Laboratory, Parks Road, Oxford, OX1 3PU, United Kingdom , \\
$^2$ Bloomberg Center for Physics and Astronomy, Johns Hopkins University, 3400 N. Charles St., Baltimore, MD 21218, USA \\
$^3$Oxford Astrophysics, Denys Wilkinson Building,  Keble Road, Oxford, OX1 3RH, United Kingdom, \\ 
$^4$ Las Cumbres Observatory, 6740 Cortona Drive, Suite 102, Goleta, CA 93117-5575, USA, \\
$^5$ Department of Astronomy, The University of Texas at Austin, 2515 Speedway, Stop C1400, Austin, TX 78712, USA\\
$^6$ Department of Physics and Astronomy, University of Alabama, Box 870324, Tuscaloosa, AL 35487 }

\date{}
\label{firstpage}
\pagerange{\pageref{firstpage}--\pageref{lastpage}}

\begin{document}

\maketitle

\begin{abstract}
    Stars in galactic centers are occasionally scattered so close to the central supermassive black hole that they are completely disrupted by tidal forces, initiating a transient accretion event.  The aftermath of such a tidal disruption event (TDE) produces a bright-and-blue accretion flow which is known to persist for at least a decade (observationally) and can in principle produce ionizing radiation for hundreds of years. Tidal disruption events are known (observationally) to be overrepresented in galaxies which show extended emission line regions (EELRs), with no pre-TDE classical AGN activity, and to produce transient ``coronal lines'', such as [FeX] and [FeXIV]. Using coupled {\tt CLOUDY}–TDE disk simulations we show that tidal disruption event disks produce a sufficient ionizing radiation flux over their lifetimes to power both EELR of radial extents of $r \sim 10^4$ light years, and coronal lines. EELRs are produced when the ionizing radiation interacts with low density $n_H \sim 10^1 – 10^3 \, {\rm cm}^{-3}$ clouds on galactic scales, while  coronal lines are produced by high density $n_H \sim 10^6 – 10^8 \, {\rm cm}^{-3}$ clouds near the galactic center. High density gas in galactic centers will also result in the rapid switching on of narrow line features in post-TDE galaxies, and also various high-ionization lines which may be observed throughout the infrared with JWST. Galaxies with a higher intrinsic rate of tidal disruption events will be more likely to show macroscopic EELRs, which can be traced to originate from the previous tidal disruption event in that galaxy, which naturally explains why TDEs are more likely to be discovered in galaxies with EELRs. We further argue that a non-negligible fraction of so-called optically selected ``AGN'' are tidal disruption events.  
\end{abstract}

\section{Introduction}
Tidal disruption events (TDEs) occur when an unfortunate star is scattered onto a near-radial orbit of the supermassive black hole in a galactic center. Upon entering its own tidal radius, the star is completely disrupted, and roughly half of the stellar debris thereafter rain down onto the black hole, powering an luminous flare \citep{Rees88}. During this process an accretion flow also forms, which dominates the observed emission roughly $\sim 1$ year after the flare, and is then known (observationally) to persist for at least a decade \citep{vanVelzen19, MumBalb20a, Mummery_et_al_2024},  producing bright UV emission, and in a subset ($f\gtrsim 40\%$) of sources also bright soft/thermal X-ray emission \citep{Guolo24}. The population of detected TDEs is growing, mainly through the advent of large optical surveys \citep{Hammerstein22}, with roughly 100 sources now known \citep{Yao2023,Mummery_et_al_2024}.  The typical peak (innermost) temperatures of these flows is observed to be $kT_{\rm obs} \sim 50-150\, {\rm eV} \sim 5-15 \, {\cal R}$, where ${\cal R}=13.6$ eV is the Rydberg energy. The bulk of this long lived accretion disk emission is emitted at photon energies which will, if they encounter gas, be highly ionizing. In a very real sense therefore a TDE results in a transient, but still long lived, active galactic nucleus. It is natural therefore to expect TDEs to produce significant ionization features in their host galaxies, similar to those more classically associated with standard, long-lived active galactic nuclei (AGN). It is the purpose of this paper to examine these possibilities in detail. %

\begin{figure}
    \centering
    \includegraphics[width=1.\linewidth]{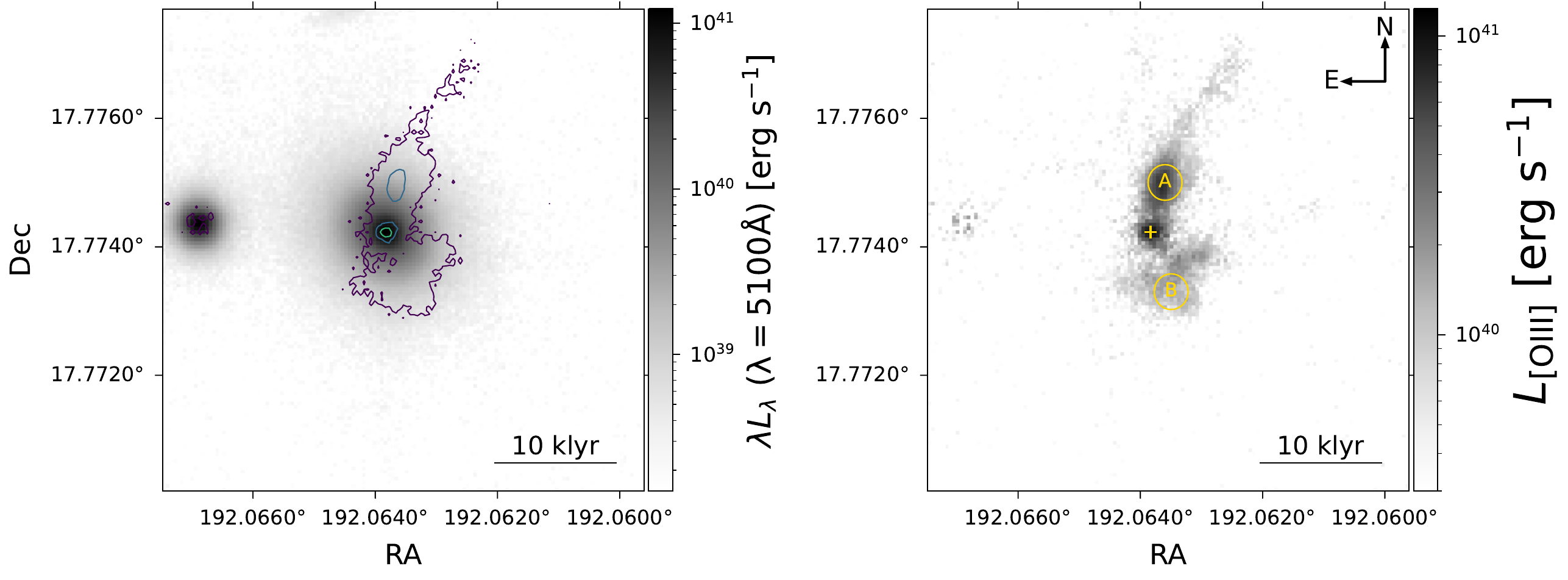}
    \includegraphics[width=1.\linewidth]{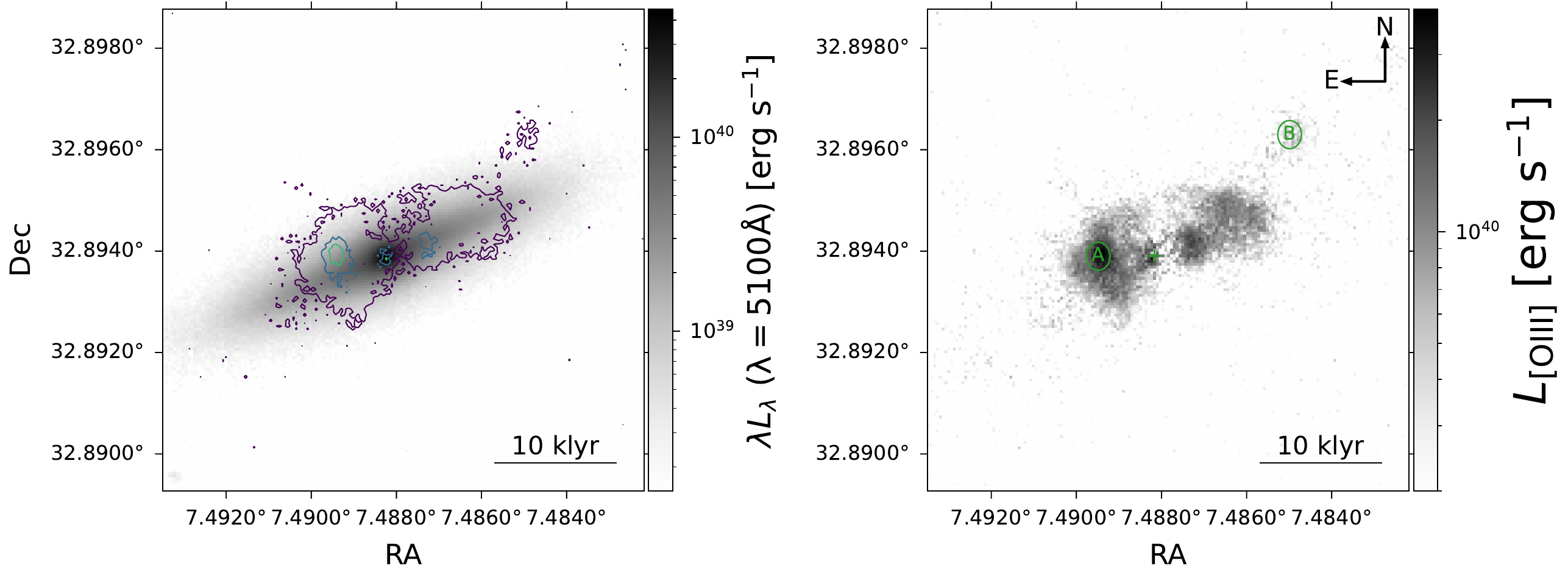}
    \caption{The extended emission line regions of two TDE hosts (upper = ASASSN-14li, lower = iPTF-16fnl). The left hand column shows maps of the continuum emission (as observed at 5100$\mathring{\rm A}$), while the right hand column shows the maps of [OIII] line emission. On the left hand column we also display contours of the [OIII] emission, for ease of comparison with the host continuum emission. A radial scale of 10,000 light years (at the distance of the host) is displayed in the lower right corner of each panel. It is clear that some strong source of ionizing radiation must have been previously present in these galaxies, although there is no evidence of an active galactic nuclei prior to the onset of the recent TDE. In the right panel, the crosses represent the location of the peak continuum emission, while the regions A/B, are spectral extractions for which line ratios are shown in Fig. ~\ref{fig:weird_tdes}. }
    \label{fig:tde_hosts}
\end{figure}

\begin{figure}
    \includegraphics[width=1.\linewidth]{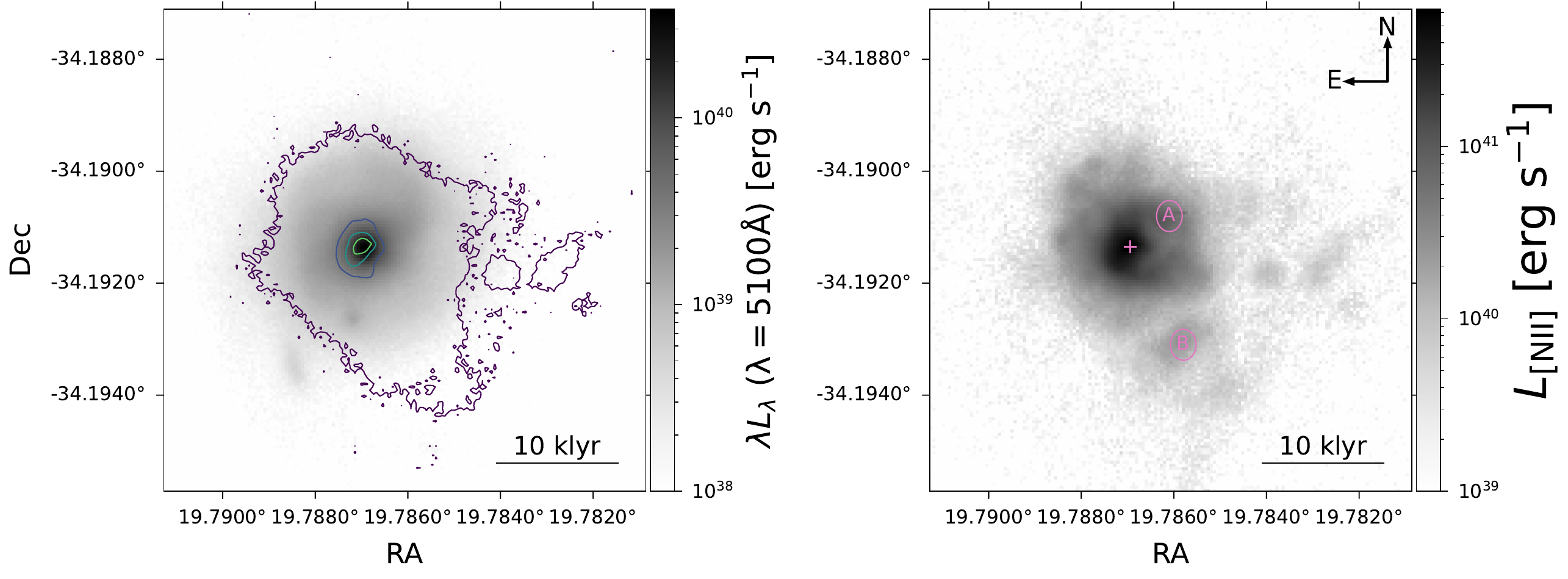}
    \includegraphics[width=1.\linewidth]{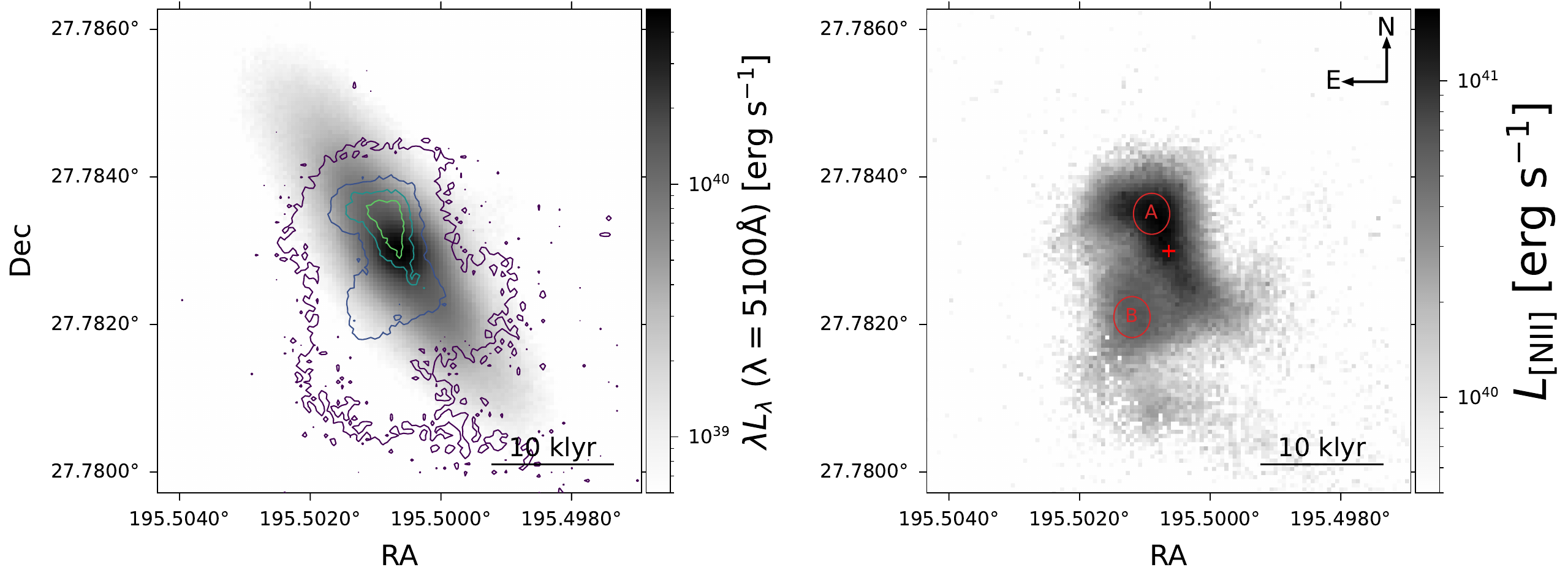}
    \caption{Similar to Figure \ref{fig:tde_hosts}, except now for the quasi-periodic eruption (QPE) hosts (upper = GSN 069, lower = RXJ 1301), with the right hand column showing [NII] emission. Note that there is strong evidence that (at least a significant fraction, if not all) QPEs  follow a TDE \citep[e.g.,][]{Nicholl24}, and it is extremely likely that at least GSN 069 was originally a TDE before becoming a QPE \citep[e.g.,][]{Guolo25}.   }
    \label{fig:qpe_hosts}
\end{figure}

As part of this analysis we shall seek to explain two interesting observational properties of TDEs and their host galaxies, which have become apparent only in the last few years. The first is that many tidal disruption events are observed to occur in galaxies which host extended emission line regions (EELRs), with no recent classical AGN activity (e.g., \citealt{{French23,Prieto2016_MUSE_14li,Wevers24EELR, WeversFrench24}} see Figures \ref{fig:tde_hosts}, \ref{fig:qpe_hosts}). By classical AGN activity we, in this paper, are referring to properties such as a bright infrared torus, a hard (power-law) X-ray corona, and a persistent and virialized broad line region (while keeping in mind that this definition may not capture every possible flavor of AGN). Each of these features are completely absent (or in the case of the hard power-law corona at most very rare) in TDE systems, but are ubiquitous to classical AGN systems.  While it is well known that classical AGN can power large scale EELRs, an interesting and important question is whether or not TDEs themselves are capable of producing EELR. This question is important  because, excluding the possibility that the over representation of EELR features in TDE host galaxies is a quirk of small number statistics, or that the presence of an EELR {\it causes} an increased rate of TDEs (for which there is no plausible physical mechanism), there are two possible mechanisms which can explain this observed over representation: (i) that there is a third process which causes {\it both} an EELR and an elevated TDE rate, or (ii) an elevated TDE rate is likely to produce an EELR. 

Much of the analysis in the literature has focused on possibility (i), suggesting that the switching off of an AGN (which would lead to an EELR with no current AGN activity) might cause an elevated TDE rate \citep[e.g.][]{Prieto2016_MUSE_14li, WeversFrench24}. While, if true, this would explain the observations, it is not clear what dynamical mechanism could result in an increased TDE rate as a result of the switching off of gas supply to the central black hole, making this a somewhat strained model. A more natural mechanism, we suggest, for explaining this link is that TDEs {\it themselves} can power EELR (i.e., suggestion (ii) above), and therefore those galaxies with high TDE rates (which will be overrepresented in a given TDE sample) will be more likely to show EELR features (which can be traced to prior TDEs). As there is no requirement for a classical AGN to have been active at any point in this mechanism, there is no expectation for these systems to show classical AGN features at the time of TDE detection. Of course, we are not intending to show that {\it all} EELR are produced by TDEs, merely that some can be. 


The second recent observational development is the discovery and categorization of so-called extreme coronal lines, detected following TDEs \citep[e.g.,][]{Neustadt2020, Onori2022, Hinkle2023, Li2023_ECLE, Short2023, Somalwar2023, Yao2023, Koljonen2024, Wang2024_ECLE, Newsome2024_22upj, Clark2025}. These coronal lines (such as [FeX] and [FeXIV]) are extremely rare in galaxies which do not host a current AGN, with only $5$ non-AGN systems found in the entire SDSS sample \citep{Komossa2008, Wang2011, Wang2012, Callow2024}, but have now been discovered in 11 galaxies in the months-to-years post TDE (see above references). This clear link between coronal lines and TDEs suggests that the disks formed in these events are sufficiently powerful ionizers to cause these short lived flares of coronal lines. This is a particularly interesting observational result as coronal lines require large $(E\gtrsim 100$ eV$)$ activation energies, energies which likely can only be reached by the innermost regions of accretion flows.

The purpose of this paper is to examine the potential of the disks formed in TDEs to produce ionization signatures in their host galaxies. We do this by coupling the TDE disk evolution code {\tt FitTeD} \citep{mummery2024fitted} to the photoionization code {\tt CLOUDY} \citep{Cloudy98}.  The key results of this work are that we demonstrate that TDE disks are sufficiently ionizing to (i) power EELR on galaxy-wide scales, with the interaction of TDE-disk emission with molecular clouds producing emission-line ratios that, when placed in BPT-type diagnostic diagrams, will fall above the \citet{Kewley01} line and therefore will appear to indicate that the galaxy hosts an ``AGN'', despite there being no classical AGN present. These same mechanisms will (in denser galactic center environments)  produce (ii) narrow line regions observable on $\sim {\cal O}(10^{-1}–10^2)$ parsec scales observed only months-decades post TDE. These dense clouds can also (iii) produce extreme coronal lines, just as observed from many TDE systems. Further, we make a series of predictions for what may be observed by the James Webb Space Telescope throughout the infrared, with a particular focus on a set of four Neon lines. We predict that bright [NeVI], [NeII], [NeV] and [NeIII] should be common in TDE host galaxy infrared spectra on $\sim$ month-to-year timescales post-TDE, particularly for those TDEs in which a infrared dust echo is also observed in the IR continuum. 

Our results therefore suggest that TDE host which show EELR may in fact be observations of (at least) {\it two} TDEs, one unfolding in real time and the second being echoed back to us from ionized clouds. We speculate that a non-negligible fraction of so-called ``faded AGN'' could in fact be the detection of TDE light echoes (we shall use the term ``light echo'' in this paper to mean the detection of light following the ionization and recombination of gas, not the scattering of light into our line of sight for which this term is also used), and indeed that tidal disruption events will represent a contaminant in optical studies of AGN quite generally. 

The layout of this paper is as follows. In section \ref{BOE} we begin with a  simple analytical analysis of these systems, showing that TDE disks should be significant sources of ionizing photons. In section \ref{geom} we discuss the observed geometry (in a distant observers frame) of the reprocessed emission resulting from a single shell of ionizing TDE emission, highlighting the broad range of galactic radii which can be probed by a single light echo. In section \ref{disk} we analyze the disk evolution in more detail, focusing on its ionizing potential. In sections \ref{cloudy1}, \ref{cloudy2}, \ref{jwst} and \ref{cloudy3} we combine the disk evolution with the photoionization code {\tt CLOUDY} \citep{Cloudy98}, to probe the emission line features of EELR (section \ref{cloudy1}), extreme coronal lines (section \ref{cloudy2}), possible line signatures detectable by the James Webb Space Telescope (section \ref{jwst}), and galactic center narrow line regions (section \ref{cloudy3}). In section \ref{MW} we speculate about the possibilities of detecting TDE light echoes within the Milky Way, before concluding with a broad discussion in \ref{conc}. Some further figures are presented in Appendix \ref{other}, while technical details regarding data reduction  are detailed in Appendix \ref{appB}.

\section{Initial plausibility analysis}\label{BOE}
\subsection{Observational constraints on the ionization lifetime of a TDE disk}
It is by now an observational fact that tidal disruption events result in long lived accretion flows. These accretion flows power thermal X-ray emission which is detected in at least $40\%$ of sources \citep{Guolo24},  but more relevantly for our purposes, these disks are also observed as bright long-lived UV sources which dominate the total TDE emission at times exceeding roughly one year from the peak of the flare \citep{vanVelzen19, MumBalb20a, Mummery_et_al_2024, MummeryVV24,GuoloMum24}. Observationally (and theoretically) TDE disks are bright and very blue, with emission detected from optical to X-ray bands. Indeed, the spectral peak of TDE emission is quite generally expected to be (as we shall show) at energies $E_{\rm peak} \sim 50-100$ eV, i.e., at photon energies which will be highly ionizing. This is a simple result of the typical temperature of Eddington-limited disks about massive black holes with masses at the typical TDE scale $M_\bullet \sim 10^6–10^7 M_\odot$ being roughly $kT \sim 50–100$ eV.  

Observational lower limits on the ionizing lifetime of TDE disks are now set exclusively by the time that has elapsed since the first tidal disruption event systems were discovered and systematically followed up at late times. The first comprehensively followed up TDEs were discovered in the period 2010–2016 with the discovery of eight sources \citep[namely SDSS-TDE2][PTF-09djl and PTF-09ge \citealt{Arcavi14}, ASASSN-14ae \citealt{Holoien14}, ASASSN-14li \citealt{Miller15, Holoien16b}, ASASSN-15oi \citealt{Holoien16}, iPTF-15af \citealt{Blagorodnova19} and  iPTF-16fnl \citealt{Blagorodnova17}]{vanVelzen10}, all of whom are still UV bright and detected at times $t \gtrsim 2000$ days post flare \citep[see the light curves presented in][the data sets of which are publicly available\footnote{\url{https://github.com/sjoertvv/manyTDE}}]{Mummery_et_al_2024}. We can therefore place an observational {\it lower bound} on the ionizing lifetime of TDE disks, which must be at least one decade to explain each of these sources. We stress that while the observed sources discussed above are all ``optically selected'' (i.e., sources discovered by optical surveys), there is nothing special about this population in terms of their disk physics \citep[when compared to ``X-ray selected'' TDEs, e.g.,][]{Guolo24, MummeryVV24}. Indeed, GSN 069 \citep{Saxton2011}, which was never observed to undergo an early optical flare (owing to a lack of observations), and was discovered by an X-ray survey {\it also} has a decade-old accretion flow detectable with detailed UV observations \citep{Guolo25}. 

\begin{figure}
    \centering
    \includegraphics[width=0.65\linewidth]{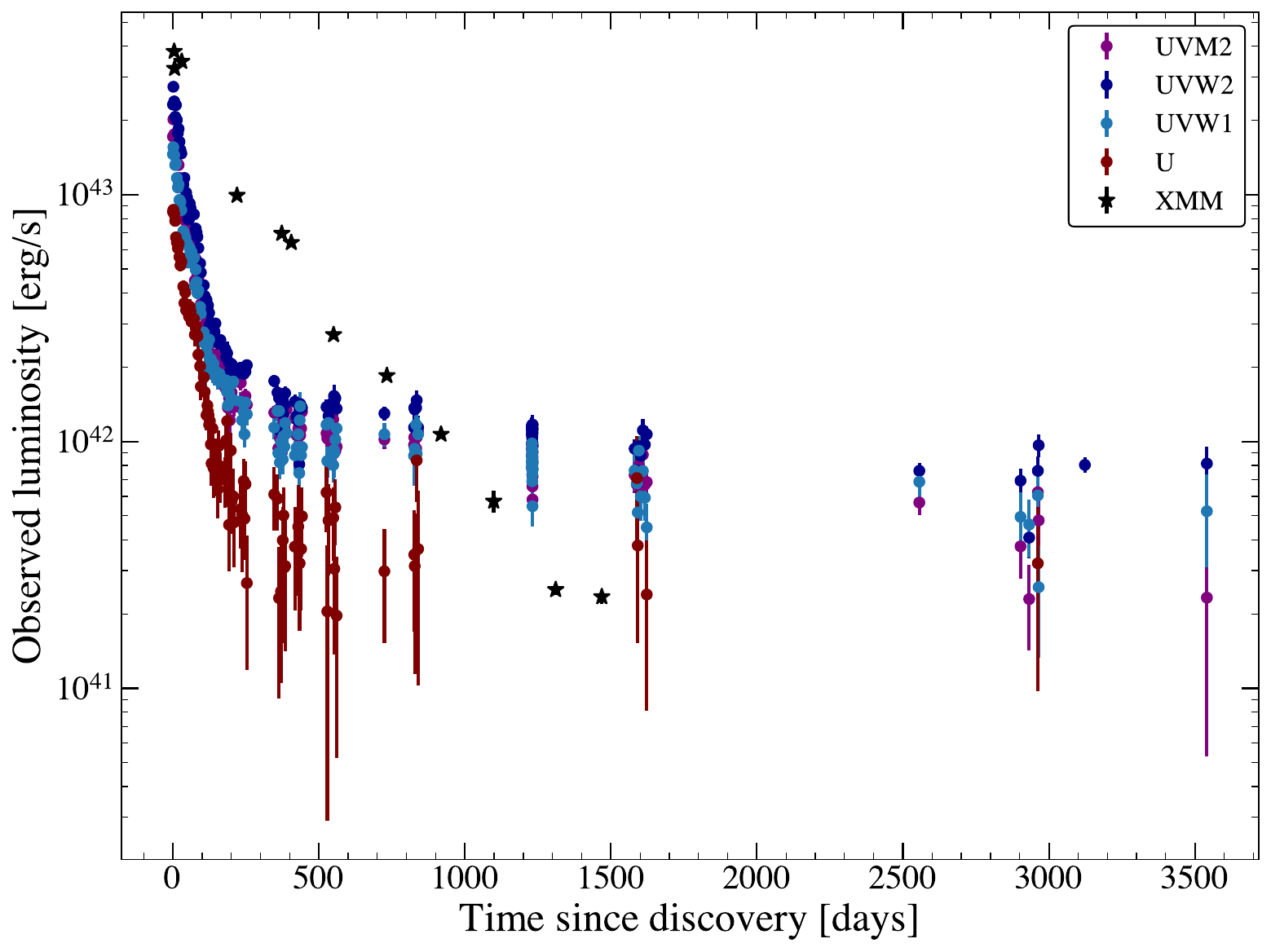}
    \includegraphics[width=0.95\linewidth]{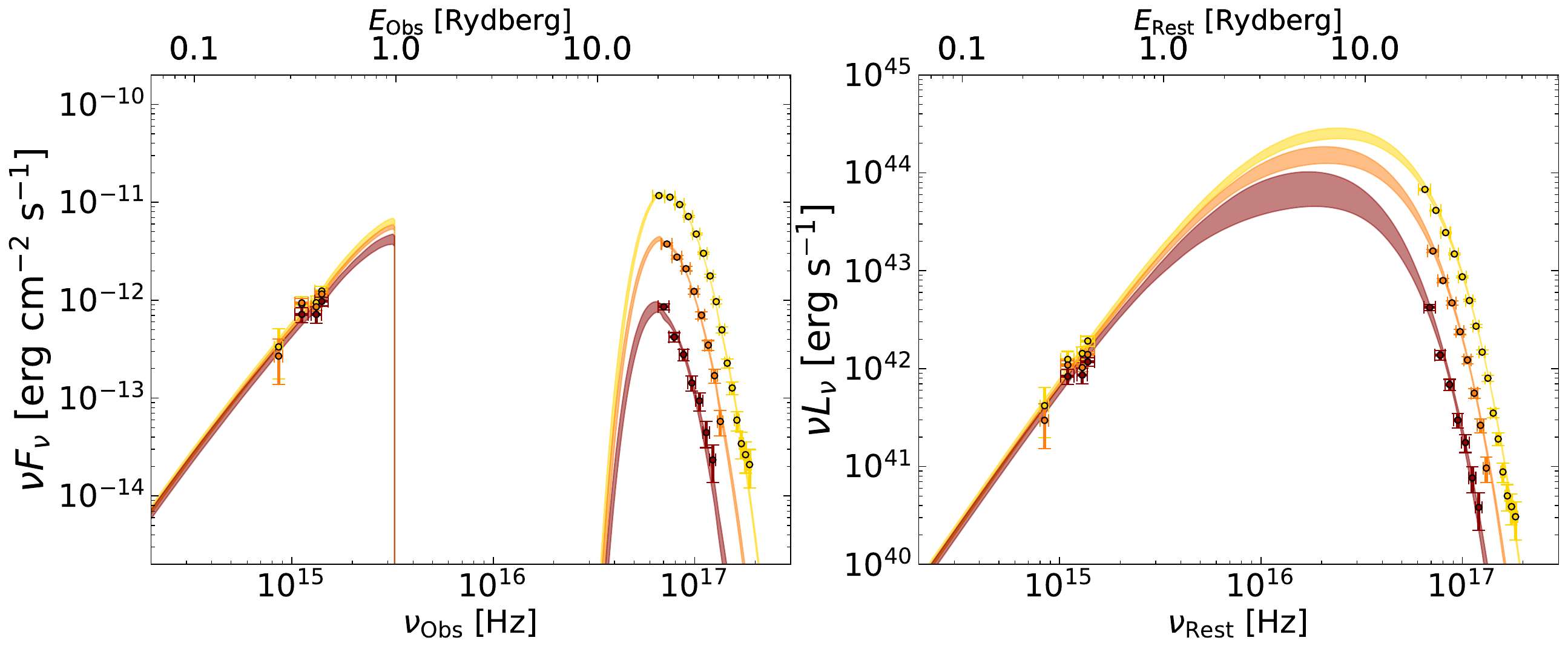}
    \caption{{\bf Upper:} The host-subtracted light curve of ASASSN-14li at X-ray (black) and UV (brown – purple, see legend) frequencies. At times beyond one year the systems remains UV and X-ray bright, with all flux resulting from an evolving accretion flow. This implies that there is a large ionizing flux which will be emitted between the UV and X-ray bands.  ASASSN-14li is a typical TDE in this respect, and long lived UV bright emission is a feature of a large number of these systems, stemming from a long lived accretion flow. {\bf Lower:} The observed spectral energy distributions {\bf (left)}, and unfolded intrinsic spectral energy distribution {\bf (right)}, for three different times in the plateau phase of ASASSN-14li's evolution. The final epoch is $\sim 4$ years post flare. At this stage the ionizing luminosity remains extremely bright $\sim 10^{44}$ erg/s, which will eventually have observable effects on the ionization structure of surrounding neutral gas. Figure adapted from \citet{GuoloMum24}.}
    \label{fig:14li}
\end{figure}

These late-time TDE disks are UV bright and have a blue spectrum, hence they produce a large amount of ionizing radiation. Indeed, in Figure \ref{fig:14li} we display the light curves and observed spectra of the TDE ASASSN-14li \citep{Miller15}, which is still producing bright UV emission to this day, more than one decade after it first was detected \citep{GuoloMum24}.  

At times greater than $\sim 1$ year post flare, all of the observed spectral energy distribution (SED) of ASASSN-14li is well described by an evolving accretion flow \citep[e.g.,][]{MumBalb20a, GuoloMum24}. The SED is shown, for three different times at the UV/optical plateau phase, in the lower panels of Fig. \ref{fig:14li}, with the dimmest epoch corresponding to $\sim 4$ years post flare. At this stage the ionizing luminosity remains extremely bright $L_{\rm ION} \sim 10^{44}$ erg/s, which will eventually have observable effects on the ionization structure of surrounding neutral gas. 

\subsection{Theoretical constraints on the ionization lifetime of a TDE disk}
In this subsection we set up the basic theoretical framework with which we shall analyze TDE accretion disks, their ionizing radiation and lifetimes. While we shall begin with an analytical scaling analysis, for later concrete calculations we shall move to full numerical solutions of the TDE disk equations. 

Theoretically, a TDE disk will continue to emit ionizing radiation until the disk transitions to an optically thin state $\Sigma_{\rm disk} \kappa_{\rm es} \sim 1$ (where $\Sigma_{\rm disk}$ is the disk surface density and $\kappa_{\rm es}$ the electron scattering opacity), or cools below the Rydberg energy $k_B f_{\rm col} T_{\rm disk} \simeq {\cal R} = 13.6$ eV. This first condition can be shown to take an extremely long time (typically $t \gtrsim 1,000$ years), and so is not relevant as the second is shorter. The second condition typically takes $\sim 100'$s of years and is therefore the relevant timescale of interest, as we now show.  

Let us assume that the compact disk which forms after the TDE will have an initial outer radial scale at the circularisation radius of the stellar orbit 
\begin{equation}
    r_c = 2 r_T \approx 2 R_\star \left({M_\bullet \over M_\star}\right)^{1/3}, 
\end{equation}
where we have introduced the tidal radius of the star $r_T \simeq R_\star (M_\bullet/M_\star)^{1/3}$, and the factor 2 results from angular momentum conservation as a parabolic stellar orbit is turned into a circular disk orbit in the disruption process. This disk is assumed to form with $M_{\rm disk} \simeq f_d M_\star/2$, where $M_\star/2$ is assumed to remain bound after the disruption \citep{Rees88}, of which a factor $0 < f_d < 1$ is assumed to form into a disk. 
Depositing this much material in a compact disk typically results in a TDE disk producing luminosity bounded at the Eddington limit at early times. There is good observational support for this statement \citep{Mummery_Wevers_23}. This implies that after one ``viscous'' timescale when the disk emission peaks that 
\begin{equation}
    L_{\rm bol}(t_{\rm visc})  \simeq 4\pi R_p^2 \sigma_{\rm sb} T_p^4 \simeq L_{\rm edd} \simeq {4 \pi G M_\bullet c\over \kappa_{\rm es}} , 
\end{equation}
and thus 
\begin{equation}
    k_B T_p(t_{\rm visc}) \simeq \left({k_B^4 c^5 \over x_p^2 \sigma_{\rm sb} \kappa_{\rm es}G M_\bullet }\right)^{1/4} \simeq 50 \, {\rm eV} \, \, \left({10\over x_p}\right)^{1/2} \left({10^6 M_\odot \over M_\bullet}\right)^{1/4}, 
\end{equation}
where $x_p \equiv c^2 R_p / GM_\bullet$ is the radial scale (in units of gravitational radii) at which the bulk of the disk emission is sourced, and $\sim 10$ is a value roughly corresponding to a Schwarzschild black hole disk. We remind the reader that this temperature is smaller than what would be observed from a TDE disk spectrum, as the emission is color corrected $(T_{\rm obs} \simeq f_{\rm col} T_p)$, with $f_{\rm col} \simeq 2.5$, owing to radiative transfer effects in the disk atmosphere \citep[e.g.][]{Davis06, Done12, Mum21L}.  At late times (exceeding the viscous timescale) material drains from the inner edge of the disk, and the temperature drops along with the disk density. The late time disk temperature follows a simple power-law decline 
\begin{equation}
    T_p(t\gtrsim t_{\rm visc}) \simeq T_p(t_{\rm visc}) \left({t \over t_{\rm visc}}\right)^{-n/4} ,
\end{equation}
where the index $n$ (also the index at which the bolometric luminosity of the disk decays) depends on some precise assumptions about the turbulent transport in the disk, but is typically around $n \approx 1.2$ \citep[note that this is a theoretical expectation of the decay rate, e.g.,][but one which is broadly insensitive to assumptions about the microphysics of the accretion process \citealt{MumBalb19a}]{Cannizzo90}. The viscous timescale in a TDE disk depends only on stellar properties 
\begin{equation}
    t_{\rm visc} = \alpha^{-1} \left({h\over r}\right)^{-2} \sqrt{r_c^3 \over GM_\bullet} = \alpha^{-1} \left({h\over r}\right)^{-2} \sqrt{8R_\star^3 \over  GM_\star} \equiv  {\cal V} t_\star,
\end{equation}
where we have defined the dimensionless nuisance parameter ${\cal V} \equiv \alpha^{-1} \left({h / r}\right)^{-2} \gg 1$, and labeled $t_\star \equiv \sqrt{8R_\star^3 / GM_\star }$, which is the dynamical timescale of the star (up to order unity constants). 

The disk will stop producing ionizing radiation when the color-corrected disk temperature drops below the Rydberg, i.e. $T_p(t_{\rm ION}) \simeq {{\cal R} / k_B f_{\rm col}}$ which implies 
\begin{align}
    t_{\rm ION} &\simeq t_{\rm visc} \left({c^5 k_B^4  f_{\rm col}^4 \over x_p^2 \sigma_{\rm sb} \kappa_{\rm es} G M_\bullet {\cal R}^4 }\right)^{1/n} \\ 
    t_{\rm ION} &\simeq 200\, {\rm yr}\, \left({{\cal V}\over 1000}\right) \left({R_\star \over R_\odot}\right)^{3/2} \left({M_\star \over M_\odot}\right)^{-1/2} \left({M_\bullet \over 10^6 M_\odot}\right)^{-1/n} \left({x_p \over 10}\right)^{-2/n}. 
\end{align}
We see that typical TDE parameters imply that a TDE disks produce ionizing radiation fluxes which will last for $\sim$ centuries. This is obviously completely in keeping with the observations which show bright disk emission lasting for at least a decade. Note that our simple calculation here is strictly describing photons which can ionize Hydrogen, and there is a $t_{\rm ION} \sim (E_\gamma/{\cal R})^{-4/n}$ dependence on this ionizing lifetime for other atomic species with activation energy $E_\gamma$ (this may be particularly relevant for lines such as the coronal line [FeX] which require activation energies $E_\gamma \sim 100$ eV and may therefore only be ionized in the first $\sim$ year of the event). 

\subsection{ Ionizing energy and photon budget }
During the time it takes for the disk to cool sufficiently to stop producing ionizing photons, the total accreted mass budget will be 
\begin{equation}
    \Delta M_{\rm acc} \equiv \int_{t_{\rm visc}}^{t_{\rm ION}} \dot M_H(t') \, {\rm d}t' \simeq {(n-1)M_{\rm disk} \over t_{\rm visc}} \int_{t_{\rm visc}}^{t_{\rm ION}} \left({t' \over t_{\rm visc}}\right)^{-n} \, {\rm d}t' \simeq M_{\rm disk} \left[1 - \left({t_{\rm ION} \over t_{\rm visc}}\right)^{1-n} \right], 
\end{equation}
where the normalisation on the horizon accretion rate $\dot M_H$ is set so that eventually ($t \to \infty$) all of the disk is accreted at times $t > t_{\rm visc}$ \citep[this is a mathematical property of the accretion disk equations, e.g.,][]{LBP74}.  This means that the fraction of accreted material in the ionizing phase is independent of disk evolutionary timescale (a fortunate result as this timescale is the least theoretically well constrained element of the disk evolution) 
\begin{equation}
    \Delta M_{{\rm acc}} \simeq M_{\rm disk}\left[1 - \left({c^5 k_B^4  f_{\rm col}^4 \over x_p^2 \sigma_{\rm sb} \kappa_{\rm es} G M_\bullet {\cal R}^4 }\right)^{(1-n)/n} \right], 
\end{equation}
and is in fact only very weakly dependent on all parameters (as $(1-n)/n \approx -1/6$). As up until this point the hottest regions of the disk (which dominate the radiated energy budget) are above the Rydberg, nearly all of this radiated energy will take the form of ionizing photons. The ionizing energy budget can therefore rather simply be estimated 
\begin{equation}
    E_{\rm ION} \simeq \eta(a) \, \Delta M_{\rm acc}\, c^2 \approx 9\times10^{52} \,\, {\rm erg}\, \left({\eta\over 0.1}\right)\left({\Delta M_{\rm acc}\over 0.5 M_\odot}\right) ,
\end{equation}
where $\eta(a)$ is the usual black hole spin dependent accretion efficiency. This leads to an average ionizing luminosity 
\begin{equation}
    \left\langle L_{\rm ION}\right\rangle \simeq {\eta(a) \, \Delta M_{\rm acc}\, c^2 \over t_{\rm ION}} \approx 3 \times 10^{43} \, {\rm erg/s} \, \left({\eta \over 0.1}\right) \left({\Delta M_{\rm acc} \over 0.5 M_\odot}\right) \left({100 \, {\rm yr} \over t_{\rm ION}}\right) .
\end{equation}
Each ionizing photon is emitted with an energy $E_{\gamma} \simeq {\cal R}$, meaning a total of 
\begin{equation}
    N_{\rm ION} \simeq {\eta(a) \, \Delta M_{\rm acc}\, c^2 \over {\cal R}} \approx 4 \times 10^{63} \, \left({\eta \over 0.1}\right) \left({\Delta M_{\rm acc} \over 0.5 M_\odot}\right) ,
\end{equation}
are emitted over this time, at an average rate  
\begin{equation}
    \left\langle \dot N_{\rm ION} \right\rangle \simeq {\eta(a) \, \Delta M_{\rm acc}\, c^2 \over t_{\rm ION} {\cal R}} \approx 1 \times  10^{53} \, {\rm s}^{-1} \, \left({\eta \over 0.1}\right) \left({\Delta M_{\rm acc} \over 0.5 M_\odot}\right) \left({100 \, {\rm yr} \over t_{\rm ION}}\right). 
\end{equation}
\subsection{The radial extent of the ionized region}
The ionizing photon flux will be emitted into a shell of radial width $\Delta r = c\, t_{\rm ION}$. These ionizing photons will ionize the gas in this region, which will then recombine. The gas will seek to reach recombination equilibrium within this photon shell. The timescale for Hydrogen recombination is 
\begin{equation}
    \tau_{\rm rec, \, H} \approx (\alpha_B \, n_e)^{-1} \simeq 20 \, \left({10^4 \, {\rm cm}^{-3}\over n_e}\right) \, {\rm yr}, 
\end{equation}
where $\alpha_B\simeq 1.43 \times 10^{-13}$ cm$^3$ s$^{-1}$ is the hydrogen case B recombination coefficient evaluated at a gas temperature $T_g \approx 20,000$ K \citep{Osterbrock06}. For $[{\rm OIII}]$ this time is much shorter, roughly $\tau_{\rm rec, \, OIII}\sim 0.1 \tau_{\rm rec, \, H}$. 

The validity of statistical steady state assumptions \citep[made by codes such as {\tt CLOUDY}][]{Cloudy98} therefore depends on the density of the neutral gas with which the photons interact (through the recombination timescale). If $\tau_{\rm rec} \gg t_{\rm ION}$ then by the time excited atoms de-excite the ionizing photon source will have passed through the cloud, leading to a system which is far from the steady state. On the contrary, for short recombination timescales $\tau_{\rm rec} \ll t_{\rm ION}$ each atom will undergo a large number of excitation-emission-excitation cycles, and the statistical steady state assumptions will be robust. As $t_{\rm ION} \sim 100$ years, for Hydrogen this boundary is roughly $n_e \gtrsim 10^3\, {\rm cm}^{-3}$, while for [OIII] this is about an order of magnitude lower. 

We shall show that to produce extended emission line regions with radii $r \sim 10^4$ lyr, the density of the clouds will likely be lower than these values.  As the timescales of recombination  of these species are comparable to (and likely larger than) the ionizing lifetime of the TDE, it is important to consider the possible effects of this out-of-steady state behavior when interpreting our results in quantitative detail. Our results for denser, galactic center, clouds will be comfortably within statistical recombination equilibrium however, as once an atom de-excites it will immediately interact with another disk photon. This is important, as those ionization features which switch on most quickly after a TDE (such as extreme coronal lines, infrared lines and transient narrow line regions) will result from dense clouds close to the galactic center, a regime for which our models are robustly valid. 

In recombination equilibrium the recombination rate within the volume swept out by the shell must equal the ionizing photon flux, namely 
\begin{equation}
    \left\langle \dot N_{\rm ION} \right\rangle \simeq {4\over 3}\pi r^3 f_V \, \alpha_B n_e n_p, 
\end{equation}
where $f_V$ is a ``volume filling factor'', representing the fraction of the sphere in which gas clouds exist, which is expected to be $f_V \ll 1$. Note that by writing this expression we are explicitly assuming the gas distribution around the galaxy is spherical, something which of course may not be the case for a disk-like galaxy. The general scaling laws we derive are insensitive to this level of detail, which will introduce order 1 corrections to numerical values but will not change the physics. We discuss galaxy morphology in more detail later. 

Inverting the above expression shows that the radial size of the ionized region can extend up to a maximum given roughly by  
\begin{equation}
    r_{\rm max} \simeq \left({3 \left\langle \dot N_{\rm ION} \right\rangle \over 4\pi \,  f_V \, \alpha_B n_e n_p }\right)^{1/3} .
\end{equation}
Substituting in rough numbers (and assuming $n_e \simeq n_H$) we find 
\begin{equation}
    r_{\rm max} \approx 14,000\, {\rm lyr} \, \left({\eta \over 0.1}\right)^{1/3} \left({\Delta M_{\rm acc} \over 0.5 M_\odot}\right)^{1/3} \left({100 \, {\rm yr} \over t_{\rm ION}}\right)^{1/3} \left({10^2 \, {\rm cm}^{-3} \over n_e}\right)^{2/3} \left({10^{-4} \over f_V }\right)^{1/3} ,
\end{equation}
which is suggestive when considering the typical properties of TDE host galaxy EELRs (e.g., Figure \ref{fig:tde_hosts}, \ref{fig:qpe_hosts}). We therefore can imagine a simple schematic of the structure of an emission line region produced by a TDE
\begin{itemize}
    \item There is a shell of ionizing radiation with width $ct_{\rm ION} \sim 100\, {\rm lyr}$ sweeping through the galaxy.
    \item Lagging behind this by a radius $c\tau_{\rm rec, H} \sim 20  (10^4 \, {\rm cm}^{-2}/n_H)\, {\rm lyr}$ is a reionization front, producing observable Hydrogen line emission, while OIII emission is slightly further forward in the galaxy, as  $\tau_{\rm rec, OIII}\sim 0.1 \tau_{\rm rec, H}$.  
    \item From the perspective of someone at the center of the TDE host galaxy, both regions will expand at the speed of light out to a maximum radius $r_{\rm max}$ (given above) 
    \item In a distant observer-frame however, both the observed geometry of the emission line regions and the inferred propagation speed of this reionization front will depend on the distribution of gas clouds in the galaxy, and the time since the TDE (this will be discussed more carefully later). However, a single shell of radiation results in an evolving ellipsoid moving through the galaxy from which the reionization emission could be observed. This is true up until the shell reaches (in the host galaxy frame) $r \sim {\cal O}(r_{\rm max})$, after which the emission should decay on timescales $t\sim {\cal O}(\tau_{\rm rec})$.  
    \item This means that if any other observation of the galaxy is taken prior to an elapsed time $\Delta t_{\rm TDE} \sim {\cal O}(r_{\rm max}/c)$, then there will be an observable emission line feature in the TDE host.  For low densities $n_H \sim 10^2\, {\rm cm}^{-3}$ this timescale can be as long as tens of thousands of years, while for higher densities $n_H \sim 10^7\, {\rm cm}^{-3}$ this can be as short as decades. 
    \item For default numbers (above), this corresponds to potentially observable EELR in galaxies with low-ish densities ($n_H \sim 10^2 \, {\rm cm}^{-3}$) and TDE rates ${\cal R}_{\rm TDE} \gtrsim c/r_{\rm str} \gtrsim 10^{-4} \, {\rm yr}^{-1}$. These rates would be at the high end of the inferred TDE rate, and therefore would be overrepresented in a sample of TDEs. The low-ish densities would mean the next TDE is more likely to be unobscured, and therefore detectable.  
\end{itemize}

\subsection{ Line intensities and ionization potentials  }
The large number of ionizing photons emitted from a TDE disk over its lifetime will result in a observable reprocessing line signature, even when superimposed onto background galaxy levels. As an explicit example, the H$_\beta$ line intensity observed from an ionized cloud is given by 
\begin{align}
    L_{H_\beta} &= h \nu_{\rm H_\beta} \left({\alpha^{\rm eff}_{\rm H_\beta} \over \alpha_B}\right) \left({\Delta \Omega \over 4\pi}\right) \int_{\cal R}^\infty {\nu L_\nu \over h\nu }\, {\rm d}\ln\nu \\ 
    &\approx \left({2\over 17}\right) \left({\Delta \Omega \over 4\pi}\right) h \nu_{\rm H_\beta} \left\langle \dot N_{\rm ION} \right\rangle  ,\\
    &\approx  6\times10^{41} \, {\rm erg/s} \, \left({\Delta \Omega \over 4\pi}\right) \left({\eta \over 0.1}\right) \left({\Delta M_{\rm acc} \over 0.5 M_\odot}\right) \left({100 \, {\rm yr} \over t_{\rm life}}\right)  ,
\end{align}
in these expressions $\Delta \Omega$ is the solid angle subtended by the cloud(s) in steradians, and $\alpha^{\rm eff}_{\rm H_\beta} / \alpha_B \approx 2/17$ is the average number of H$_\beta$ photons produced per hydrogen recombination. Even for moderate covering fractions $\Delta \Omega / 4\pi \sim 0.01$ this would be observable above typical background star forming galaxy levels $L_{\rm H_\beta} \sim 10^{39}$ erg/s. 

The (dimensionless) ionization parameter $U$ is defined by 
\begin{align}
U &\equiv {1 \over 4\pi r^2 cn_H} \int_{\cal R}^\infty {\nu L_\nu \over h\nu }\, {\rm d}\ln\nu , \\
&\approx {1 \over 4\pi r^2 cn_H} \left\langle \dot N_{\rm ION} \right\rangle , \\
&\approx 4\times10^{-3}\,  \, \left({1\, {\rm kpc} \over r}\right)^2 \left({10^2 \, {\rm cm^{-3}} \over n_H}\right) \left({\eta \over 0.1}\right) \left({\Delta M_{\rm acc} \over 0.5 M_\odot}\right) \left({100 \, {\rm yr} \over t_{\rm life}}\right) ,
\end{align}
which physically represents the local (at a distance $r$ from the ionizing source) ratio of the number densities of photons to electrons. As a point of reference, the ionization parameter $U$ is typically $-3.2 < \log_{10} U < -2.9$ for local HII regions and star forming galaxies. 

Indeed, for any given atomic line, reprocessed emission will peak at a radius corresponding to a (line-specific) characteristic value of $U$, which we denote $U_{\rm ion}$. As TDE emission is well approximated, on a galactic scale, by a narrow radial shell of photons, this will result in different lines switching on at different radii at which these conditions are satisfied, with a rough scaling $r_{\rm on} \sim n_e^{-1/2}$, i.e., denser clouds will be more efficient re-processors if they are located nearer to the galactic center. As different lines have different characteristic values of $U$ this will correspond to different times post-TDE at which different lines become observerable $t_{\rm on} \sim r_{\rm on}(U)/c \sim U_{\rm ion}^{-1/2} \, n_e ^{-1/2}$. This will be examined in more quantitative detail in future sections.


\section{The observed geometry of emission from an ionizing shell}\label{geom}
In this section we compute the observer-frame geometry which a distant observer would ``see'' when detecting reprocessed emission from an originally narrow shell of ionizing radiation moving through a distant galaxy. The host galaxy may be being observed as a TDE has just been detected in the galactic center, in which  case we are interested in the observed geometry of previous emission from a past TDE which may have occurred within the last $\sim 10^4$ years (i.e., the EELR hosts in Figures \ref{fig:tde_hosts} and \ref{fig:qpe_hosts}), or more generally this may be a galaxy in an optical survey which has relatively recently hosted a TDE. 

\begin{figure}
\centering 
\includegraphics[width=0.6\linewidth]{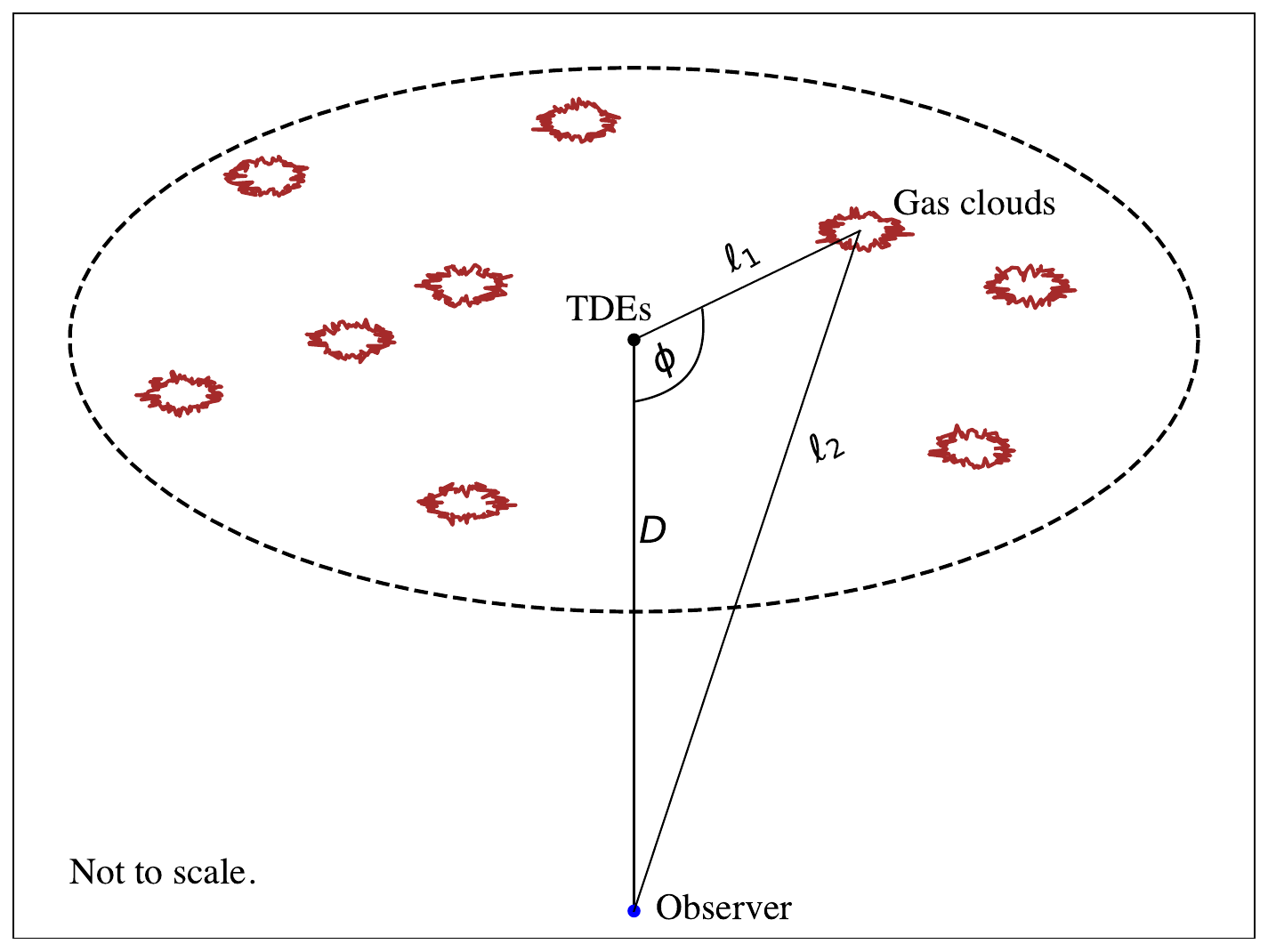}
\caption{A schematic diagram of the photon trajectory geometry. A photon emitted from an earlier TDE can be reprocessed in a gas cloud (by ionizing the gas which then re-combines) and arrive at the location of a distant observer coincident with photons directly emitted from the galactic center. This geometry can be rotated around the plane displayed here, as the different light paths are symmetric about the line joining the observer to the galactic center.   }
\label{schematic}
\end{figure}

\begin{figure}
\centering 
\includegraphics[width=0.48\linewidth]{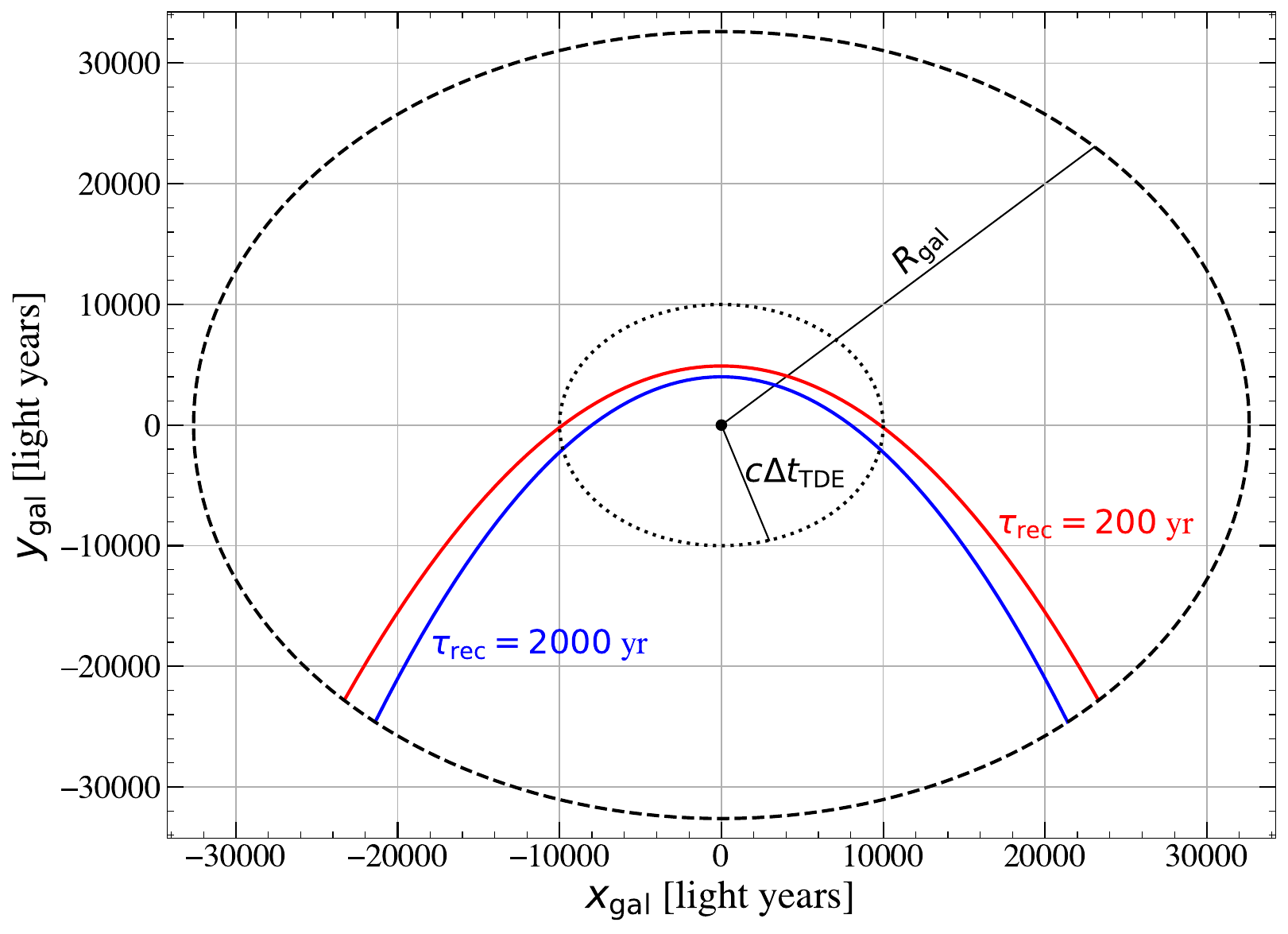}
\includegraphics[width=0.48\linewidth]{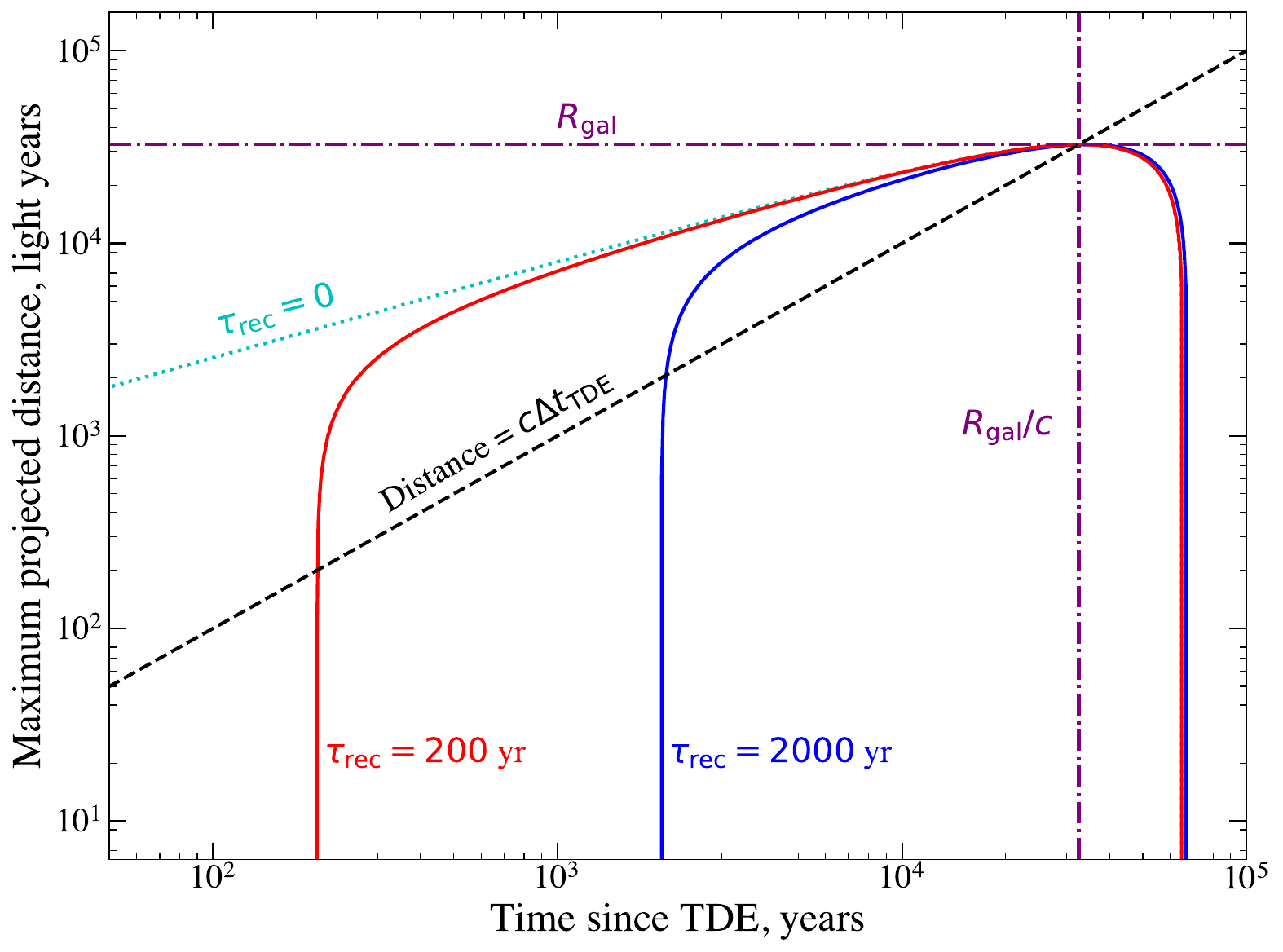}
\caption{{\bf Right:} The locus of points $(x(\phi), y(\phi))$ at which photons from a TDE could be received coincident with direct emission from the galactic center, observed at a time $\Delta t_{\rm TDE} = 10^4$ years later (i.e., the {\it isodelay surface}, which is a paraboloid in these coordinates). The size of the galaxy is taken to be $R_{\rm gal} = 10$ kpc (we assume a spherical galaxy meaning this appears as a circle in this plane). The observer is placed a distance $D = 100$ Mpc below the point $(0, 0)$. Two curves are shown for two different recombination times, $\tau_{\rm rec} = 2000$ years (blue), and $\tau_{\rm rec} = 200$ years (red). Both contours can be rotated by $2\pi$ about the line $x_{\rm gal} = 0$ while remaining a valid isodelay contour, meaning the true three dimensional shape is a cone.  Shown by a dotted curve is the distance light has traveled in the time since the TDE $c \Delta t_{\rm TDE}$, which is significantly smaller than the projected location at which emission could be detected. {\bf Left:} The maximum possible inferred projected distance (from the host galactic center) at which reprocessed emission could be detected coincident with direct emission from the galactic center, as a function of time since the TDE. These inferred sizes are bounded by the assumed size of the galaxy (all parameters are the same as in the top panel), but are typically $X_{\rm max} \gg c \Delta t_{\rm TDE}$. In fact, for nearly all times at which reprocessed emission is detectable, it will be inferred to be at $\sim 10^4$ light years.  The turnover at large  times since the TDE simply results from the assumed finite size of the galaxy, as eventually no reprocessed emission can be detected as the original shell of radiation has passed through the galaxy and all gas clouds have recombined.     }
\label{sizes}
\end{figure}

Consider a shell of radiation emitted at time $t = 0$, which expands at $c$. We then observe the galaxy at a later time $t = \Delta t_{\rm TDE}$, i.e., this defines the time since the onset of the tidal disruption event. Photons emitted from the galactic center at this time will be detected at Earth a distance $D$ from the location of the disruption. Reprocessed emission from earlier in the evolution of the disruption can be detected coincident with direct emission from the galactic center if this previous TDE emission is first absorbed in the TDE host galaxy, ionizing the gas, which then recombines (after a delay of $\tau_{\rm rec}$), emitting photons. The constraint that the time of both journeys must be equal implies 
\begin{equation}\label{constraint}
{\ell_1 + \ell_2 \over c} + \tau_{\rm rec} = {D \over c} + \Delta t_{\rm TDE} , 
\end{equation}
where $\ell_1$ is the distance across the galaxy from the location of the TDE to the gas cloud, and $\ell_2$ is the subsequent distance to the observer from this cloud (Figure \ref{schematic} for a schematic). The left hand side of this expression is the time taken for the reprocessed emission to reach Earth, while the right hand side is the time taken for the direct galactic center emission to reach Earth, with an offset set by the time since the TDE occurred. We need only consider a planar geometry, as the constraints we derive can be rotated by $2\pi$ around the line joining the observer and galaxy center.  Simple trigonometry then relates $\ell_1, \ell_2$ and $D$, namely 
\begin{equation}\label{trig}
\ell_2^2 = \ell_1^2 + D^2 - 2 D \ell_1 \cos\phi, 
\end{equation}
where $\phi \in [0, \pi]$ is the angle between the lines joining the observer to the TDE and the TDE to the cloud (see Figure \ref{schematic}). Combining equations \ref{constraint} and \ref{trig}, and defining the lagged light travel time $T \equiv \Delta t_{\rm TDE} - \tau_{\rm rec}$, we find 
\begin{equation}
\ell_1 = { c  T(c T + 2D) \over 2D(1 - \cos\phi) + 2cT } \approx {c T \over 1 - \cos\phi} .
\end{equation}
The solution $l_1(\phi)$ is known as an {\it isodelay contour}, and in its second approximate form is used in the reverberation mapping community \citep[e.g.][]{Peterson93}. To reach the second approximate solution we have assumed that $D \gg c \Delta t_{\rm TDE}, c \tau_{\rm rec}$, appropriate for all TDE systems which are typically observed at distances ($D \sim 100$ Mpc), far larger than a typical galaxy size ($R_{\rm gal} \sim 10$ kpc) which will bound the typical distance to gas clouds. Note that this expression provides formal negative solutions for $\ell_1$ when $T < 0$ (i.e., the time between two TDEs is shorter than the recombination time), these solutions are unphysical and should be disregarded. We plot the isodelay contours in projected galaxy coordinates in Figure \ref{sizes}\footnote{Note that in these projected galaxy coordinates the isodelay contour is well approximated (when $D\gg cT, R_{\rm gal})$ by the paraboloid $Y + X^2 / 2cT = cT/2$.}. The projected distance from the location of the TDE an observer at distance $D$ would perceive these reprocessed photons to have been emitted from is then given by 
\begin{equation}
X_{\rm proj} = \ell_1 \sin \phi \approx c T \left({ \sin\phi \over 1 - \cos\phi}\right) .
\end{equation}
It is interesting, and important for our purposes, to note that this expression implies that the projected distance can be (substantially) larger than the actual lagged light travel time, provided that the gas cloud is in between the TDE and the observer (i.e., $\phi \lesssim \pi/2$). This means it is  non-trivial to infer the time which has elapsed since the previous flare from spatially resolved emission line regions. Assuming that there are clouds which could be illuminated out to a maximum size $R_{\rm cloud}$ (i.e., $0 \leq \ell_1 \leq R_{\rm cloud}$ for observable reprocessed emission), then the maximum projected size that will be inferred as a function of time is in fact given by 
\begin{equation}
    X_{\rm proj, max} \approx \sqrt{2 R_{\rm cloud} cT} \left[1 - {cT \over 2 R_{\rm cloud}}\right]^{1/2} ,
\end{equation}
where we have again simplified by assuming $D \gg R_{\rm cloud}$. Note that this (i) does not scale linearly with $cT$, but rather primarily with $(cT)^{1/2}$, and (ii) is typically of the order of the cloud-galaxy center distance, {\it independent of $cT$}.  This is an example of apparent superluminal motion \citep{Rees66}, an effect which also occurs in radio observations of blazars \citep{Davis91} and X-ray binaries \citep{Mirabel94}, and has also been observed in galactic optical light echoes \citep[an analogous process to what we are considering here on a smaller scale,][]{Bond03}. 

The above expression for $\ell_1$ also demonstrates the wide range of galactic radii which can be emitting reprocessed emission (at different times in the distant galaxy frame) which are all then observed simultaneously (in the Earth frame) with direct emission from the galactic center emitted at a later point. Indeed, the smallest radii which can be observed simultaneously is $\ell_{1, {\rm min}} = cT/2$. This is a trivial solution, which describes photons which traveled directly away from the observer for half the time lag, before traveling  towards the observer for the second half of the time lag and passing the galactic center at the time at which the direct emission is emitted. Formally the largest possible distance observable is $D + cT/2$, which is another trivial solution corresponding to photons illuminating a cloud $cT/2$ behind the observer, which then travel back to the observer coincident with the direct emission. More relevantly however this implies that, provided $\Delta t_{\rm TDE}>\tau_{\rm rec}$, gas clouds at any galactic distance satisfying $r > cT/2$ from the site of the flare can produce reprocessed emission which is received coincident with direct emission. Therefore a single shell of ionizing radiation will result in reprocessed emission being observed which probes radial scales in the local galaxy ($r_{\rm emit}$), with observed projections from the galactic center ($X_{\rm proj}$), in the ranges 
\begin{equation}
    r_{\rm emit} \in \left[{1\over 2}cT,\,\, R_{\rm gal}\right], \quad X_{\rm proj} \in \left[0,\,\, \sqrt{2 R_{\rm gal} cT} \left(1 - {cT \over 2 R_{\rm gal}}\right)^{1/2} \right] .
\end{equation}
This constraints on $r_{\rm emit}$ therefore bound the maximum time between flares at which detectable reprocessed emission will still be observed coincident with direct emission, namely 
\begin{equation}
    \Delta t_{\rm max} = 2 R_{\rm gal}/c + \tau_{\rm rec} \sim {\cal O}({\rm few} \times 10^4 \,\, {\rm years}) .
\end{equation}
Note that this maximum time is comparable to one over the observed TDE rate. The fact that the typical size of a galaxy in light years is comparable to one over the TDE rate measured in years means that reprocessed emission from the previous TDE should still be detectable when the next TDE goes off in a non-trivial fraction of TDEs.  This is precisely in keeping with observations, although we stress that this requires the presence of clouds of gas with the right properties (i.e., density) so that the emission is still bright on times $t \sim \Delta t$.

We stress that in the last few expressions, and in Figure \ref{sizes}, we have bounded the maximum radial scale of relevant gas clouds with $R_{\rm gal}$, the size of the galaxy. While this is likely the relevant radial scale for low density clouds which may produce EELR, really the relevant scale for any one emission line is set by a combination of $U$, the ionization potential at which the line is brightest, and $n_H$, the density of the clouds. For high density gas clouds near the host galactic center, which will produce transient narrow line regions and extreme coronal lines, the relevant maximum radial scale may be only a few pc, as we shall discuss in later sections.  All of the solutions in this past section are self-similar in the ratio $cT/R_{\rm cloud}$, and so while reprocessed emission from low density clouds may be detectable for $\sim 10^4$ years, high density near-nuclear clouds may only be detectable for $\sim 10$'s of years. 

\section{ Disk evolution and ionizing radiation }\label{disk}
The simple scaling arguments of section \ref{BOE} suggest that tidal disruption event disks will act as sources of significant ionizing radiation for hundreds of years, and the ultimate ionizing energy budget will be set by the incoming stellar mass and will be otherwise broadly independent of system parameters. To test these arguments we use the {\tt FitTeD} code \citep{mummery2024fitted}, which solves the time-dependent relativistic accretion disk equations for the disk temperature, and then computes the rest-frame disk spectrum of the system. These calculations includes all relevant relativistic effects (for both the disk evolution and the propagation of photons through curved spacetimes). 
 
\begin{figure}
    \centering
    \includegraphics[width=0.65\linewidth]{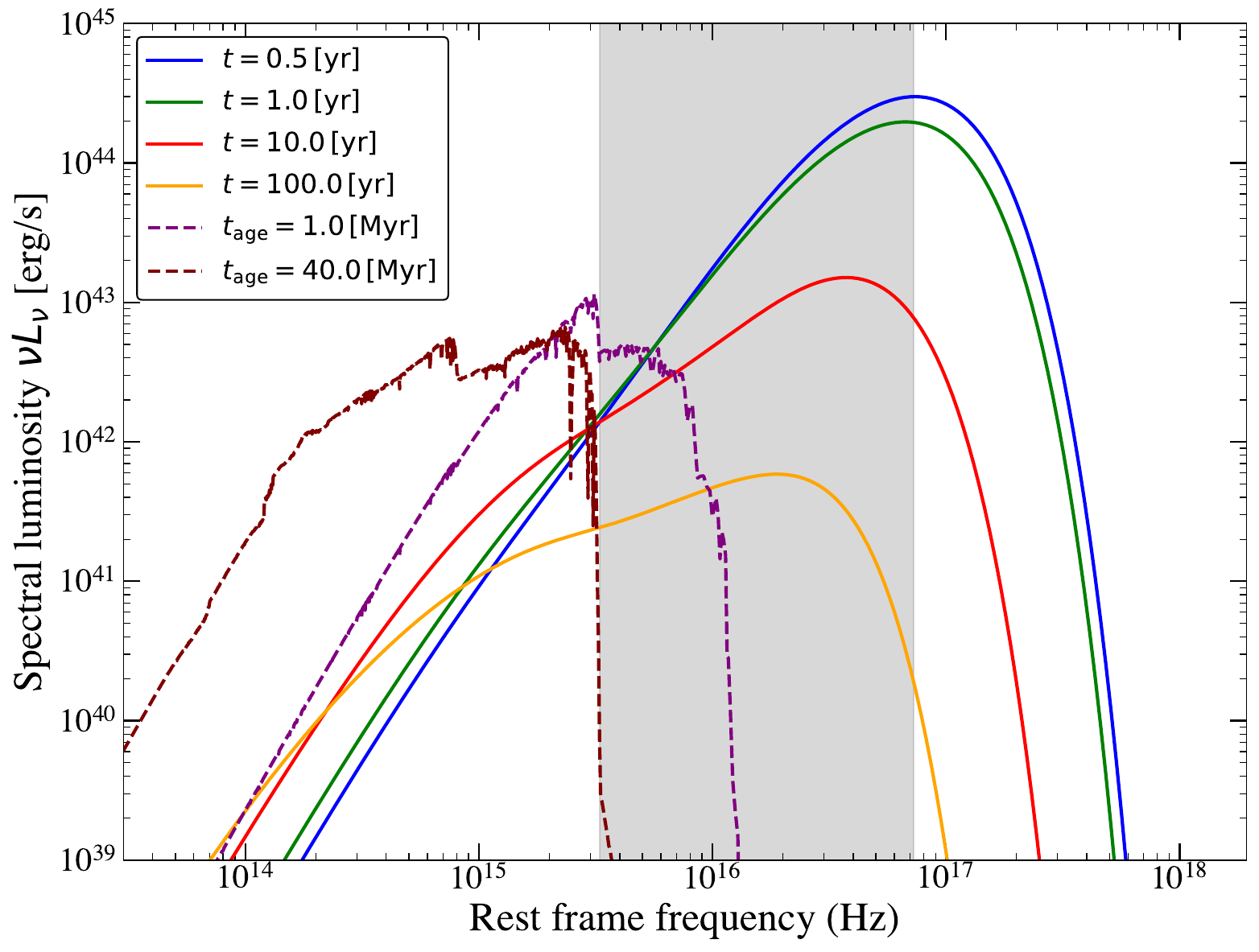}
    \caption{The rest-frame spectra of an evolving TDE disk (solid curves) and stellar clusters of different ages (dashed curves). The luminosity of the disk model is set physically, while the stellar cluster models are normalized to have an integrated luminosity $L_c = 10^{43}$ erg/s. The gray shaded region denotes the energy range of ``ionizing radiation'' and spans  rest-frame photon energies from one Rydberg ($E = 13.6$ eV) up to $E = 300$ eV. It is clear to see that even after 100 years of evolution, TDE disk spectra are harder than even young stellar clusters, and will therefore produce significant ionization signatures in their host galaxies. This is despite the ``viscous'' timescale of this disk being equal to one year in this model.   }
    \label{fig:disk_spec}
\end{figure}

We begin with the spectral energy distribution (SED) of evolving TDE disks. We consider a standard set of TDE parameters, namely a black hole with mass $M_\bullet = 3 \times 10^6 M_\odot$, and spin parameter $a_\bullet = 0.9$, which disrupts a solar-type star $(M_\star = M_\odot, R_\star = R_{\odot})$, of which $M_{\rm disk} = 0.1 M_\odot$ forms into a disk at the circularisation radius. We set the viscous timescale of the disk to be one year, which corresponds to a viscosity nuisance parameter ${\cal V} = 7000$. The SED of this system, which we display in $\nu L_\nu$ units, is plotted against rest-frame frequency for four different times in Figure \ref{fig:disk_spec}. The relevant point of comparison, when considering the ionizing flux of this system, are the SEDs produced by a population of  UV-bright young stars, produced during a period of star formation. To contrast a TDE disk spectrum with the spectrum produced by a star cluster or an HII region, we use the simple stellar population models of \cite{Maraston05}. While the amplitude of the TDE disk spectrum is set self-consistently by the parameters of the system, we normalize these stellar population spectra so that they have an integrated bolometric luminosity of $L_c = 10^{43}$ erg/s, for ease of visual comparison. We define a range of photon energies which we deem ``ionizing radiation'' which spans rest-frame photon energies from one Rydberg ($E = 13.6$ eV) up to $E = 300$ eV. At these photon energies effectively all radiation is absorbed by galactic neutral gas (for example, see e.g., Figure \ref{fig:14li}). This radiation band is plotted as a gray shaded region in Figure \ref{fig:disk_spec}. 

There are a number of important points highlighted in Figure \ref{fig:disk_spec}. Firstly, it is clear to see that even 100 years post disk formation, TDE disk spectra are significantly harder than even young ($\sim 1$ Myr) stellar populations, and will therefore produce significant ionization signatures in their host galaxies. Second, the bulk of radiation produced by a TDE disk is emitted into the ionizing region of the spectrum over its lifetime, meaning that the majority of the liberated accretion energy will power ionization features. Third, the evolution of this ionizing component is slow, meaning that ionizing radiation is still being amply produced a century into the disk evolution, even though the ``viscous'' timescale of the disk is only one year. 

\begin{figure}
    \centering
    \includegraphics[width=0.47\linewidth]{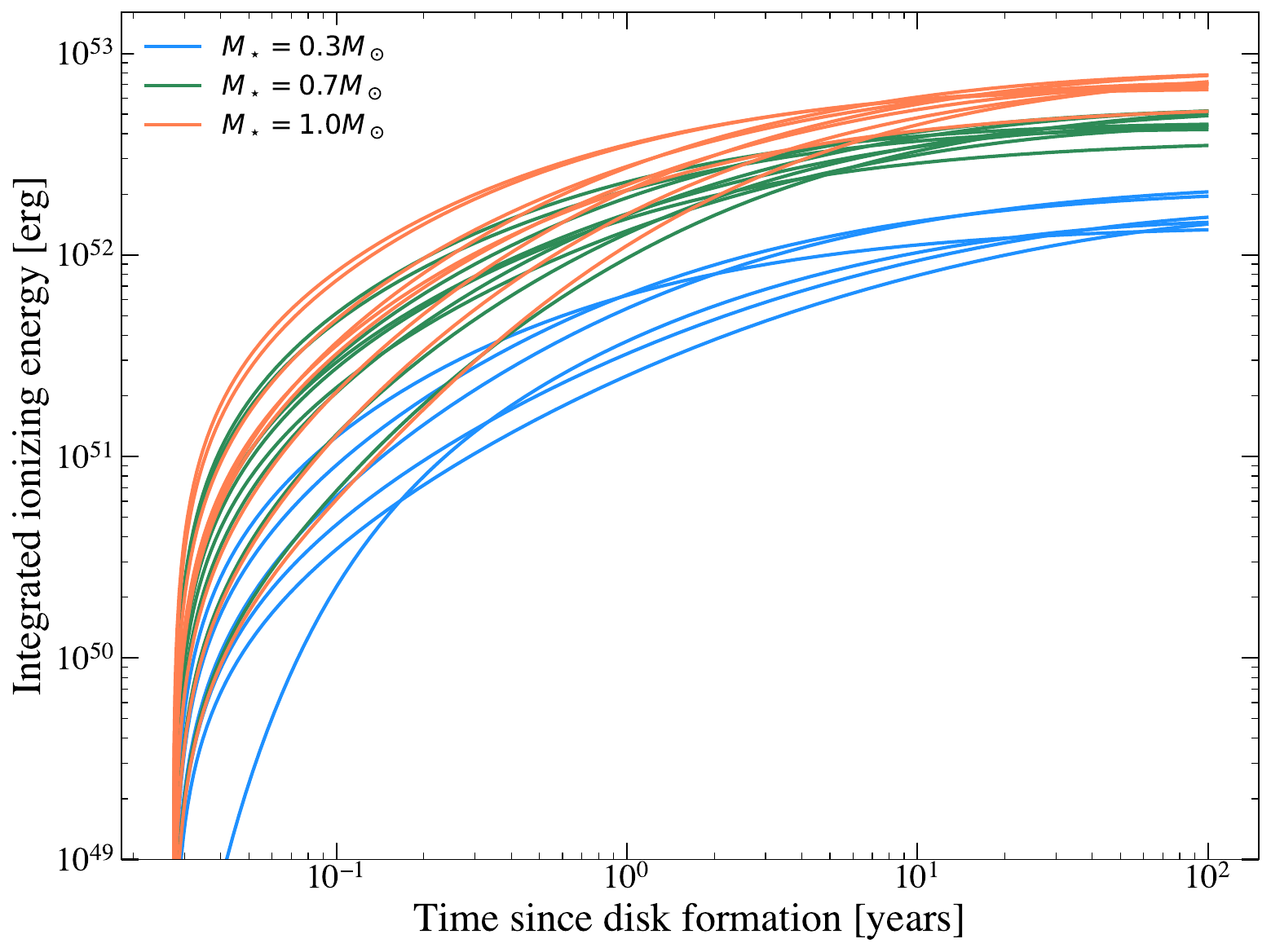}
    \includegraphics[width=0.52\linewidth]{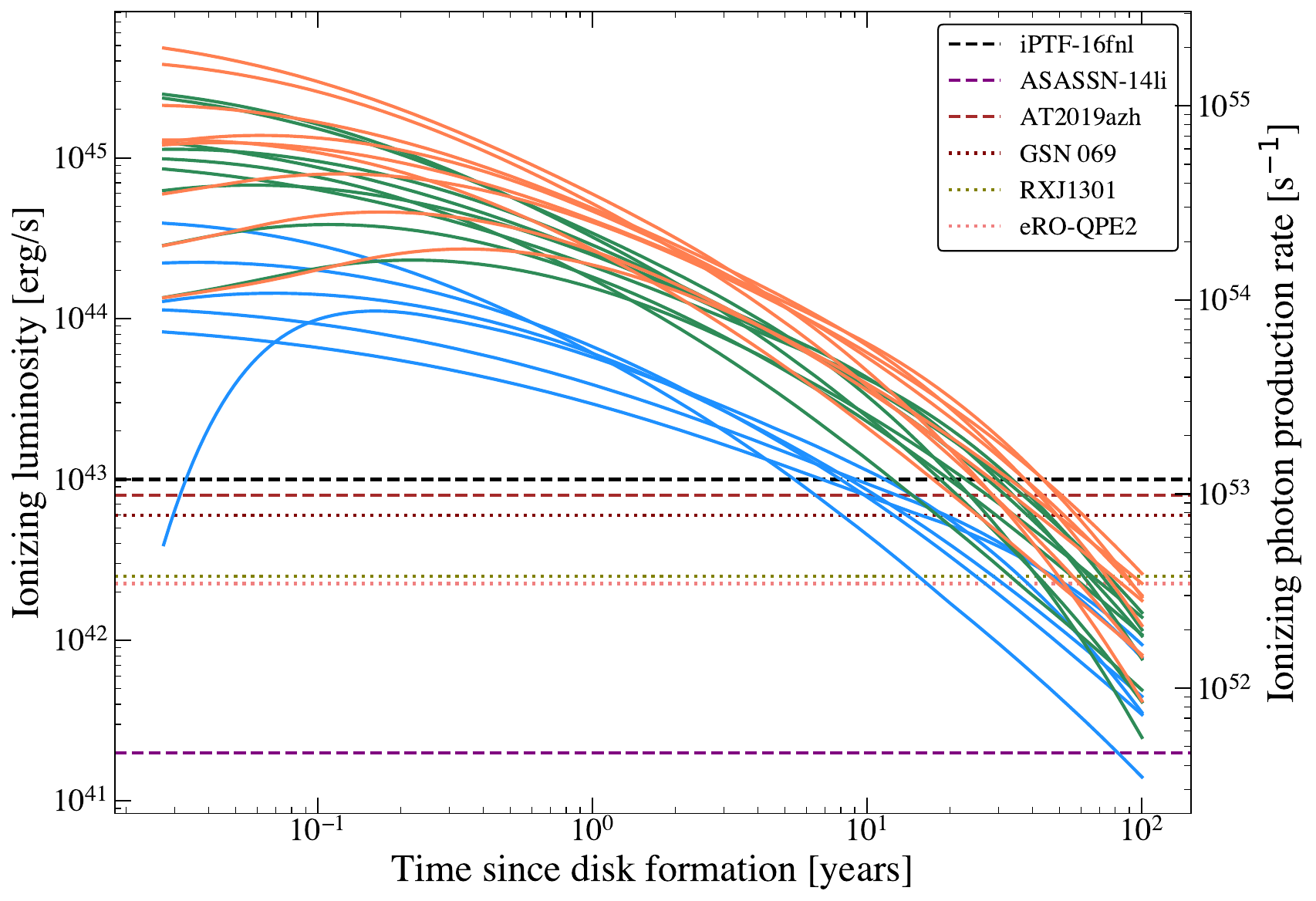}
    \caption{{\bf Left:} the integrated ionizing energy budget of the different TDE disks, as a function of time since disk formation, for a grid of different models. The grid is made up of systems with differing black hole masses, stellar masses and viscous timescales. Different stellar masses are denoted by different colors, with blue representing $M_\star = 0.3 M_\odot$, green representing $M_\star = 0.7 M_\odot$ and orange representing $M_\star = 1M_\odot$. While each system evolves uniquely for the first $\sim$ decade, on the longest timescales the integrated ionizing energy emitted from each system is set primarily by the incoming stellar mass.  {\bf Right:} the ionizing luminosity (left axis) and ionizing photon production rate (right axis) as a function of time for the same grid of disk models. Shown by horizontal dashed (dotted) lines are the minimum ionizing luminosities required to explain observations of EELR in TDE (QPE) host galaxies. Typical TDE disks lie above these minimal luminosities for decades.   }
    \label{fig:disk_evol}
\end{figure}

The second key element of the analytical analysis (section \ref{BOE}) was the prediction that, to leading order, the ionizing energy budget of TDE systems is set by the incoming stellar mass, with all other parameters playing a sub-leading order role. To demonstrate this rather simple fact, we display in Figure \ref{fig:disk_evol} the integrated ionizing energy, defined as 
\begin{equation}
    E_{\rm ION}(t) \equiv \int_0^t L_{\rm ION}(t')\, {\rm d}t',
\end{equation}
for a grid of disk models. In this expression the time-dependent ionizing luminosity of the disk is 
\begin{equation}
    L_{\rm ION}(t') = \int_{E_1}^{E_2} L_E(E, t')\, {\rm d}E, 
\end{equation}
where $E_1, E_2$ are the lower and upper energies of the ionizing energy band (taken to be 13.6 eV and 300 eV respectively in this work), and $L_E(E, t)$ is the SED of the disk system in units of erg/s/eV. We perform a simple grid of models for $M_\star/M_\odot = [0.3, 0.7, 1], M_\bullet/(10^6M_\odot) = [1, 5, 10]$ and ${\cal V}/1000 = [1, 5, 10]$ to verify this result. In the left panel of Figure \ref{fig:disk_evol} we display the integrated ionizing energy budget of each disk system, colored by incoming stellar mass. It is clear to see that, although on the typical observing timescales of an individual system ($\sim $ years) each TDE disk evolution will be unique, after long integration times ($\sim 100$'s of years), the ultimate radiated ionizing energy budget is set almost exclusively by the stellar mass, with very weak dependence on other system parameters. 

We note an important point here, which we shall discuss in more detail in later sections. While the ionizing energy is only weakly dependent on black hole mass once a TDE has occurred (see Figure \ref{fig:disk_evol}), the {\it possibility} of a TDE occurring has an extremely strong suppression for black hole masses above $M_\bullet \gtrsim 10^8 M_\odot$, for which the tidal radius of a typical star is within the event horizon and therefore the disruption is unobservable \cite[e.g.,][]{Hills75}. There may therefore be black hole mass-dependent signatures in the population statistics of TDE-powered ionization signatures, despite this weak energy dependence. We will return to this point later, 

While it is clear that the total radiated energy in ionizing photons is substantial, and the disk spectrum is hard (Fig. \ref{fig:disk_spec}), it is key to verify that the typical amplitude of the ionizing disk luminosity is sufficient to power the observed EELR features in known TDE/QPE host galaxies. A minimum ionizing luminosity can be estimated from an observed EELR by measuring its size and current line emission strength. Such estimates are provided by \cite{Wevers24EELR} and \cite{WeversFrench24} for six host galaxies of TDEs and QPEs (some of whom are displayed in Figures \ref{fig:tde_hosts}, \ref{fig:qpe_hosts}), and are reproduced as horizontal dashed (TDE) and dotted (QPE) lines in the lower panel of Figure \ref{fig:disk_evol}. The ionizing luminosity of the grid of disk solutions is also displayed (again the colors indicate different stellar masses, other parameter values are not explicitly displayed). We see that again there is significant variance in the emission over the first few $\sim$ years, but that all systems comfortably exceed the observed constraints on the ionizing luminosity, and the late time emission is set primarily by stellar mass, with all systems producing emission of $\sim 10^{43}$ erg/s (or equivalently $\sim 10^{53}$ ionizing photons per second) for at least a decade. 

As it is clear that TDE disks produce SEDs with the requisite properties to power various observed features of TDE host galaxies (EELR, coronal lines, etc.), we now examine the spectral signatures which would be expected from reprocessed TDE disk emission for a wide range of different emission lines, cloud parameters and times since emission.

\section{ Extended emission line regions }\label{cloudy1}
In this section we first compute the observer-frame luminosities of the optical light echoes of TDE disks, before determining whether these TDE-powered ionization signatures  are sufficiently bright so as to be observable in addition to typical galaxy emission. 

\subsection{Optical light echoes of TDE disks}
To examine the possible line signatures produced from gas clouds ionized by a previous TDE disk,  we use the {\tt CLOUDY} \citep{Cloudy98} code to simulate the reprocessing of the accretion disk spectrum by neutral gas assumed to pervade the host galaxy. {\tt CLOUDY} assumes that the gas and radiation fields are in a statistical steady-state, something we have discussed is unlikely to be true for all of the relevant atomic species for {\it low density} clouds $n_H \lesssim 10^3 \, {\rm cm}^{-3}$ (owing to the comparable lengths of the TDE disks ionizing lifetime and the recombination of some atomic species in low density clouds which are required for EELR observations, for example). We stress that the higher density clouds we consider later are robustly in statistical steady state.   As such we do not attempt to directly match to observations in this work, but simply examine what broad properties the reprocessed emission may have. A fully time dependent study (which includes a time varying input spectrum and full time dependent ionization calculations) is of real interest, and will be postponed to future work. Only with fully time dependent analyses can comparisons to specific observations be made. 

We begin this study by determining the effects of varying four parameters (which we believe will be the principle parameters of interest) on the line luminosities of the reprocessed emission. Two of these parameters, namely the average density of hydrogen in the galaxy $n_H$ and the volume filling factor $f_V$, describe properties of the neutral gas in the galaxy, while two describe properties of the disk spectrum (namely its luminosity and spectral shape). 

\begin{figure}
    \centering
    \includegraphics[width=0.49\linewidth]{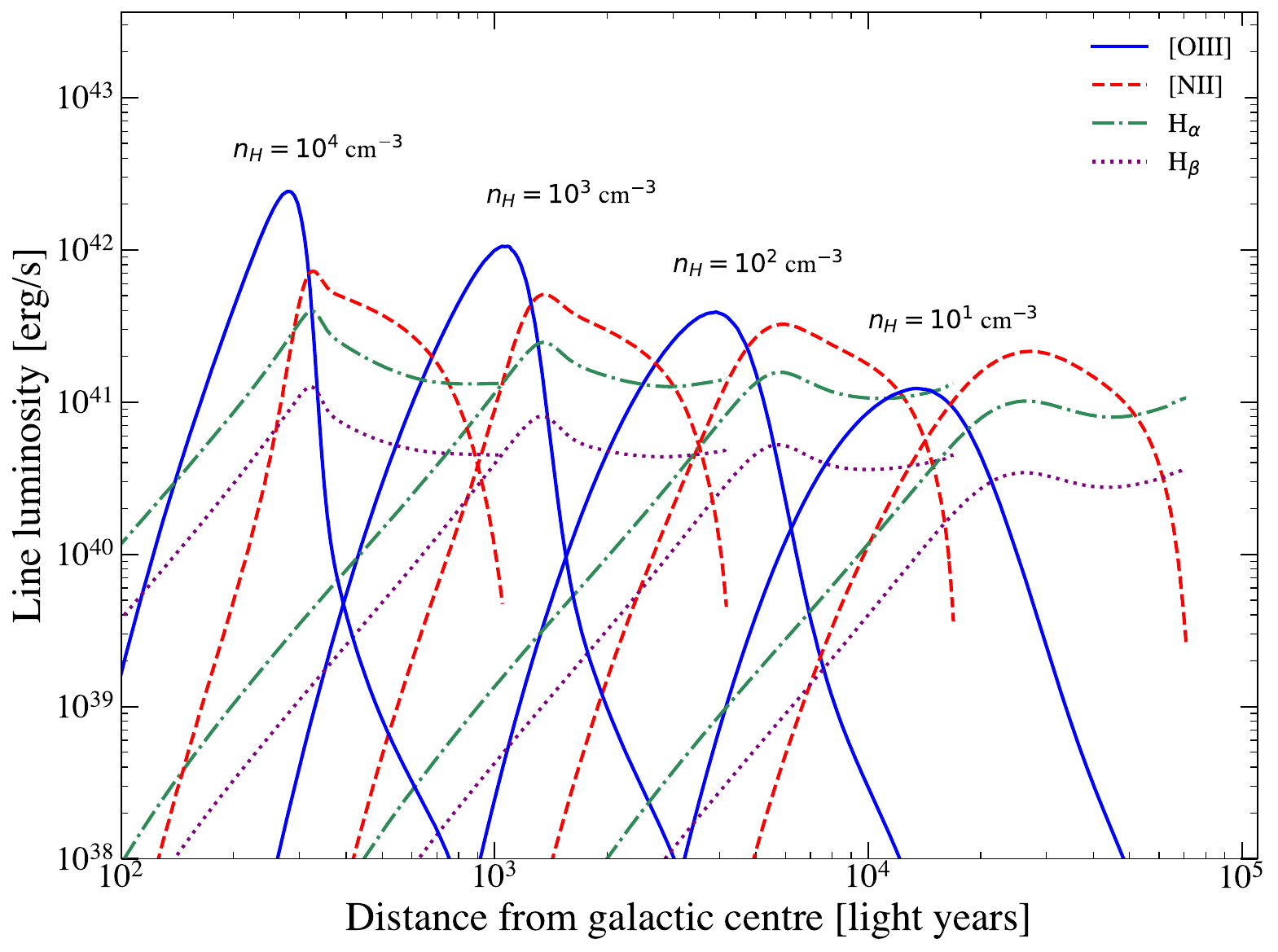}
    \includegraphics[width=0.49\linewidth]{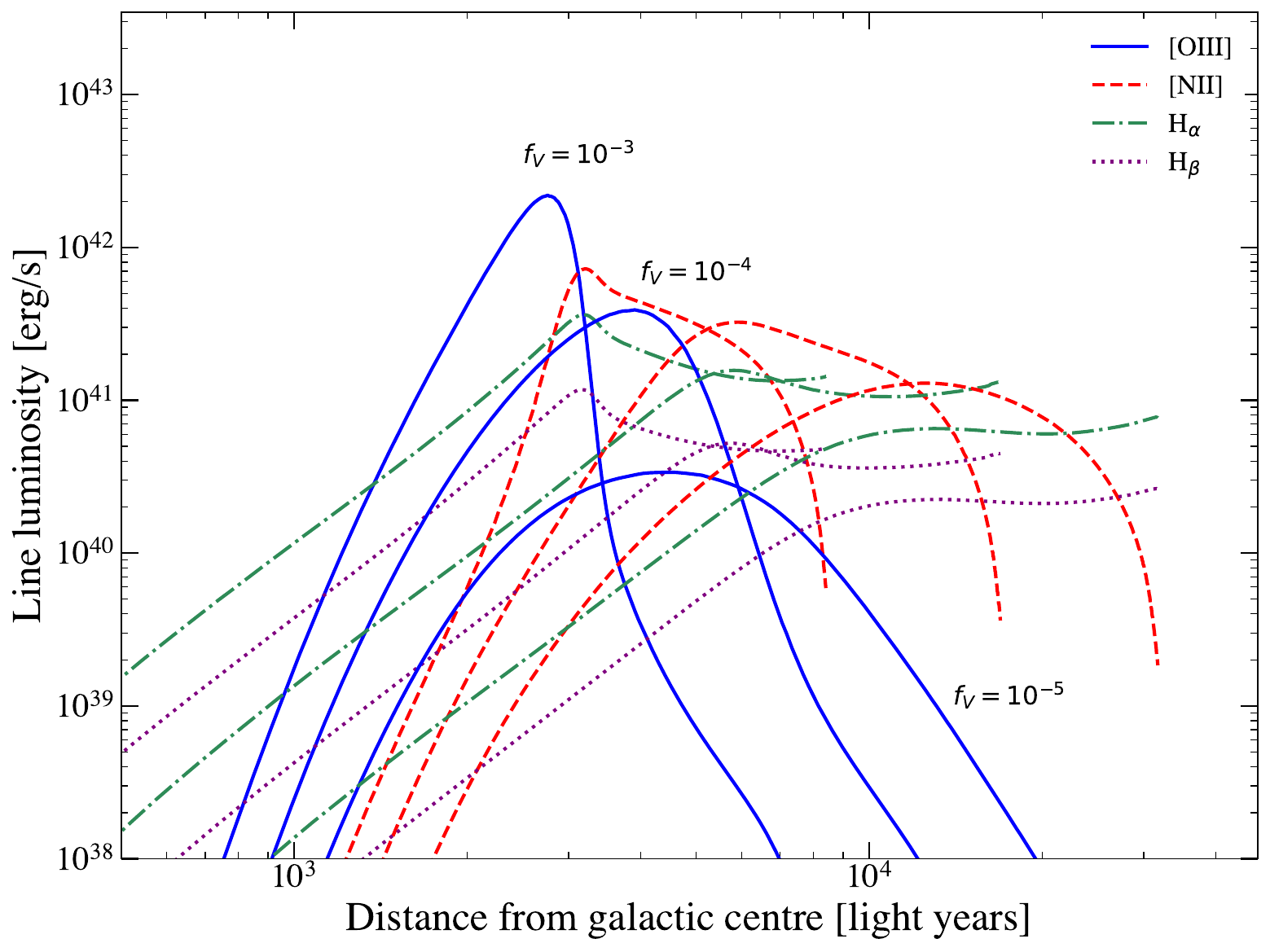}
    \includegraphics[width=0.49\linewidth]{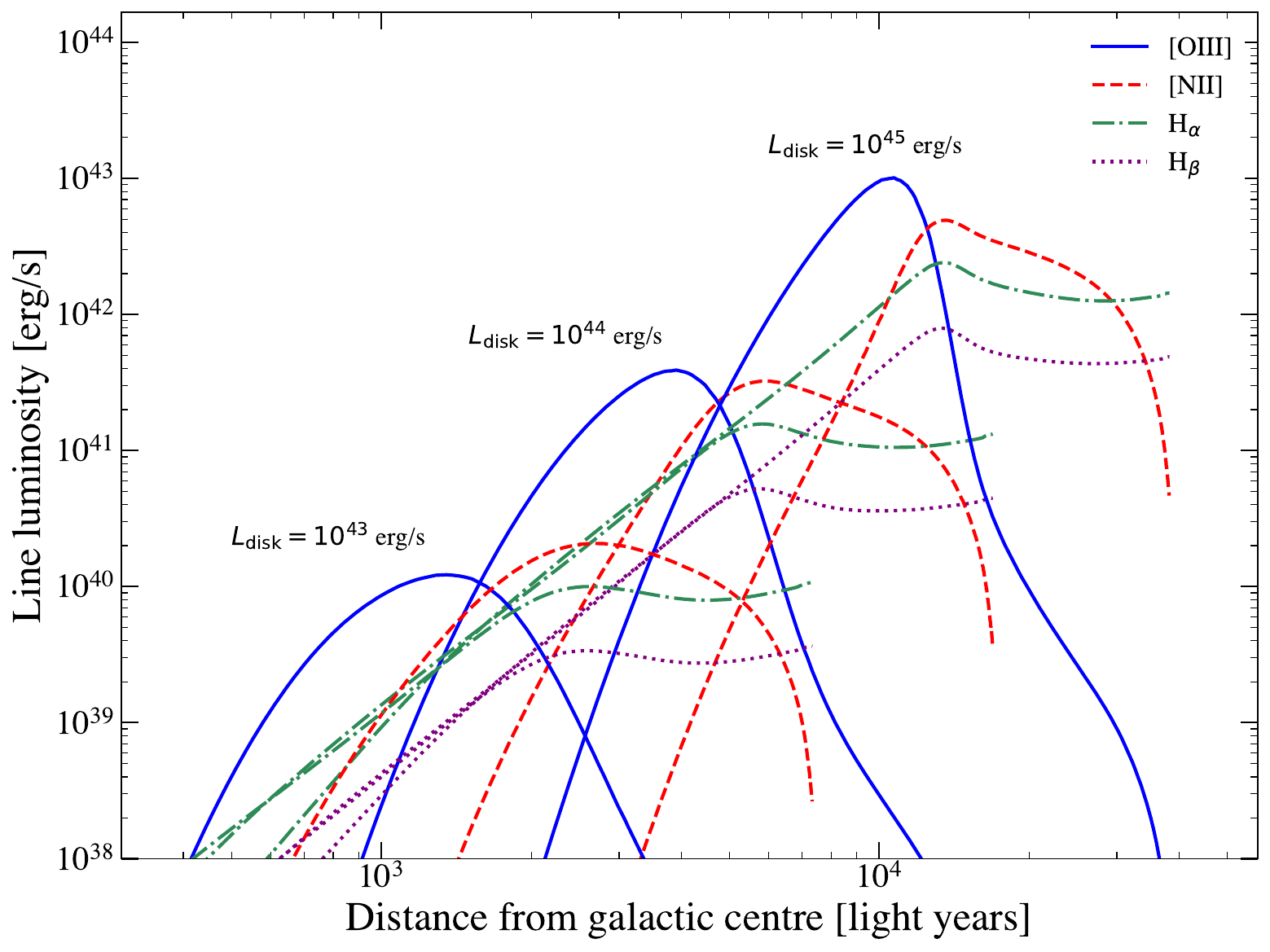}
    \includegraphics[width=0.49\linewidth]{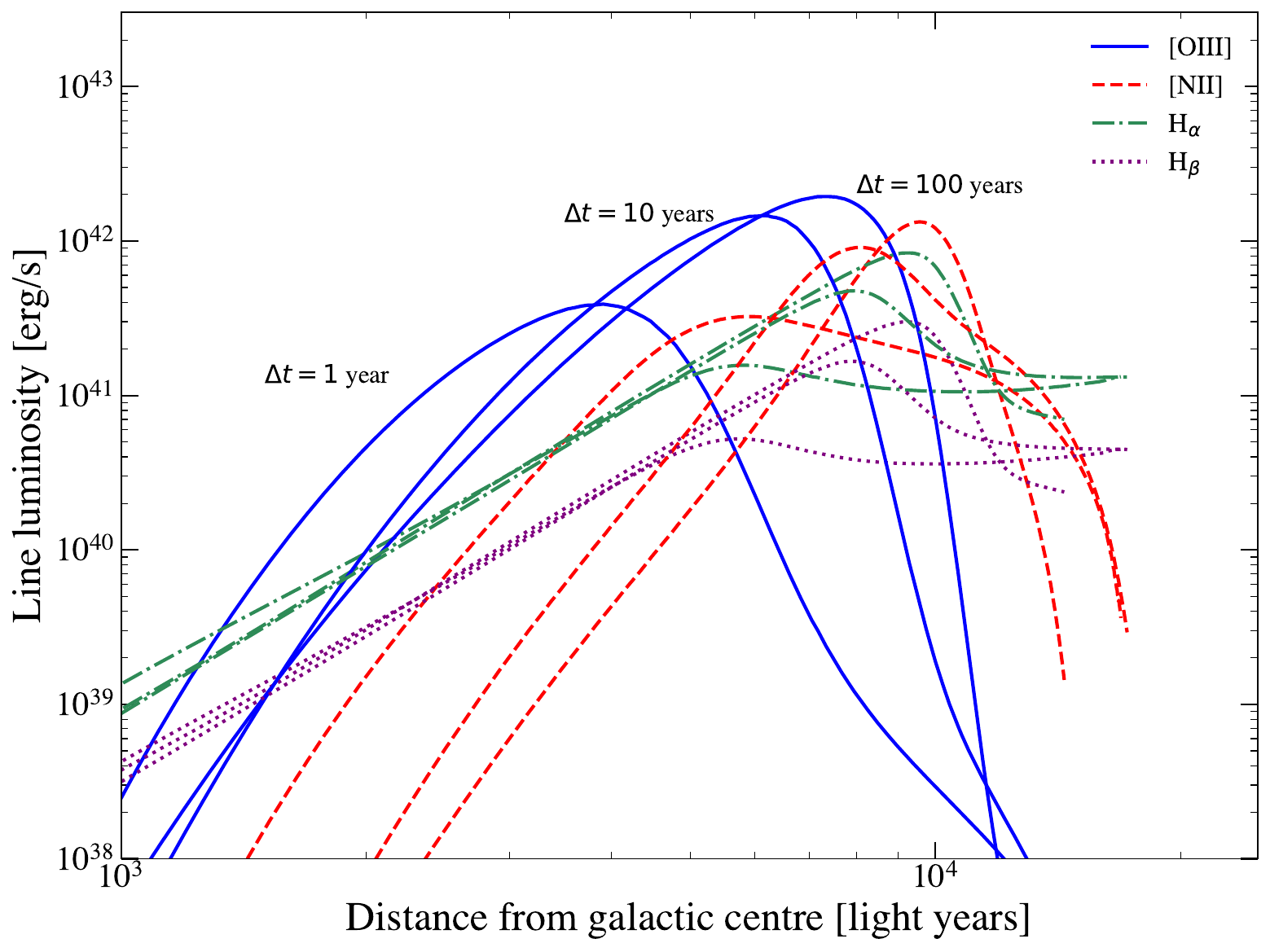}
    \caption{Line luminosities of relevant atomic species, as a function of distance from the galactic center for differing disk and cloud parameters, computed using the {\tt CLOUDY} code. Unless otherwise specified the default parameters in each plot are $f_V = 10^{-4}, \,n_H = 10^2 \,{\rm cm}^{-3}, \, L_{\rm disk} = 10^{44} \, {\rm erg/s}$ and $\Delta t = 1$ year. {\bf Upper left:} the effects of varying the hydrogen density $n_H$ on the line luminosities, denser clouds produce brighter line emission which is contained within a closer neighborhood of the galactic center. {\bf Upper right:} the effects of varying the volume filling factor $f_V$. Sparser populations of clouds (lower $f_V$) produce less emission which is slight further out in the galaxy. {\bf Lower left:}  the effects of increasing disk luminosity. The brighter the disk, the brighter and further out in the galaxy the line emission which is observed. {\bf Lower right:} the effect of disk spectral shape. As TDE disks age their spectrum changes shape, and disks in the later stages of TDE evolution ionize gas out to larger radii. Note this final plot is made with {\it fixed disk luminosity}, a choice made to to isolate this specific effect. In reality the evolution of a single system will be a combination of the lower two panels (cooling and dimming). We stress that the luminosities here are naive $L = 4\pi f_V r^3 j/3$ computations, which is {\it not} what a distant observer would see (see text for more discussion of this point).   }
    
    \label{fig:cloudy1}
\end{figure}

As we are first interested in general trends across different parameter spaces, we vary these four parameters independently. This of course is not realistic in a real TDE disk system (where, for example, the luminosity and spectral shape of the disk will be fundamentally coupled by the disk evolutionary equations, i.e., mass and angular momentum conservation), but it will aid in gaining intuition. In Figure \ref{fig:cloudy1} we display the line luminosities of four relevant lines ([OIII], [NII], H$\alpha$ and H$\beta$) as a function of distance from the galactic center (or more specifically the ionizing source which we assume is located in the galactic center). For a given input spectrum, luminosity and cloud parameters $n_H, f_V$ {\tt CLOUDY} computes the emissivity $j_{\rm line}$ of various atomic lines as a function of radius. The line luminosity is then 
\begin{equation}
    L_{\rm obs, \, line} = \iiint_{{\cal V}_{\rm obs}(t)} \, f_V \, j_{\rm line}(r) \, {\rm d}^3V,
\end{equation}
where ${\cal V}_{\rm obs}(t)$ is the total emitting volume of the clouds (as observed by the distant observer which in our case is a function of time), ${\rm d}^3V$ is a volume element, and $j_{\rm line}$ is assumed to vary only with the distance from the illuminating source $r$. In a strictly steady state system, with spherical symmetry, this would simply be an integral over a sphere, and for simplicity the curves displayed in Figure \ref{fig:cloudy1} are computed from $L_{\rm line} = 4\pi f_V r^3 j_{\rm line}/3$. This highlights the broad luminosity structure of the emitting regions, but is {\it not} what a distant observer would see in a time dependent system.  A more careful calculation will be performed later. 

In each subplot of Figure \ref{fig:cloudy1} we vary one of the four parameters mentioned above, with the remaining parameters fixed to $n_H = 10^2 \, {\rm cm}^{-3}$, $f_V = 10^{-4}$, $L_{\rm disk} = 10^{44}$ erg/s (the integrated bolometric luminosity of the disk), and a spectral shape given by the disk spectrum of Figure \ref{fig:disk_spec} one year post disk formation. We display each line luminosity with a curve of the same color and line style for each parameter, and label on the plot the value of the parameter which has been varied. These labels are placed close to the [OIII] line, and each set of four curves always have the same qualitative features and simply shift, as a group, as parameters are changed.

The radial dependence of the line luminosities is simple to understand as a function of the free parameters of the model. For example, denser clouds (larger $n_H$, upper left panel) produce brighter line emission which is contained within a neighborhood which is closer to the galactic center (as they have smaller Stromgren radii), while sparser populations of clouds (lower $f_V$, upper right panel) produce less emission which is slightly further out in the galaxy. The effects of disk properties are also simple to analyze, the brighter the disk, the brighter and further out in the galaxy the line emission which is observed (lower left). Finally, as TDE disks age their spectrum changes shape (the spectra soften as the inner disk temperature drops in response to mass draining through the inner disk edge), as can be seen in Figure \ref{fig:disk_spec}. In the lower right plot of Figure \ref{fig:cloudy1} we use the different disk spectra produced at different stages in the TDE evolution as inputs into {\tt  CLOUDY} to probe the implications of this evolution. We find that disks in the later stages of TDE evolution ionize gas out to larger radii (lower right, i.e., softer TDE spectra typically ionize gas at larger radii). Note this final plot is made with {\it fixed disk luminosity}, a choice made to to isolate this specific effect. In reality the evolution of a single system will be a combination of the lower two panels (cooling and dimming).   

Examination of Figure \ref{fig:cloudy1} shows that if we are to explain the observed EELR with radial scales $\sim 10^3 - 10^4$ light years (projected) distance from the galactic centers, the relevant parameter space to consider is hydrogen densities in the range $\log_{10}n_H \sim [1,3]$, and volume filling factors in the range $\log_{10}f_V \sim [-4, -5]$. These hydrogen column densities correspond to typical molecular cloud densities, and should be common in many galaxies.  The luminosities and spectral shapes are of course not free parameters here, but are set self consistently by TDE disk theory. 

In this first analysis of radiative feedback from TDEs, we shall restrict ourselves to a general analysis of the properties of these systems, and as such will use time-averaged spectra and luminosity values, postponing a full time-dependent treatment to a future work. We define the average SED of the TDE disk as 
\begin{equation}
    \left\langle \nu L_\nu \right\rangle \equiv {1\over \Delta T}\int_0^{\Delta T} \nu L_\nu(\nu, t')\, {\rm d}t',
\end{equation}
and normalize the amplitude of this spectrum (in erg/s in {\tt CLOUDY}) by the average luminosity of the disk over the same time period
\begin{equation}
    \left\langle L\right\rangle \equiv {1\over \Delta T} \int_0^{\Delta T} L_{\rm bol}(t')\, {\rm d}t',
\end{equation}
(i.e., we set the amplitude so that $\int \left\langle \nu L_\nu\right\rangle {\rm d}\ln \nu = \left\langle L \right \rangle$). We take $\Delta T = 10$ years, which is a conservative observationally-driven lower bound on the lifetime of these disks, although we remind the reader that significant additional radiation is likely  to be emitted for hundreds of years.

\begin{figure}
    \centering
    \includegraphics[width=0.65\linewidth]{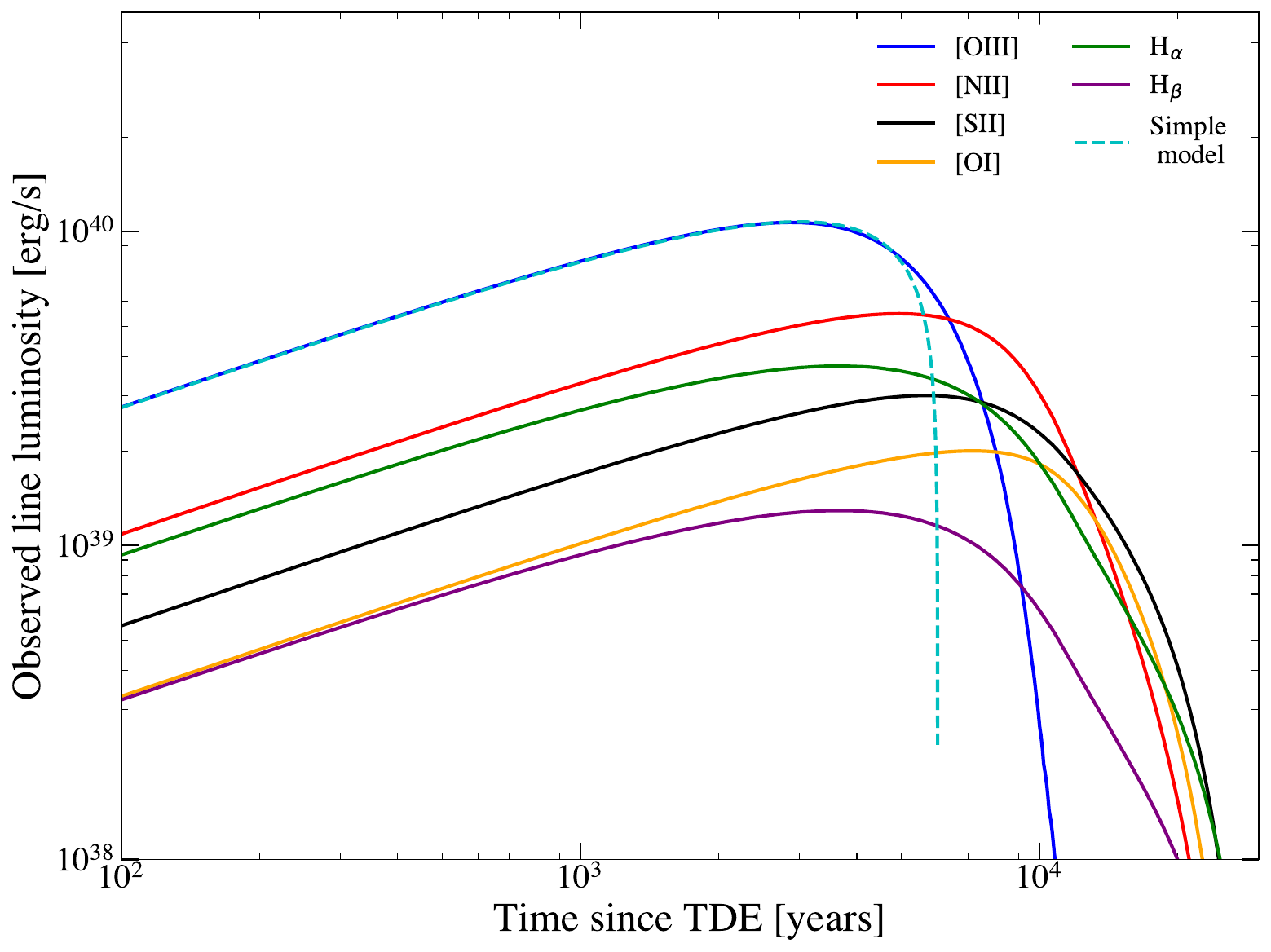}
    \caption{Evolving line luminosities, as seen by a distant observer, as a function of time since the TDE occurred (i.e., time since the first light from the TDE passed the observer in their frame, see eq. \ref{full}).  This figure is made assuming $n_H = 10^2 \, {\rm cm}^{-3}$ and $f_V = 10^{-4}$. Both the amplitude and duration of the light echo are  effected by these two parameters, and the assumed luminosity of the original disk, and so this figure should only taken to be illustrative of the general light echo evolution. By a dashed curve we show an example of the simple analytical model derived here (eq. \ref{simple}).   }
    \label{fig:light_echo}
\end{figure}

Taking the disk model which produced the spectra seen in Figure \ref{fig:disk_spec}, which corresponds to an average disk luminosity over the first decade of evolution of $\log_{10} \left\langle L \right\rangle = 43.85$ (erg/s), we use {\tt CLOUDY} to compute the radius-dependent line emissivities $j_{\rm line}$ for [OIII], [NII], [SII], H$\alpha$ and H$\beta$. These line luminosities are used as diagnostics for the properties of the ionizing source, and can be used to distinguish stellar from non-stellar continua. 

To compute the luminosities which a distant observer would detect for a given TDE-induced line emissivity profile, we must integrate this emissivity over the volume from which a distant observer would detect reprocessed photons simultaneously, emitted over the entire host galaxy. In other words we need to compute the volume bound by the interception of the three-dimensional iso-delay contour and the host galaxy. When considering the properties of EELR we shall assume that the TDE that ionized the gas on large scales happened sufficiently long ago (in the host galaxy frame) that all of the disc emission has reached large scales in the host galaxy. Using the results derived earlier, namely that the distance from the galactic center to the iso-delay contour is $\ell = cT/(1 - \cos \phi)$, and that the three dimensional iso-delay curve is described by two coordinates $x = \ell \sin\phi$, $y = -\ell \cos \phi$ and is symmetric about the observer-galaxy center axis, the observed luminosity is 
\begin{align}
    L_{\rm obs, \, line} &=  \int_{x(T)}^{x(T+\Delta t_{\rm ION})} \int_{y_{\rm min}}^{y_{\rm max}} 2\pi \, x' \, f_V \, j_{\rm line}(\ell) \, {\rm d} y \, {\rm d}x' , \\ 
    &\approx 2\pi c \Delta t_{\rm ION} \int_{y_{\rm min}}^{y_{\rm max}} \, x \, f_V \, j_{\rm line}(\ell) \, {\rm d} y  , 
\end{align}
where we have assumed that the shape of the iso-delay contour is broadly unchanged at the inner and outer edge of the shell of radiation emitted from the TDE (i.e., $x(T+\Delta t_{\rm ION}) \approx x(T) + c\Delta t_{\rm ION}$, where $c\Delta t_{\rm ION} \ll x(T)$) and the rest of the expression follows from a standard cylindrical coordinate system (where in our notation $x$ is the radius out from the axis of rotation $y$). Making the variable substitution $x = cT \sin \phi /(1- \cos\phi) $ and $ y = -cT\cos\phi/(1-\cos\phi)$ leads to 
\begin{equation}\label{full}
    L_{\rm obs, \,  line} = 2\pi c^3 T^2 \Delta t_{\rm ION}  \int^{\phi_\star(T)}_\pi {\sin^2 \phi \over (1 - \cos\phi)^3} \,f_V\, j\left({cT \over 1-\cos\phi}\right) \, {\rm d}\phi , 
\end{equation}
where $\phi_\star$ corresponds to the angle on the iso-delay contour made between the outer edge of the galaxy and the observers line of sight, namely 
\begin{equation}
    {cT \over 1 - \cos\phi_\star } = R_{\rm gal} \to \phi_\star = \arccos\left(1 - {cT \over R_{\rm gal}}\right) . 
\end{equation}
Note that in principle $f_V$ should also be considered a function of $\phi$ as, for example, a constant $f_V$ is only strictly valid for a spherical galaxy, and the interception between a disc galaxy and the three-dimensional iso-delay contour would have regions of zero emissivity which could in principle be well modeled in each specific case by $f_V(\phi)$. 

This integral can even be evaluated in closed form, under the approximation that the emissivity is strongly peaked at a characteristic radius (this can be seen to actually be the case for the {\tt CLOUDY} simulations in Figure \ref{fig:cloudy1}). Approximating this peaking with a delta function $j_{\rm line}(\ell) \approx j_0 \delta(\ell/\hat R - 1)$, the integral simplifies down to 
\begin{equation}\label{simple}
    L_{\rm  obs, \, line}(T) \approx 2\pi f_V j_0 \hat R^3 \left({c\Delta t_{\rm ION} \over \hat R} \right)\left({2 c T \over \hat R}\right)^{1/2} \,  \left[1 - {cT \over 2 \hat R}\right]^{1/2} ,
\end{equation}
i.e., a characteristic $\propto T^{1/2}$ rise, before sharply cutting off at $T_{\rm max} \approx 2 \hat R/c$. Note that both the line luminosity and projected distance from the center of the host galaxy have the same functional form $f(t) \propto t^{1/2} (1- t/t_0)^{1/2}$, meaning that the time at which the observed line emission is brightest (and therefore naturally most likely to be detected) is also the time at which it will appear to be furthest from the galactic center. This may be a natural explanation for why EELR are typically inferred to have large projected offsets from their host galactic centers. 

Note that this result is {\it very} different to the time evolution one would infer by not taking into account the result that the observed luminosity is restricted to points in the galaxy lying along the iso-delay contour. Indeed, if one were to substitute $r \propto cT$ into the naive luminosity of the line $L_{\rm line} \propto r^3 j(r)$ one would infer line luminosities which grow as strongly as $\sim T^{10}$ \citep[e.g.][]{Patra24}.  The finite speed of light has incredibly important implications for the time evolution of the observed ionization fronts. 

In Figure \ref{fig:light_echo} we display the evolution of a typical TDE disk light echo, made assuming $n_H = 10^2 \, {\rm cm}^{-3}$ and $f_V = 10^{-4}$. Note that this light echo is only intended to be illustrative, as both the amplitude and duration of the light echo are effected by the assumed cloud parameters, and the luminosity of the original disk. We see that individual lines have a range of both brightnesses and evolutionary properties, with some remaining brighter for a factor $\sim 2$ longer than [OIII], with varying levels of peak luminosity.  By a dashed curve we show an example of the simple analytical model derived here (eq. \ref{simple}), which describes the evolution well until the emission fades. The discrepancy at these late times is due to the finite radial width of the real radial emissivity profiles, not captured in the simple model. 


\begin{figure}
    \centering
    \includegraphics[width=0.65\linewidth]{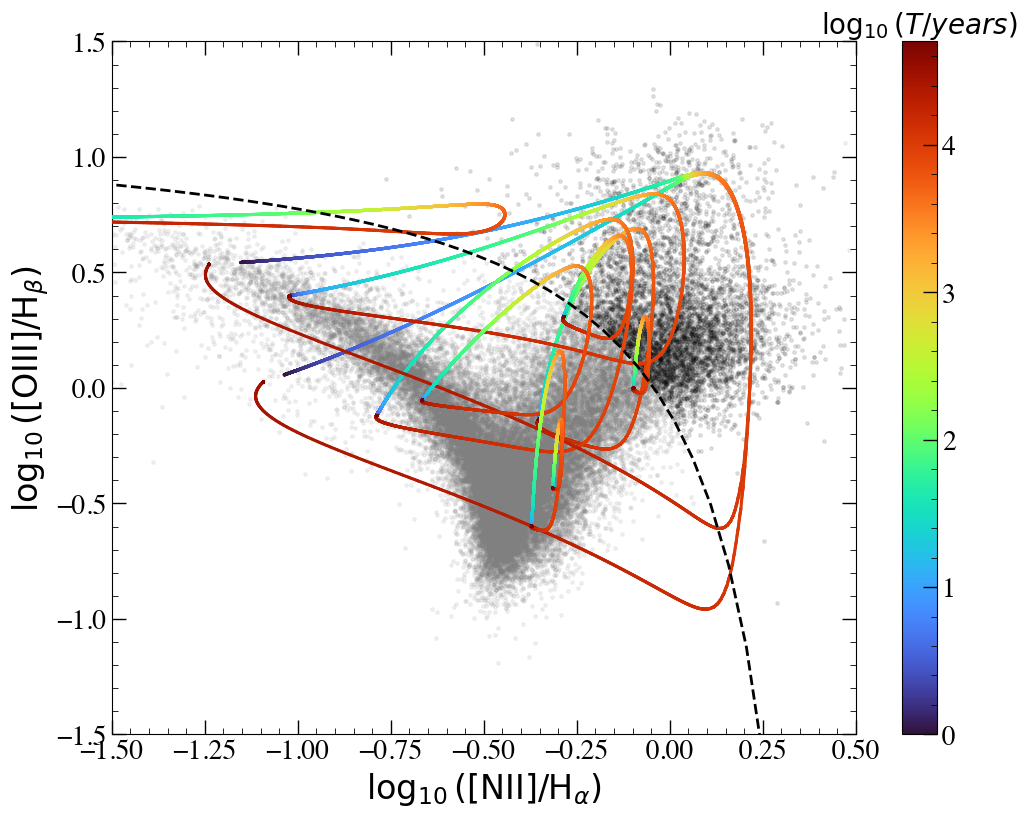}
    \caption{The evolution of the BPT diagram of 10 galaxies in the SDSS survey as a function of time since a TDE produced an ionizing radiation field. The evolution of each galaxy is displayed by a coloured line, with the colour denoting the ($\log_{10}$ of the) time since the TDE (see colourbar). Also plotted are 100,000 SDSS galaxies, with those lying under the Kewley line plotted in gray while those above the Kewley line plotted in black. The introduction of an ionizing radiation field sourced from a TDE results in dynamic evolution of the galaxy on (astronomically) short timescales, and for a few thousand years (for this particular choice of parameters) results in a galaxy entering the ``AGN'' region of the BPT diagram. These ten galaxies have been chosen to show a range of possible behaviors, and are not representative of the average (or a random) galaxy.   }
    \label{fig:bpt_tracks1}
\end{figure}
\begin{figure}
    \centering
    \includegraphics[width=0.48\linewidth]{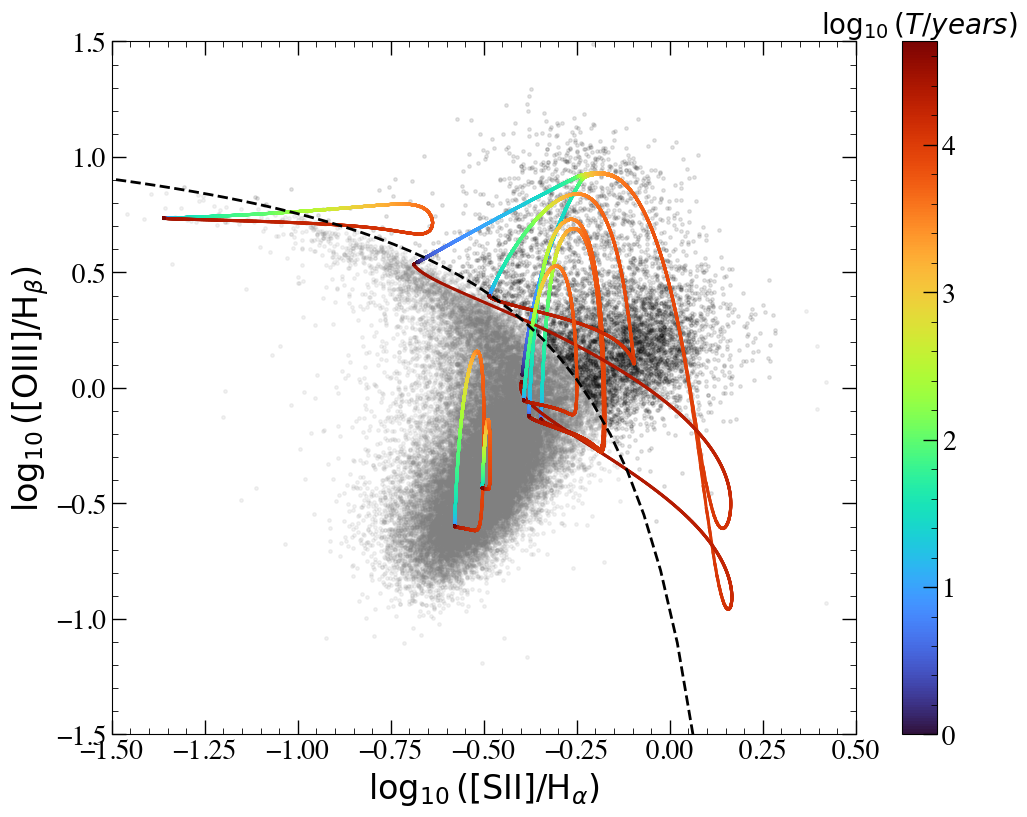}
    \includegraphics[width=0.48\linewidth]{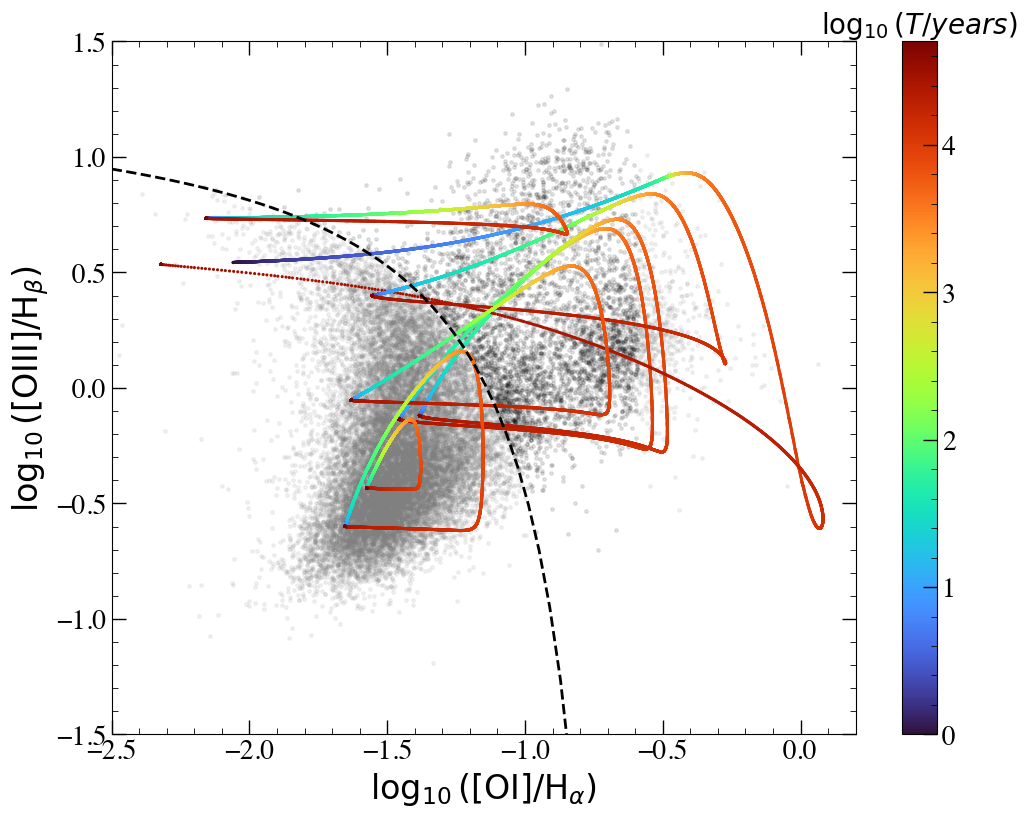}
    \caption{Same as Figure \ref{fig:bpt_tracks1} but for different diagnostic diagrams, namely the [SII] (upper) and [OI] (lower) diagrams. As in Figure \ref{fig:bpt_tracks1}, SDSS host galaxies are displayed in grey if they lie below the (relevant) Kewley line, and by black if they lie above that line. We see that the introduction of an ionizing radiation field sourced from a TDE results in dynamic evolution of the galaxy on (astronomically) short timescales, and for a few thousand years (for this particular choice of parameters) results in a galaxy entering the ``AGN'' region of the BPT diagram, and that this result is independent of the diagnostic diagram used. These ten galaxies have been chosen to show a range of possible behaviors, and are not representative of the average (or a random) galaxy.    }
    \label{fig:bpt_tracks2}
\end{figure}

\subsection{The detectability of TDE optical light echoes}
The evolution of these light echoes is interesting, but it is important to determine what would be observed {\it on top of} typical galaxy emission in these lines (i.e., does the addition of a single TDE make any difference to the observed line strengths in a typical galaxy). To determine what line features would then be observed (i.e., on top of typical galaxy line ratios) we use the SDSS catalog of galaxies\footnote{\url{https://wwwmpa.mpa-garching.mpg.de/SDSS/}} \citep{York00, Abazajian09}, to which we add the time-dependent line emission determined from {\tt CLOUDY}. We assume a \cite{Planck2018} cosmology for turning fluxes into luminosities in the SDSS catalog. We then compute what line ratios would be observed if molecular clouds of given density $n_H$, distributed with volume filling factor $f_V$, were to lie on the isodelay contour of a future TDE observer at a series of times since the TDE was initiated (e.g., we compute the integral given in eq. \ref{full}). We can then place each of the TDE + SDSS galaxy combination onto the so-called BPT \citep{Baldwin81} diagram, a commonly used diagnostic which seeks to characterize the main ionization source of observed emission lines from galaxies or regions within galaxies. 
The location of a spectrum in the BPT diagram(s) is computed by determining the values of $\log_{10}($[OIII]/[H$_\beta$]$)$ and $\log_{10}($[X]/[H$_\alpha$]$)$, where [X] is a given reference line (which defines the diagram), typically [NII], [SII] and [OI]. These ratios determine (roughly speaking) the hardness of the input spectrum which produced the ionization, and thus acts as a simple proxy for accretion vs star-formation powered continua (although other ionization sources, such as shocks and evolved stellar populations, can also imprint ionization features in the diagram). 

As a first analysis we display, in Figure \ref{fig:bpt_tracks1}, the evolution of 10 galaxies in the SDSS survey on the [OIII]-[NII] BPT diagram as a function of time since a TDE injected ionizing radiation into the galaxy. The evolution of each galaxy is displayed by a coloured line, with the colour denoting the ($\log_{10}$ of the) time since the TDE (see colourbar). Also plotted are 100,000 other SDSS galaxies, with those below the Kewley line shown in gray and those above the \citet{Kewley01} line in black. The Kewley line (displayed as a black dashed curve) represents the `maximum starburst line', i.e., the theoretical upper limit for line ratios that are produced solely by star formation (primarily from {\it O} stars).
It is common in the literature to assume, perhaps naively, that every galaxy spectrum above the Kewley line—or at least in the upper-right portion of the diagram—is necessarily powered by an AGN. However, this is, of course, a misinterpretation of the Kewley line, given that it is derived from young stellar population models it cannot indicate what is an AGN, but rather, simple what is not powered by star formation (i.e., those spectra above the line). Furthermore, numerous non-accretion-related ionization sources can also produce line ratios exceeding the Kewley line, including shocks \citep{Allen2008, Rich2015}, evolved stellar populations \citep{Stasinska2008, Cid_Fernandes_2011}, and other stellar remnants \citep{Kopsacheili2020}. As we shall demonstrate, promptly identifying every spectrum above the Kewley line as evidence for a classical long-lived AGN \citep[e.g.,][]{Wevers2022}, particularly in the context of TDE/QPE science, may lead to erroneous interpretations.
Nevertheless, when considered in the context of the observations, the Kewley line remains a reasonable and widely used initial proxy for identifying accretion-driven ionization features—such as those shown here to be produced by TDEs.

We see from Figure \ref{fig:bpt_tracks1}  that the introduction of an ionizing radiation field sourced from a TDE results in dynamic evolution of the galaxy on (astronomically) short timescales, and for a few thousand years (for this particular choice of parameters) results in a galaxy crossing the Kewley line towards the upper region of the BPT diagram. 
This is expected, because of course a TDE accretion disk {\it is} an Active Galactic Nucleus, just not one which is like those typically meant when the term ``AGN'' is used. We note that we have chosen some particularly interesting examples of BPT diagram evolution in Figure \ref{fig:bpt_tracks1}, so as to demonstrate the full range of possible evolution that can occur upon the injection of TDE disk emission. These SDSS host galaxies were not chosen at random, and we shall determine the average evolution of a SDSS host galaxy following the injection of TDE disk emission shortly. 

In Figure \ref{fig:bpt_tracks2}, we repeat this analysis, but now using the [SII] and [OI] diagnostic diagrams. These lines represent other common diagnostic tests, which for a given system  may be easier to observe. We see that, broadly speaking, the results of our analysis are unchanged and that TDE disk emission can instigate global changes in a host galaxies location on a BPT diagram, independent of the diagram used. Indeed, broadly the same fraction of systems cross the Kewley line upon the addition of TDE emission, independent of diagram chosen. 

A second question one might ask, in addition to the evolution of the locations of TDE host galaxies on characteristic diagrams, is {\it what would an observer actually see?} This question can be answered precisely, under the assumption that the gas clouds in the host are spherically distributed. If a more complicated geometry is assumed (as is almost certainly the case in any given system), then such an analysis can be repeated on a case-by-case basis. We will, for this analysis, assume spherical symmetry for simplicity. 

What an observer would actually see is, in a given pixel of their detector, the local contribution to the luminosity of a given element of the galaxy in their image plane. This is then, up to an order unity projection factor, simply given by 
\begin{equation}
    L_{\rm obs, \, pixel} \propto {{\rm d} L_{\rm obs, \, {\rm total}}\over {\rm d}\phi}, 
\end{equation}
where ${{\rm d} L_{\rm obs, \, {\rm total}}/ {\rm d}\phi}$ is given by the integrand of equation \ref{full}, and each host galaxy location corresponds to an observers pixel coordinate by $X=R/(1-\cos \phi)$ (note that the situation is more complicated if the time since the TDE is comparable to the time at which the observation is taken, which is not the case in this particular calculation). In a spherically symmetric setup we can then compute ${{\rm d} L_{\rm obs, \, {\rm total}}/ {\rm d}\phi}$ for each ${\rm d}\phi$, associate with it an $X$ pixel coordinate, and rotate that one dimensional profile around $2\pi$ in the image plane. A non-spherically symmetric distribution of emission then cannot be rotated around $2\pi$, but otherwise would have the same broad radial (image plane) features, but potentially with gaps where there is no gas clouds (of the right constitution). 

\begin{figure}
    \centering
    \includegraphics[width=0.48\linewidth]{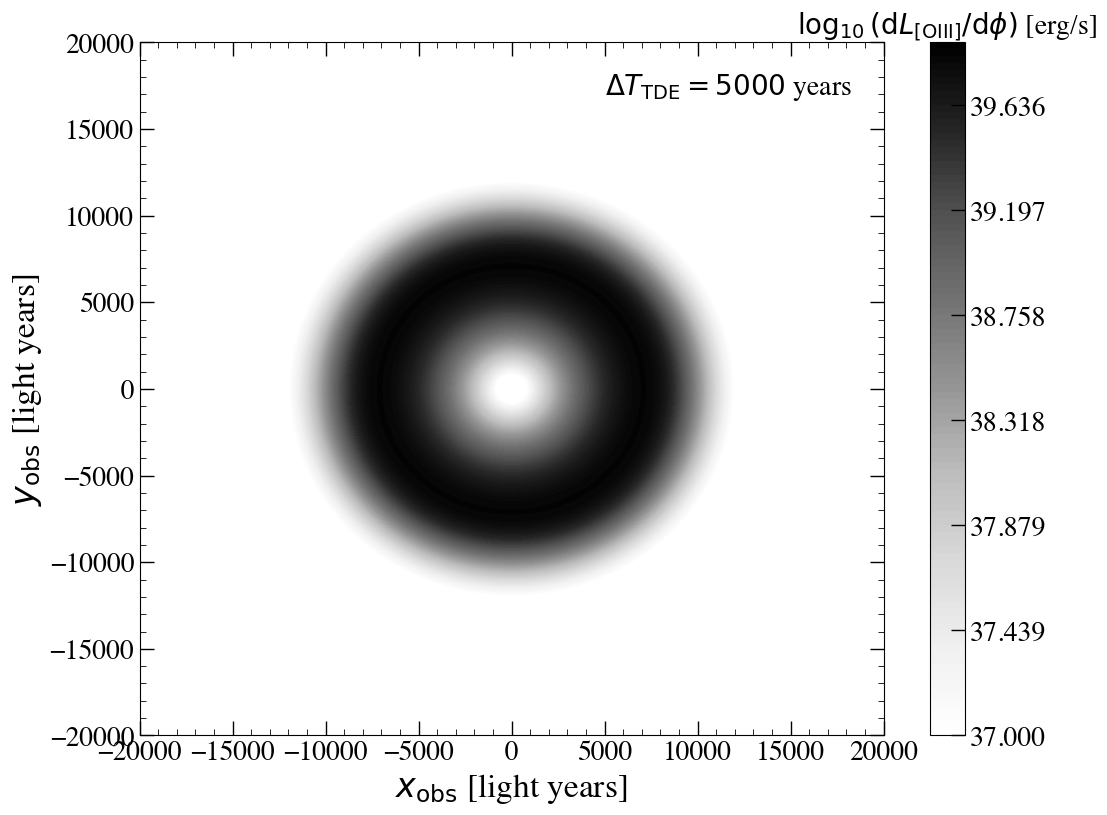}
    \includegraphics[width=0.48\linewidth]{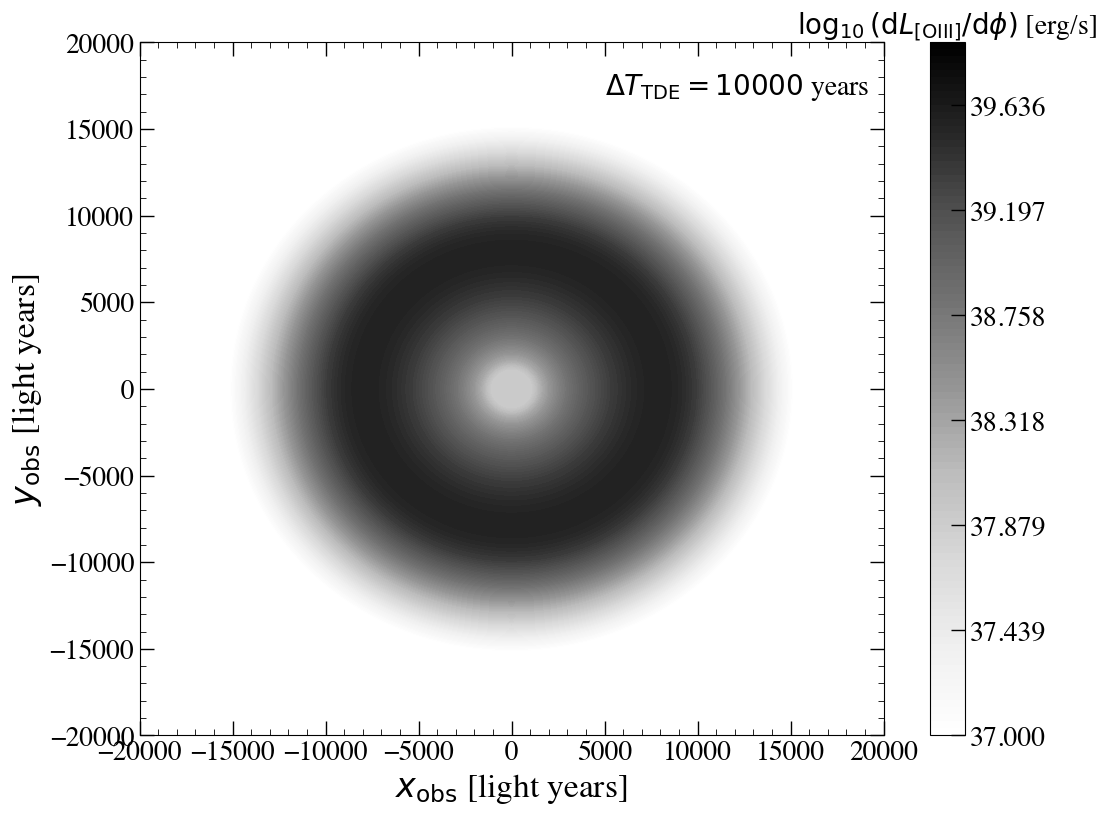}
    \includegraphics[width=0.5\linewidth]{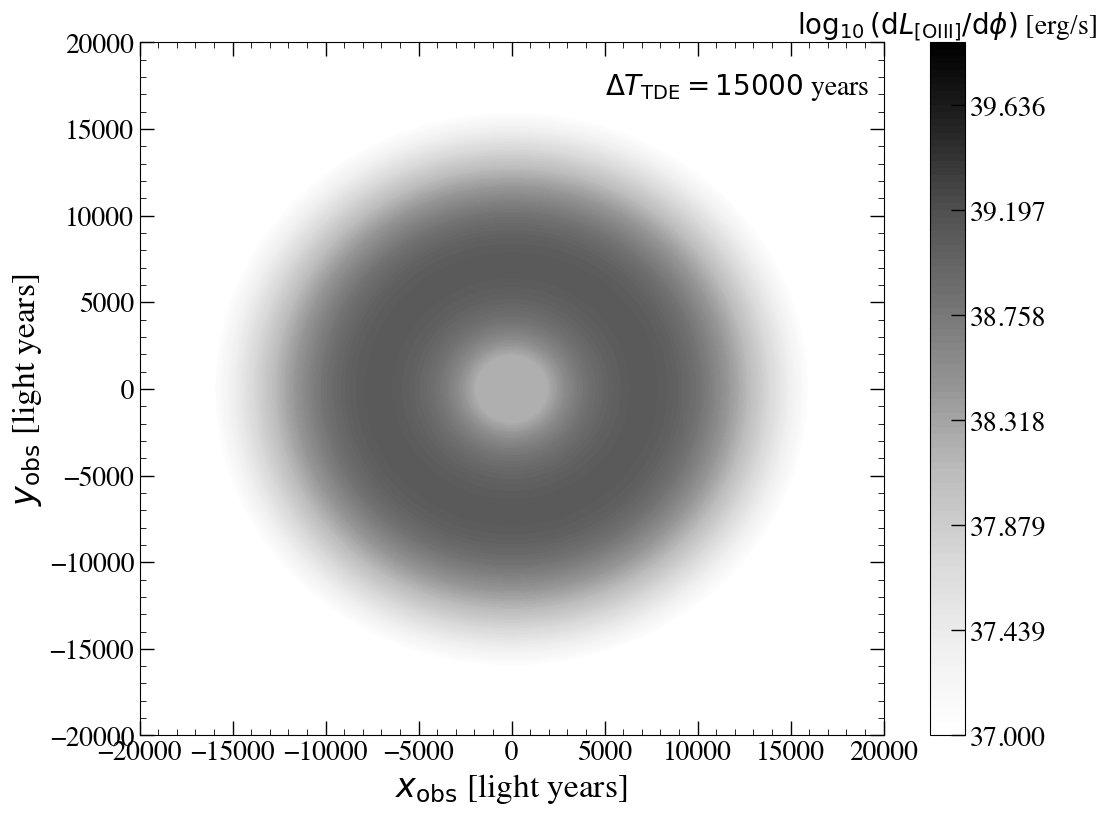}
    \caption{The observer plane projection of the reprocessed emission from {\tt CLOUDY}, highlighting how a single narrow ($c \Delta t \sim {\cal O}(100 \, {\rm lyr})$) ionizing shell of radiation produces an observer-plane signal which has a significantly broader ($r \sim {\cal O}(10^4\, {\rm lyr})$) features. This is a simple to understood result of the finite propagation speed of light, although is often counterintuitive at first. While simple, this result has an important consequence, namely that any observational campaign, even at infinite spatial resolution, hoping to resolve shell-like features of $\sim{\cal O}(10-100)$ light years across in the hope of confirming a TDE origin of these features will not succeed. In a real system one must also take into account the non-spherical symmetry of the galaxy gas distribution, which will mean that in general only a sub-set of this feature is observed.  The image plane luminosity of each pixel is displayed by the colourbar, which saturates for the brightest regions in the top panel.   }
    \label{fig:obs_maps}
\end{figure}

We display a set of image plane structures, computed at three different times $(\Delta T_{\rm TDE} = 5,000, 10,000$ and $15,000$ years) post TDE in Figure \ref{fig:obs_maps}. For this figure we display the [OIII] luminosity maps, and plot the pixel coordinates in inferred physical distance from the galactic center. The observed luminosity is displayed on a logarithmic colour scale (with colourbar next to each plot).  Note that in the upper (brightest) panel this colourbar is saturated (i.e., the emission is brighter than the upper limit). This Figure highlights the important result that a single narrow ($c \Delta t \sim {\cal O}(100 \, {\rm lyr})$) ionizing shell of radiation produces an observer-plane signal which has a significantly broader ($r \sim {\cal O}(10^4\, {\rm lyr})$) features. 

This is a simple to understood result of the finite propagation speed of light, although is often counterintuitive at first so it is worth spelling out the physics of this here. Due to the finite speed of light, a distant observer can in principle simultaneously detect reprocessed emission from all radii in a galaxy in the range $r\in [cT/2, R_{\rm  gal}]$  at a time $T$ after the TDE. These different radii are merely orientated either behind, or between, the TDE disk from the perspective of the observer. The three-dimensional isodelay contour (Figure \ref{sizes}) describes which of these points in the galaxy an observer can see at a given time, and then the observed emission is merely the projection of the three-dimensional isodelay contour onto a two-dimensional image plane. This two-dimensional projection can have (in principle) extremely broad features, merely reflecting the underlying emissivity profile of the TDE galaxy. Any observational campaign, even at infinite spatial resolution, hoping to resolve shell-like features of $\sim{\cal O}(10-100)$ light years across in the hope of confirming a TDE origin of these features {\it will fail}, a simple result of relativity. 

\begin{figure}
    \centering
    \includegraphics[width=0.49\linewidth]{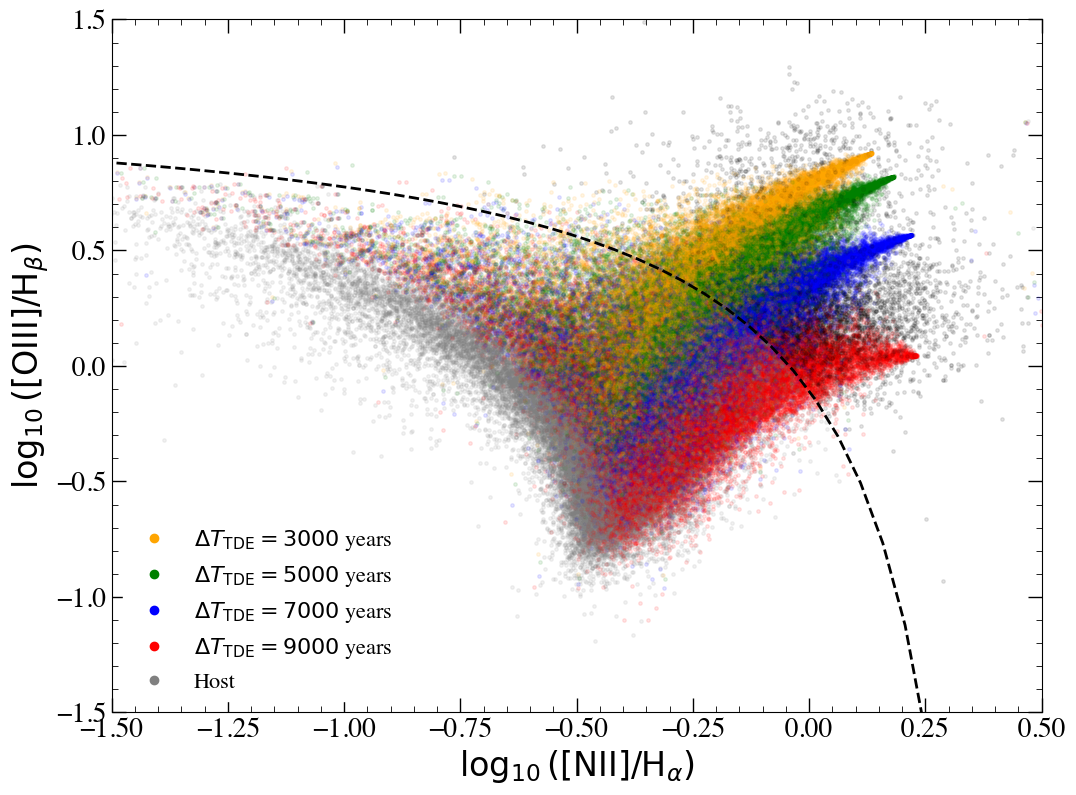}
    \includegraphics[width=0.49\linewidth]{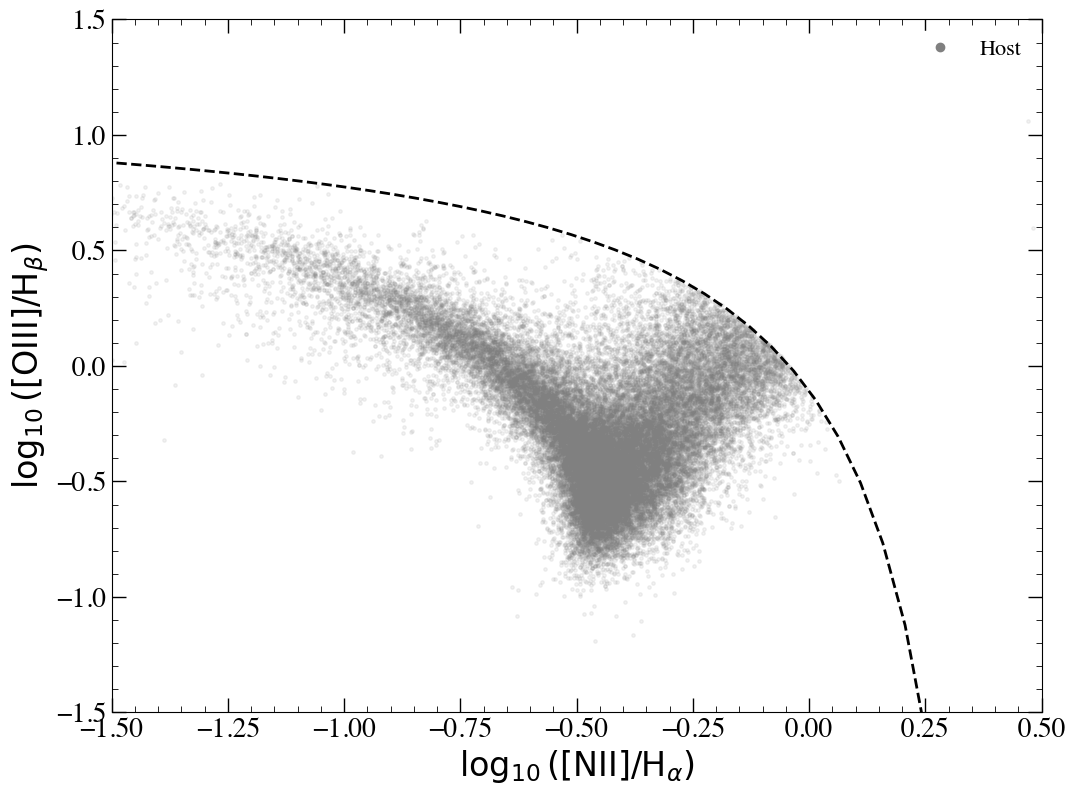}
    \includegraphics[width=0.49\linewidth]{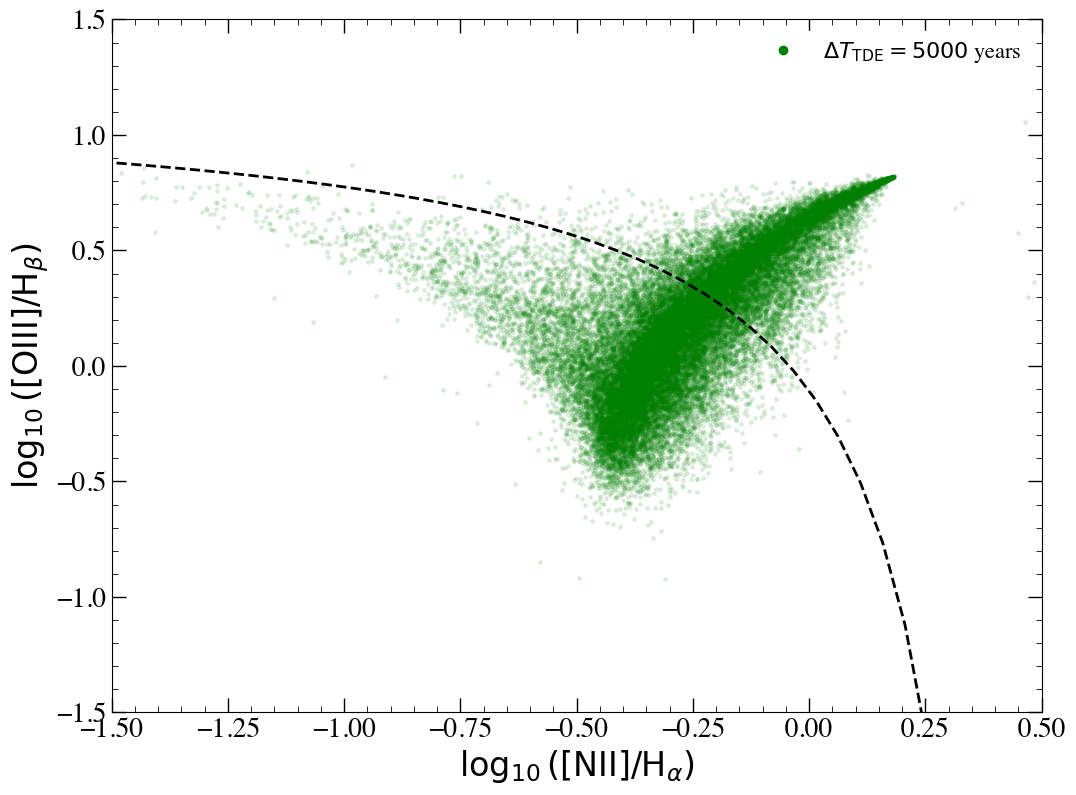}
    \includegraphics[width=0.49\linewidth]{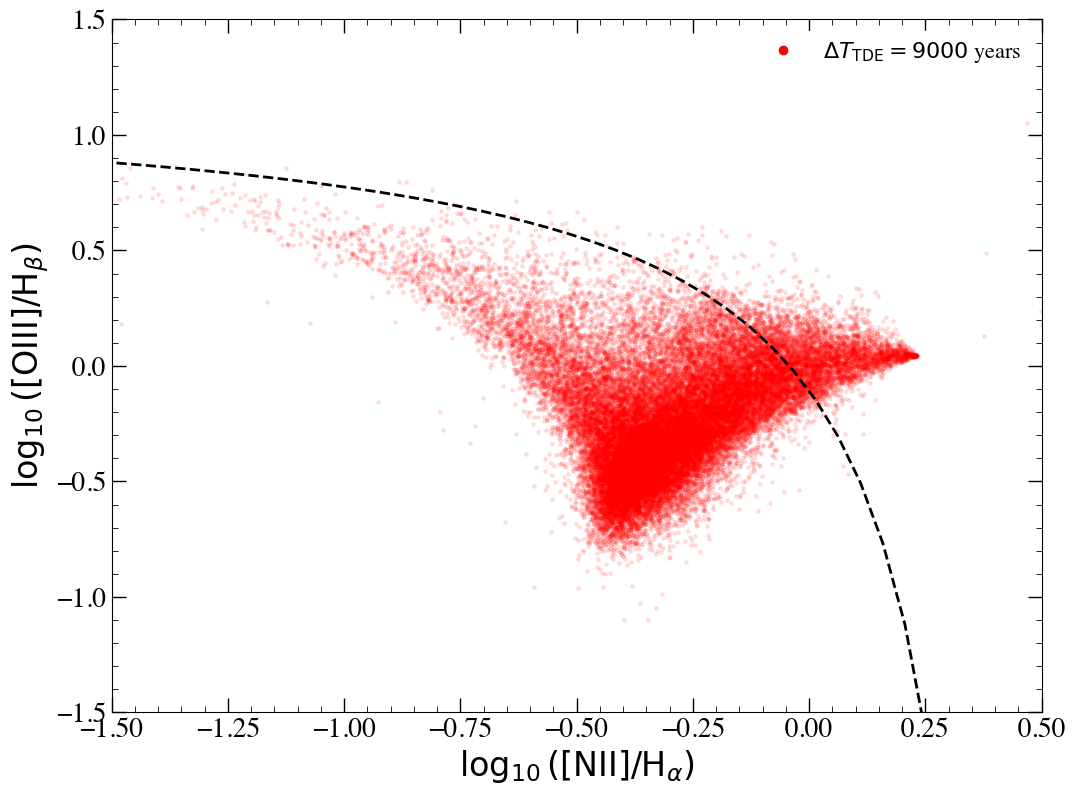}
    \caption{{\bf Upper left:} The [OIII]-[NII] BPT diagram, with 100,000 SDSS galaxies shown by gray points. In each SDSS galaxy we add the reprocessed TDE emission, assuming $n_H = 10^2 \, {\rm cm}^{-3}$ and $f_V = 10^{-4}$. The TDE disk properties are discussed in the main body of the text. By each colored point we plot the galaxy+reprocessed emission, for different times since the TDE occurred (in the observer frame). For a wide range of observationally relevant times the galaxy+ionized cloud emission lines lie above the \citealt{Kewley01} line (black dashed curve) for many SDSS host galaxies. {\bf Upper right:} the sample of ``non-AGN'' SDSS host galaxies which we add TDE reprocessed emission to, which results in the movement of the galaxy on the BPT diagram to the locations shown in the {\bf lower left} plot after 5000 years, and the locations denoted in the {\bf lower right} plot after 9000 years.  We stress that the exact period of time for the galaxy+TDE system to move above the Kewley line, and the period of time spent in this region, are a function of both TDE disk properties, $n_H$ and $f_V$. }
    \label{fig:BPTNII}
\end{figure}

Finally, a third interesting question one can ask regards the change in the emission line structure of an {\it average} galaxy following the injection of ionizing radiation from a TDE disk. This can be simply analyzed by adding reprocessed TDE disk emission on top of a wide range of SDSS host galaxies, and determining the median evolution. The results of one such analysis are shown in Figures \ref{fig:BPTNII}. In the upper left panel we plot 100,000 SDSS galaxies, onto which we TDE reprocessed emission observed at a given time post TDE (denoted by colored points). Note that we do not add TDE emission to an SDSS galaxy which already have line-ratios above the Kewley line, 
only those which fall in the star forming region of these diagrams.  

As expected from our earlier analysis, for a wide range of observationally relevant times the galaxy+ionized cloud emission lines lie above the \cite{Kewley01} line (black dashed curve) for many SDSS host galaxies. To isolate this evolution more clearly, in the upper right plot we show the sample of ``star forming'' galaxies to which we add the reprocessed TDE disk emission, which results in the movement of the galaxy population on the BPT diagram to the locations shown in the lower left plot after 5000 years, and the locations denoted in the lower right plot after 9000 years.  Clearly many SDSS host galaxies cross the Kewley line, and remain there for many thousands of years (for these disk+cloud parameters). Similar results are displayed for different diagnostic diagrams in Figures \ref{fig:BPTSII} and \ref{fig:BPTOI} in Appendix \ref{other}.

\begin{figure}
    \centering
    \includegraphics[width=0.6\linewidth]{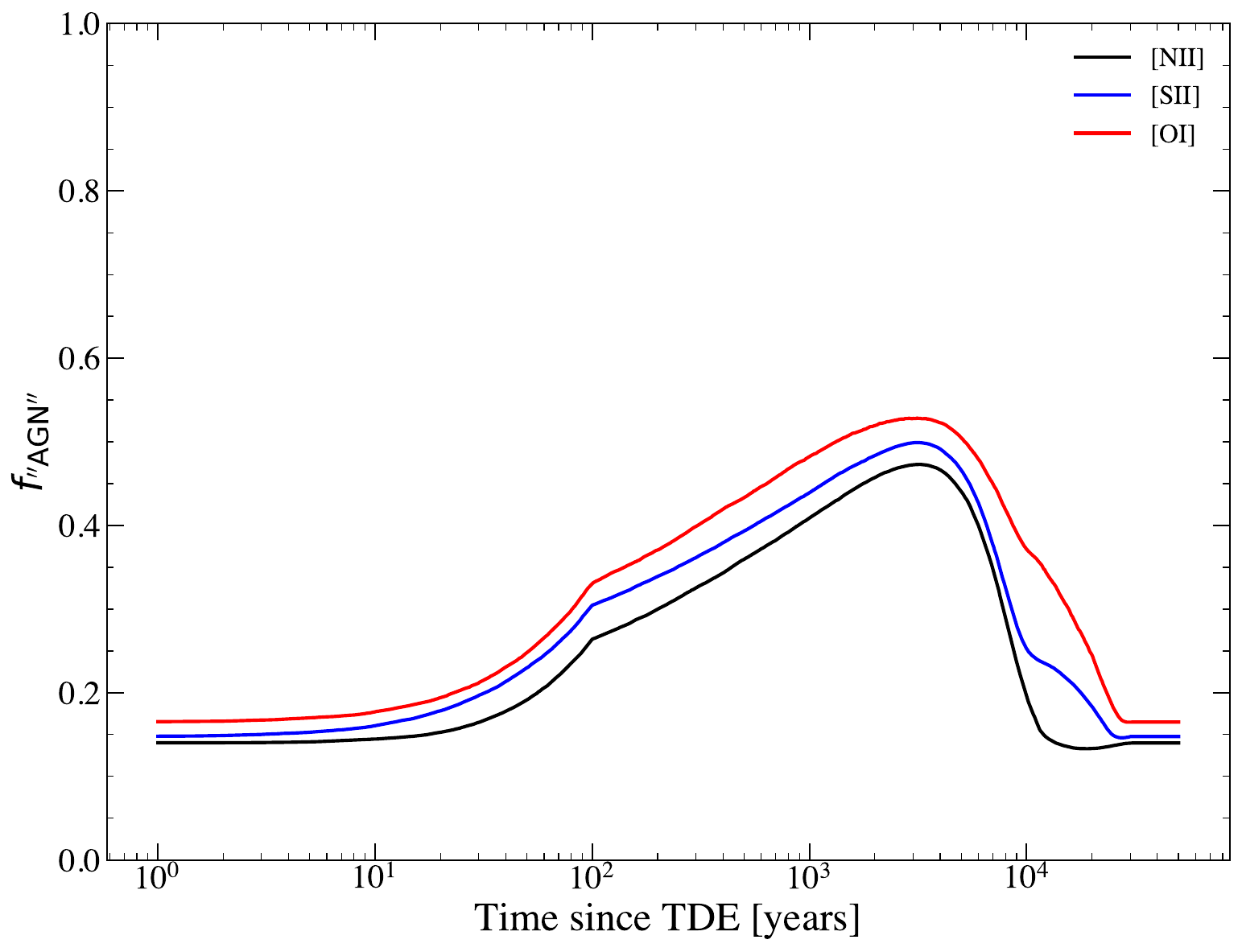}
    \caption{The total fraction of SDSS galaxies which lie above the Kewley line of different diagnostic diagrams (see text for more details) as a function of time since a TDE disk was formed and injected a shell of ionizing radiation into that galaxy. A galaxy lying above a Kewley line would be classified as an AGN, and so we denote this fraction $f_{\rm ``AGN"}$. We note that the injection of ionizing radiation from a TDE can move a significant fraction of SDSS galaxies above the Kewley line, on timescales $\Delta t\sim {\cal O}(10^3 - 10^4)$ years post TDE. This therefore implies that a non-negligible fraction of galaxies which we categorize as AGN may in fact be TDE light echoes.  The 100,000 galaxies plotted here include systems already above the Kewley line, the fraction of which can be inferred from the value of $f_{{\rm ``AGN"}}$ at $t \to 0$.   }
    \label{fig:fagn}
\end{figure}

\begin{figure}
    \centering
    \includegraphics[width=0.65\linewidth]{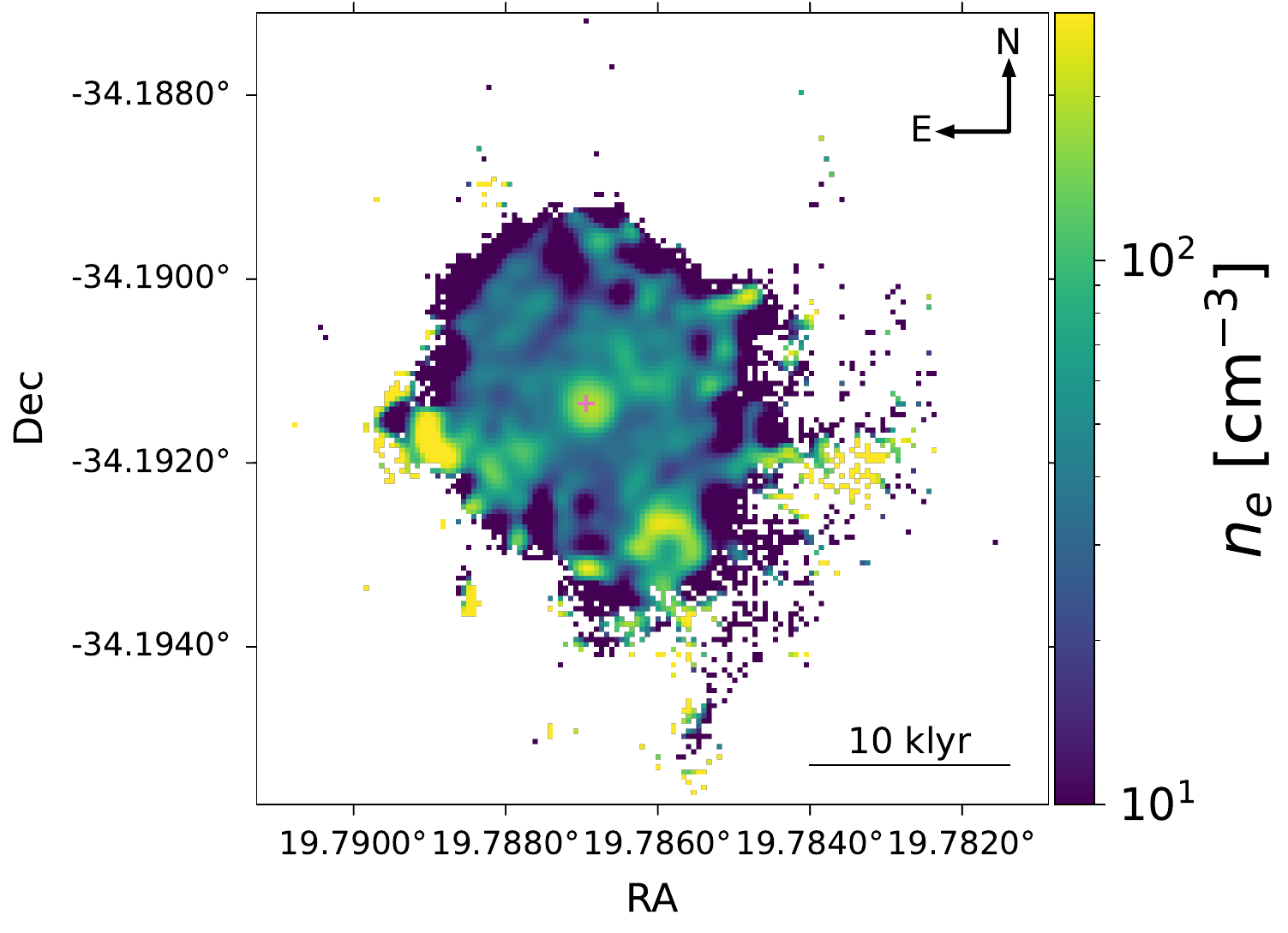}
    \caption{The inferred electron density within the EELR of the TDE/QPE host galaxy of GSN 069 (see also Figure \ref{fig:qpe_hosts} for the structure of the EELR). This density estimate was computed using the [SII] lines in the MUSE data cube (GSN 069 has highest S/N lines of all TDE/QPE hosts, see text for more details). The EELR is filled with gas with density $n_e \sim 30-300 \, {\rm cm}^{-3}$ precisely in keeping with what would be required for a previous TDE disk to power the emission lines. This result is supportive of the paradigm put forward in this work.  }
    \label{fig:gsn_ne}
\end{figure}

In Figure \ref{fig:fagn} we denote the fraction of the 100,000 SDSS galaxies which lie above the Kewley line, as a function of time since a TDE disk was formed and injected a shell of ionizing radiation into that galaxy.A galaxy lying above a Kewley line would naively be classified as an AGN, and so we denote this fraction $f_{\rm ``AGN"}$. The 100,000 galaxies plotted here include systems already above the Kewley line, the fraction of which can be inferred from the value of $f_{{\rm ``AGN"}}$ at $t \to 0$.  We assume that each TDE disk produces ionizing radiation for $t_{\rm ION} = 100$ years, after which it stops emitting. This induces a slight discontinuity in the gradient of the curves in Figure \ref{fig:fagn} at $t = 100$ years.  We note that the injection of ionizing radiation from a TDE can move a significant fraction of SDSS galaxies above the Kewley line, on timescales $\Delta t\sim {\cal O}(10^3 - 10^4)$ years post TDE (for this set of parameters). This therefore implies that a non-negligible fraction of galaxies which we categorize as AGN may in fact be TDE light echoes.   After $\sim 20,000$ years each of the host galaxies returned to their original location on the BPT diagram. An important result of this analysis highlighted by Figure \ref{fig:fagn} is that the fraction of galaxies ``perturbed'' above the Kewley line by the addition of a TDE is sufficiently large that it seems plausible that the TDE rate can be constrained (at least on an upper bound level) by preexisting surveys of galaxy emission line properties.

A prediction of the model put forward in this work, and one which can be qualitatively tested, is that the density of the gas in  TDE-powered EELR, should lie in the range $n_{H} \sim 10^1-10^3\, {\rm cm}^{-3}$. The reason for this is that only these densities result in the reprocessed TDE disk emission having an emissivity which peaks at the radial scales  ($r \sim 10^4$ light years) at which these EELR are observed (see e.g., Figure \ref{cloudy1}). This can be tested in sufficiently high signal to noise (S/N) observations of the EELR, as the ratio of [SII]$\lambda6716$/[SII]$\lambda6732$ lines allow local measurements of the electron density to be performed \citep[e.g.,][]{Osterbrock06,Proxauf14}.  

In Figure \ref{fig:gsn_ne} we perform such a test, and measure (see Appendix \ref{appB} for details) the local electron density of the GSN 069 host galaxy throughout its EELR. GSN 069 was chosen as it has the highest S/N lines of all TDE/QPE hosts, such that the fainter [SII] lines (as compared e.g. [OIII] and [NII] ) can be measured at large projected distance from the nucleus . As can be seen in Figure \ref{fig:gsn_ne}, the EELR is filled with gas with electron density $n_e \sim 30-300 \, {\rm cm}^{-3}$ precisely in keeping with what would be required for a previous TDE disk to power the emission lines (Figure \ref{fig:cloudy1}). While this is only the early stages of an analysis of real TDE/QPE host galaxies, this result is supportive of the paradigm put forward in this work. 

We reiterate that the precise times at which these emission lines are produced is in no way fixed, and by varying, e.g., the hydrogen density or filling factor (or indeed the disk luminosity by assuming different TDE parameters) a variety of different parameters can produce similar ionization features on a range of different timescales (although, as we stress above, galaxy-wide signatures require low density gas). Future analyses on a case-by-case basis (including full time dependent ionization physics) will be performed for the different TDE and QPE host galaxies which have been observed.

\section{Extreme coronal line emitters}\label{cloudy2}
The analysis of the previous section examined the reprocessed emission resulting from the illumination of low density clouds at large radial scales from the galactic center. Of course, it is entirely possible (and in fact likely) that much more dense clouds $(n_H \sim 10^6 - 10^8 \, {\rm cm}^{-3})$ may be illuminated near to the event itself. 

\begin{figure}
    \centering
    \includegraphics[width=0.6\linewidth]{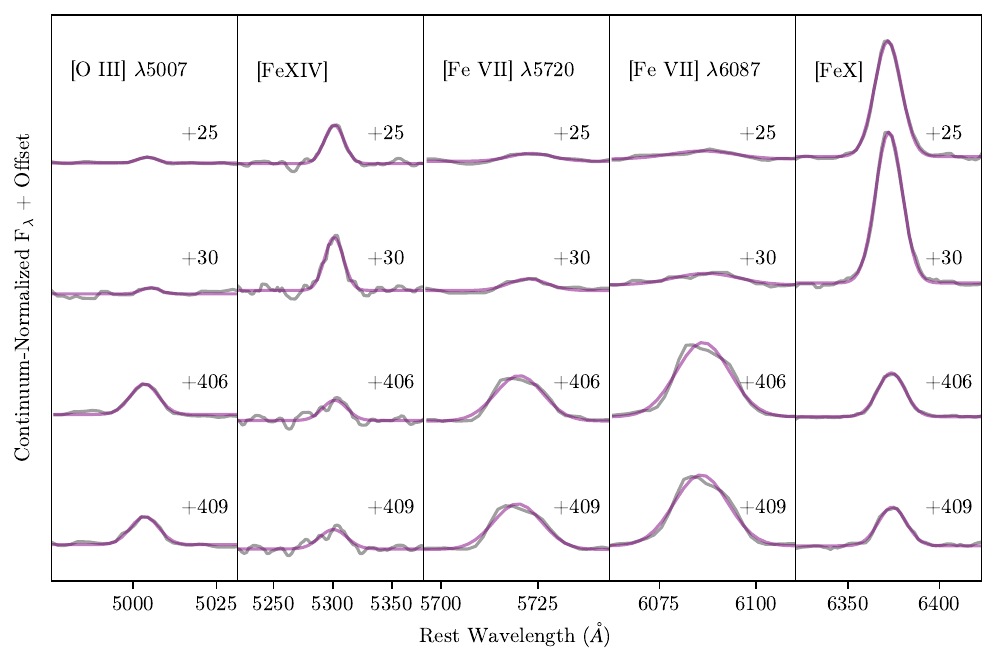}
    \caption{Evolving coronal lines, taken from the TDE AT2022upj. Here different spectral lines are separated by the left-right panels, and the vertical plots are separated by the time since the discovery of the flare (in days). We see that coronal lines such as [Fe X] and [Fe XIV] switch on quickly (within a month) of the flare, while [O III] and [Fe VII] switch on later ($\sim 1$ year post flare). This behavior is entirely in keeping with the results of {\tt CLOUDY} simulations, which demonstrate that [Fe X] and [Fe XIV] are generated by reprocessing in dense clouds closer to the galactic center than [Fe VII] and [O III]. Figure adapted from \citet{Newsome2024_22upj}.}
    \label{fig:22upj}
\end{figure}

In these situations we should expect reprocessed lines to switch on rapidly following the tidal disruption process, as even clouds at radial scales of $\sim $ pc can produce observable reprocessed emission detected only a few weeks post flare for angles close to the line of sight $\phi \sim 10^\circ$ (see section \ref{geom}). Indeed, many interesting lines do switch on rapidly following a TDE. An example of one such system is shown in Figure \ref{fig:22upj} \citep[data taken from][for the TDE AT2022upj]{Newsome2024_22upj} where, for example, [OIII] switches on within $\sim 1$ year of the TDE being detected.

One particularly interesting observational class of objects for which these reprocessed signatures are likely to be relevant are so-called extreme coronal line emitters (ECLEs). The extreme coronal lines are species such as [FeX], [FeXIV] and [FeVII], and have high activation energies (requiring photons with energies $E\gtrsim 100$ eV). These coronal lines are rare in any random sample of galaxies (once obvious AGN are removed from the sample), with only $5$ non-AGN systems found in the entire SDSS sample \citep{Komossa2008, Wang2011, Wang2012, Callow2024}. They are, however, relatively common following UV-optical selected tidal disruption events, with 11 known examples \citep[]{Neustadt2020, Onori2022, Hinkle2023, Li2023_ECLE, Short2023, Somalwar2023, Yao2023, Koljonen2024, Wang2024_ECLE, Newsome2024_22upj, Clark2025}. One of the most recent examples is shown in Figure \ref{fig:22upj}, where three coronal lines, and [OIII], are shown at four different times post the discovery of the TDE AT2022upj. We see that coronal lines such as [FeX] and [FeXIV] switch on quickly (within a month) of the flare, while [OIII] and [FeVII] switch on later ($\sim 1$ year post flare). 

It is a natural question to ask whether the reprocessing of TDE disk emission by dense clouds near the galactic center can explain these observed features. We again turn to {\tt CLOUDY} simulations to determine the reprocessed spectra of illuminated clouds. As we are analyzing lines which switch on rapidly post disk formation, we use a TDE disk spectrum (from section 4, Figure \ref{fig:disk_spec}) at peak bolometric disk luminosity, which we input into {\tt CLOUDY} with a luminosity (measured from the model) of $\log_{10}L = 44.5$ (erg/s). Note that owing to the much higher cloud densities considered here, these {\tt CLOUDY} simulations are much more likely to be reliably described by statistical steady state solutions, owing to the much reduced recombination timescales of different species at high densities. 

\begin{figure}
    \centering
    \includegraphics[width=0.49\linewidth]{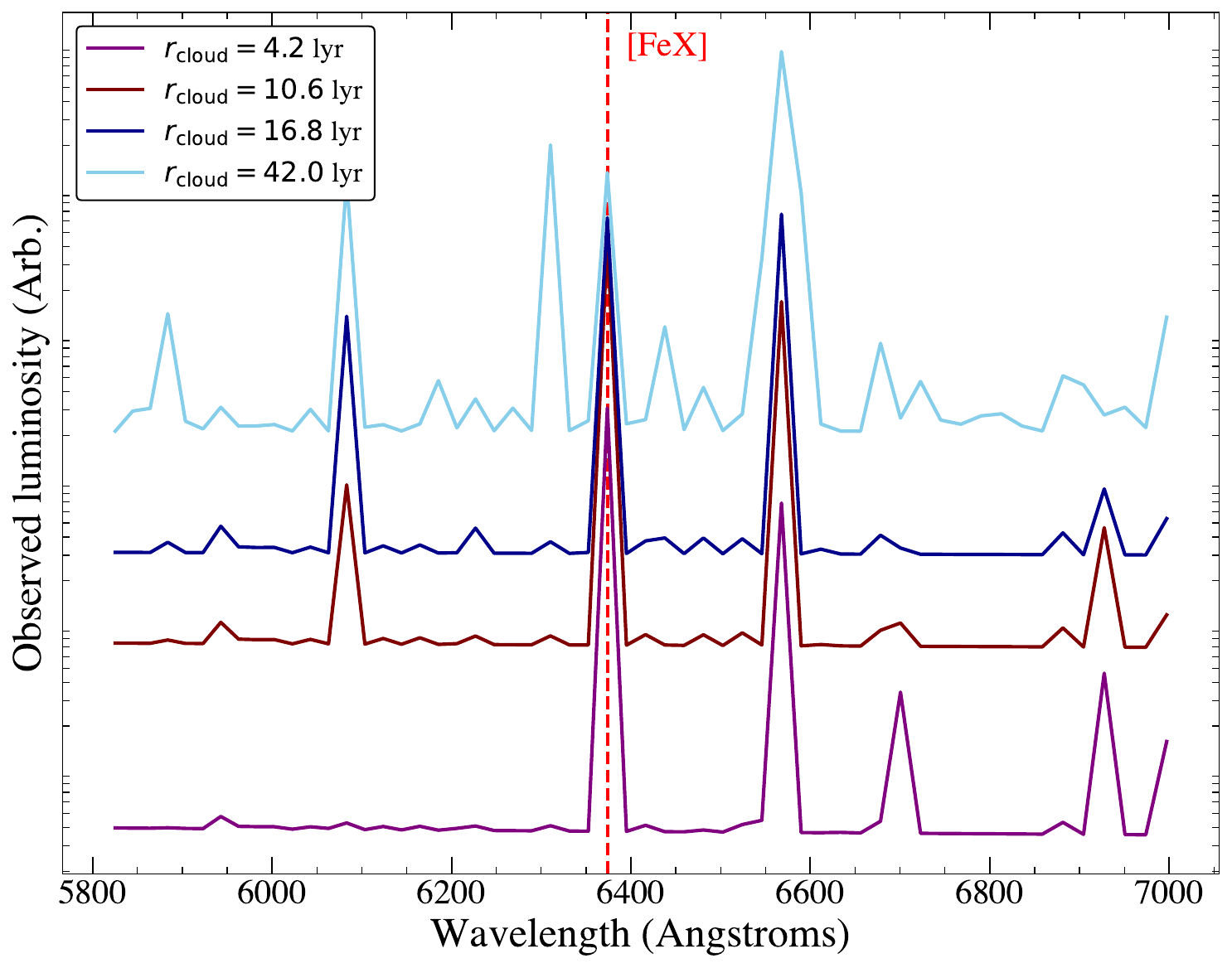}
    \includegraphics[width=0.49\linewidth]{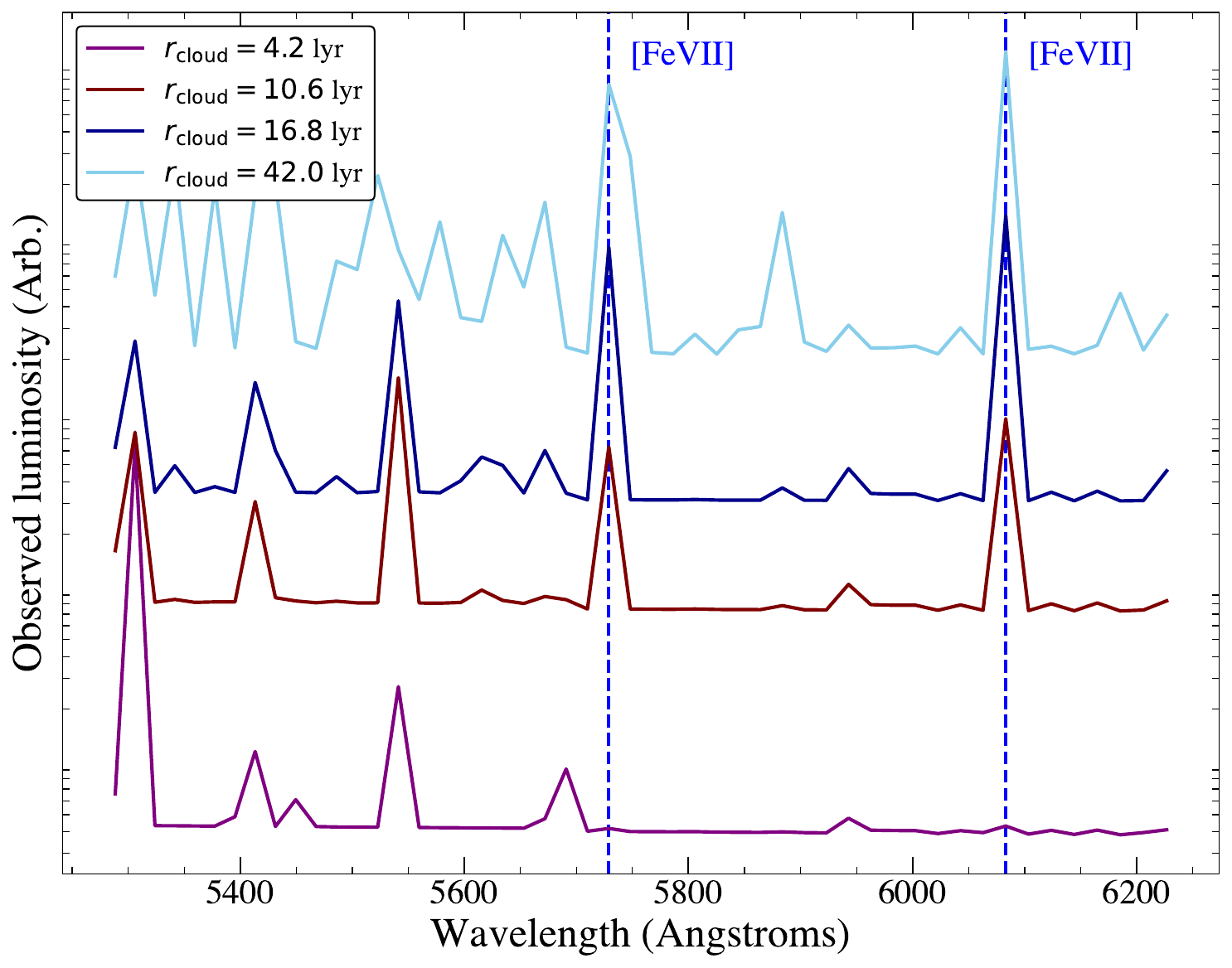}
    \includegraphics[width=0.49\linewidth]{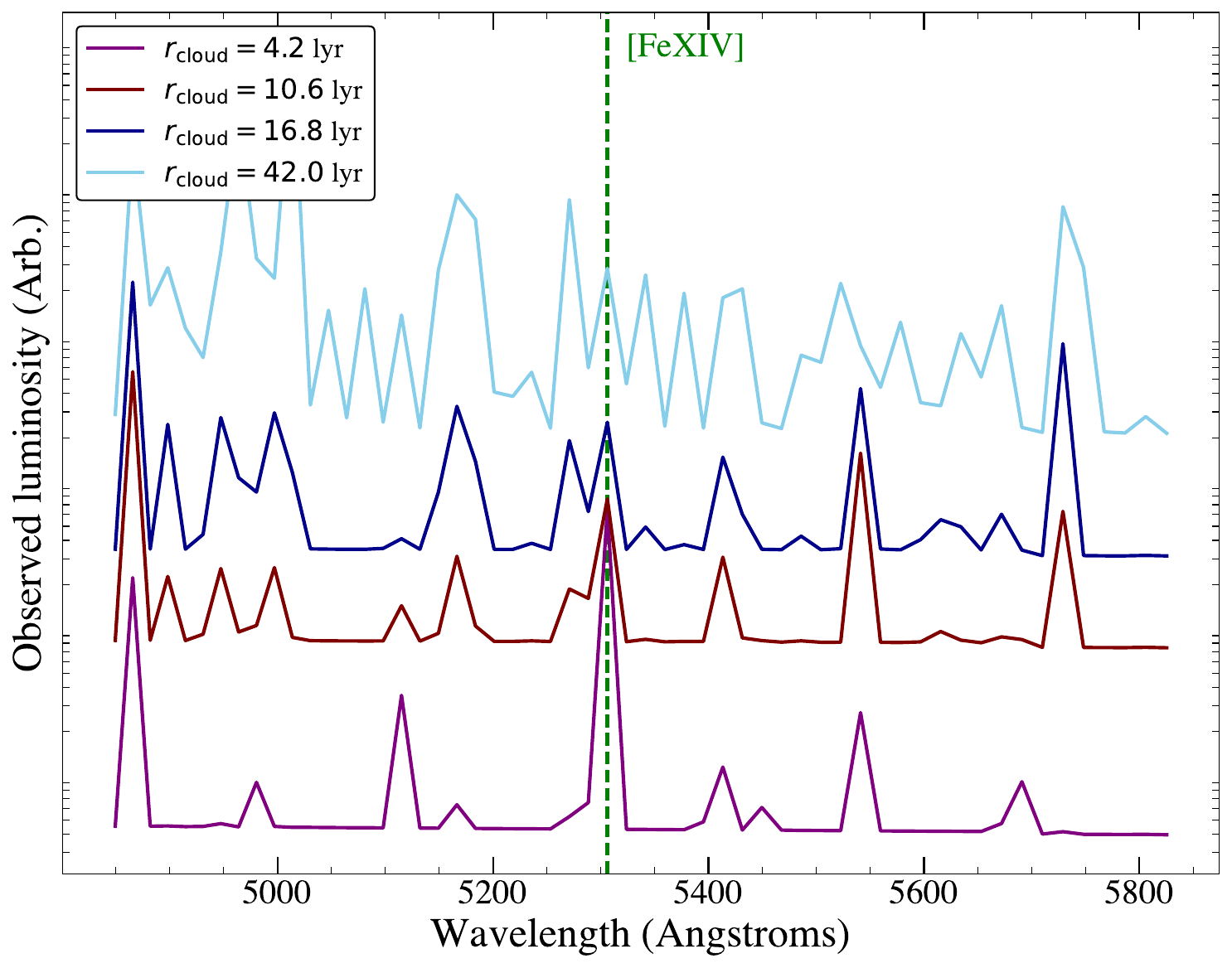}
    \includegraphics[width=0.49\linewidth]{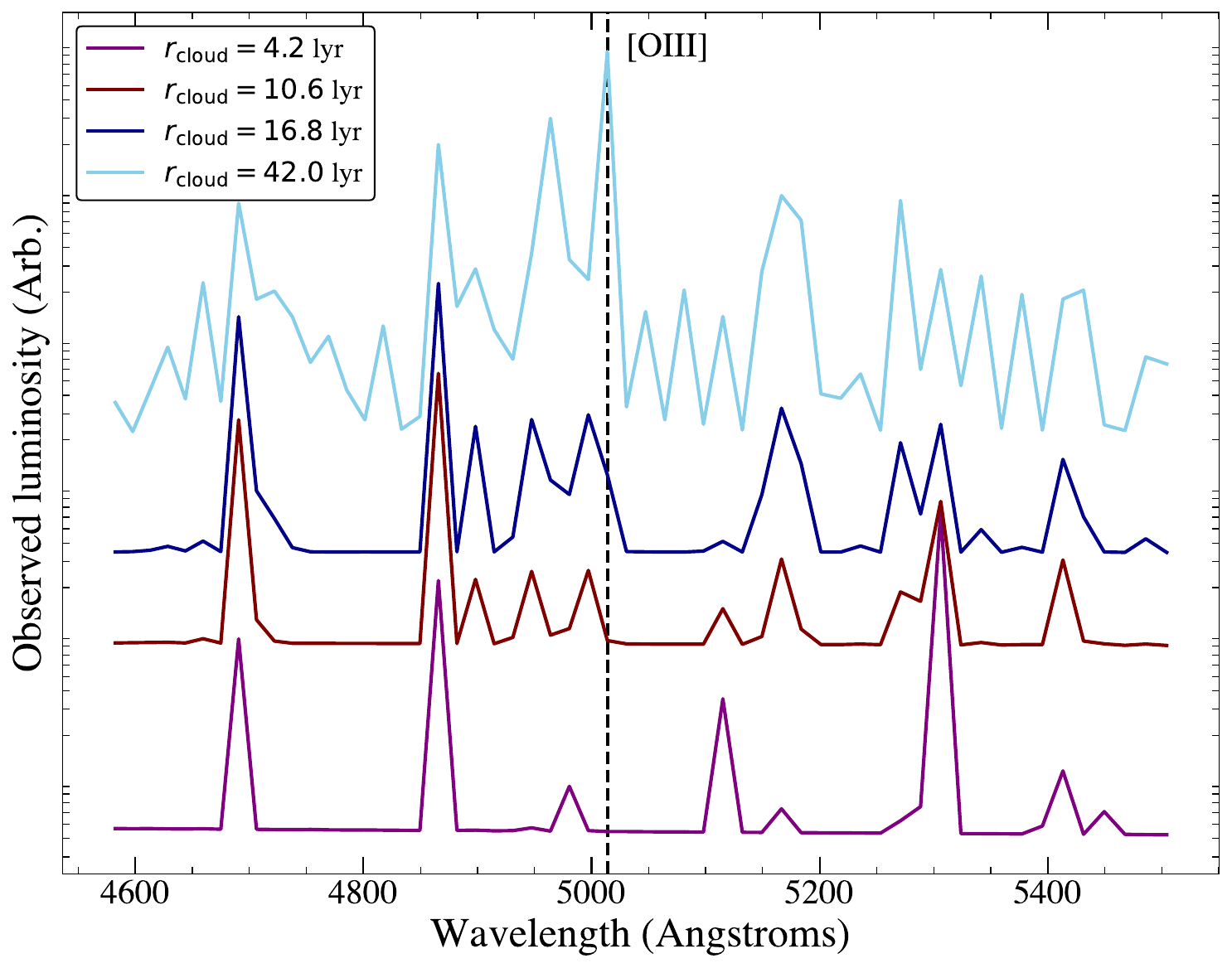}
    \caption{{\tt CLOUDY} spectra, resulting from the reprocessing of a TDE disk spectrum produced at peak bolometric luminosity, from clouds at different radii from the galactic center. The clouds are described by $n_H = 10^6\, {\rm cm}^{-3}$ and $f_V = 10^{-4}$. We show four panels, focusing on the spectral regions around various coronal lines seen in observations (denoted on plot). We note that prominent [FeX] lines can be observed from clouds illuminated at a range of radii, while [FeXIV] can only be excited on very small scales. Clouds on large radial scales can produce bright [OIII] and [FeVII] emission, this is consistent with the different times at which line emission is observed to switch on in TDE systems (e.g., Figure \ref{fig:22upj}).     }
    \label{fig:22upj_cont}
\end{figure}

In Figure \ref{fig:22upj_cont} we show the spectra resulting from the reprocessing of clouds with density $n_H = 10^6 \, {\rm cm}^{-3}$, with volume filling factor $f_V = 10^{-4}$ located at four different radii from the galactic center, $r = 4.2, 10.6, 16.8$ and $42$ light years. Owing to the different activation potentials of the different atomic species, different coronal lines are excited at different radii from the ionizing source. We note that prominent [FeX] lines can be observed from clouds illuminated at a range of radii, while [FeXIV] can only be excited on very small scales. For all of the parameter space we explored we found that [FeXIV] was fainter than [FeX] when both were excited. Clouds on large radial scales can produce bright [OIII] and [FeVII] emission. This is entirely consistent with the different times at which line emission is observed to switch on in TDE systems (e.g., Figure \ref{fig:22upj}), although we remind the reader that going between observer times and cloud distances is degenerate with the unknown angle between the galactic center and the cloud $\phi$ (section \ref{geom}).  

\begin{figure}
    \centering
    \includegraphics[width=0.49\linewidth]{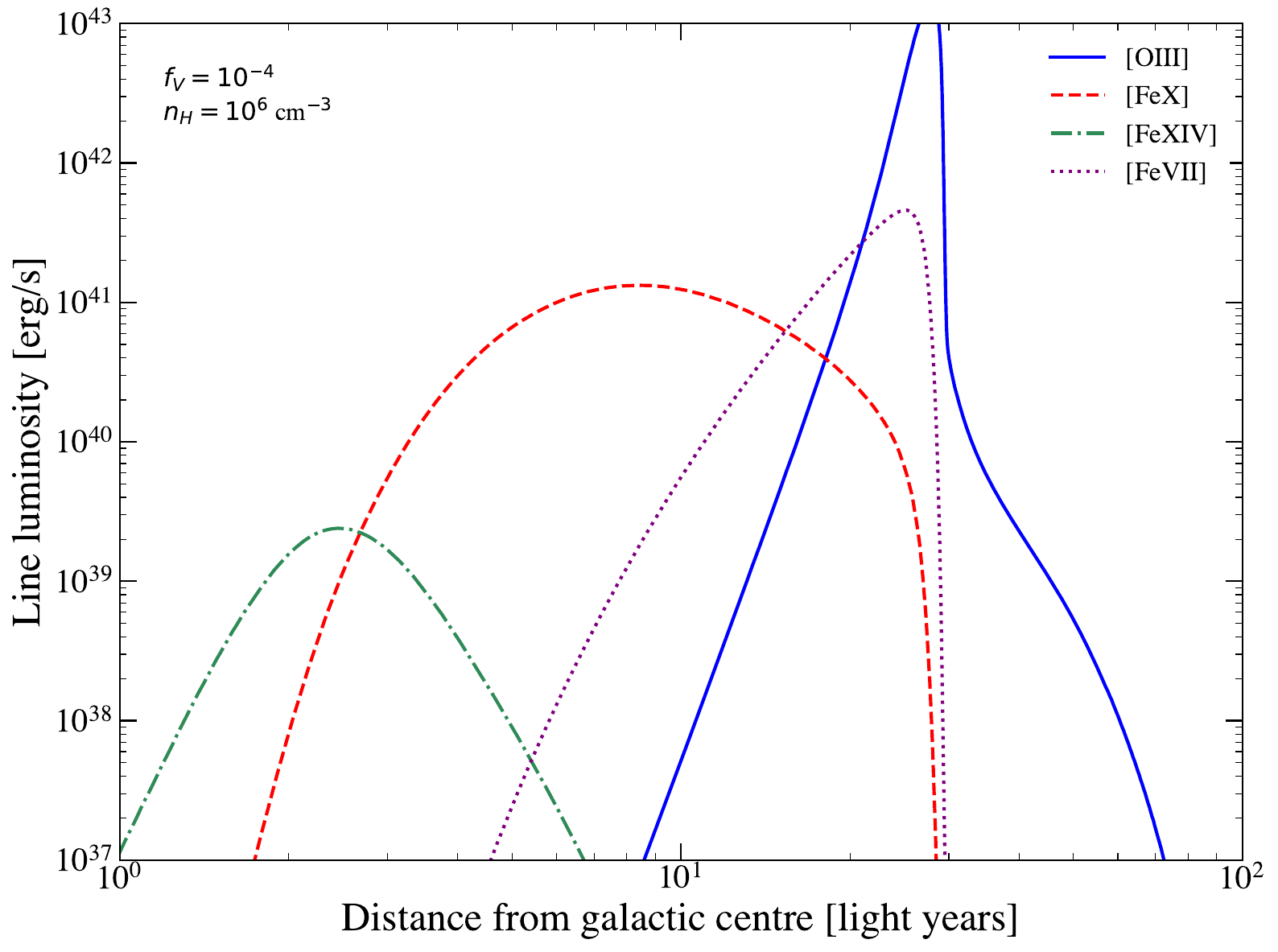}
    \includegraphics[width=0.49\linewidth]{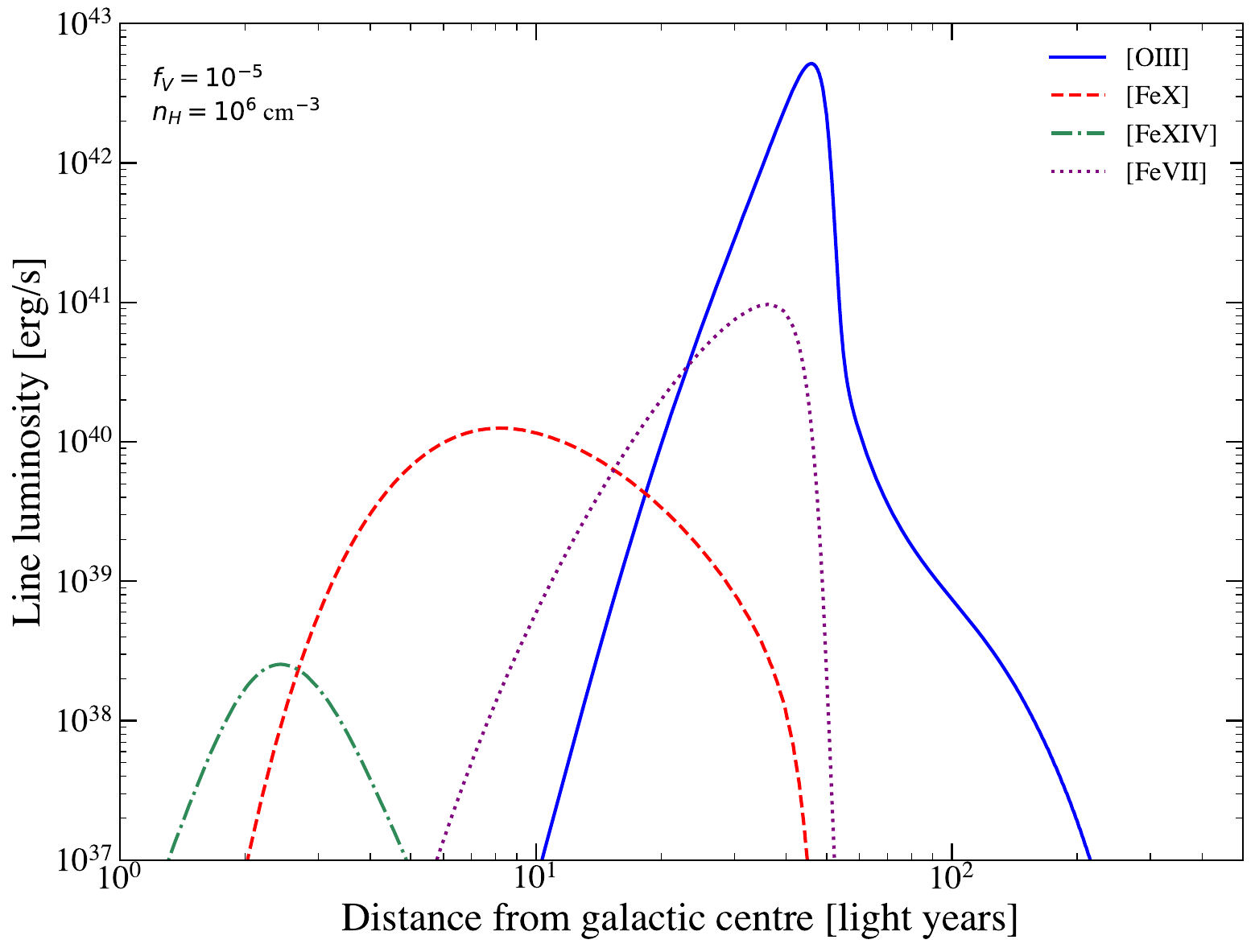}
    \includegraphics[width=0.49\linewidth]{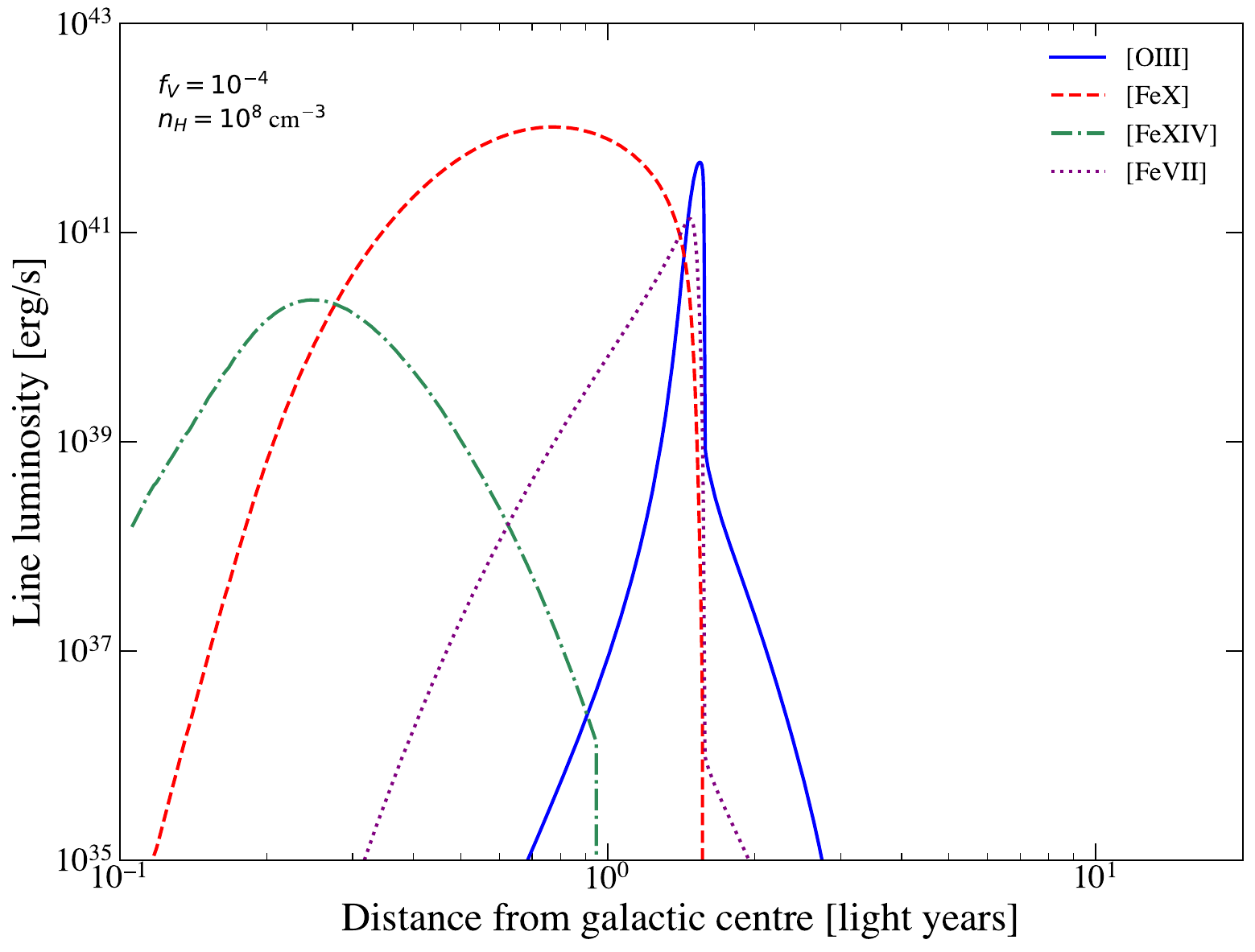}
    \includegraphics[width=0.49\linewidth]{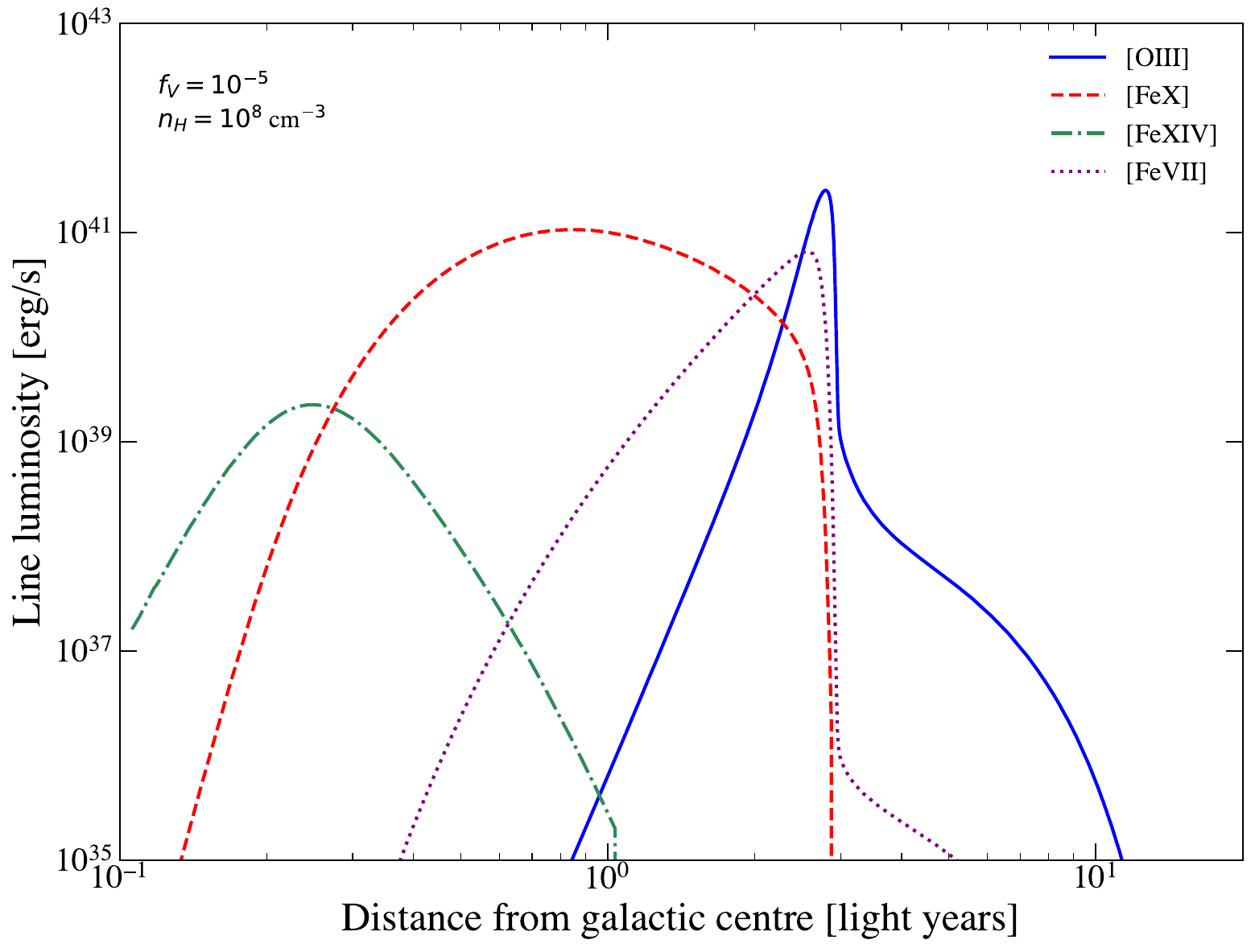}
    \caption{Luminosities of various coronal lines, as a function of distance from the galactic center for four different cloud parameters, computed using the {\tt CLOUDY} code. Cloud parameters are displayed on each panel. It is clear from this figure that dense $n_H \sim 10^6-10^8 \, {\rm cm}^{-3}$ galactic center $r\sim $ pc clouds can, when illuminated by a TDE disk, produce observable coronal line emission, with [FeX] and [FeXIV] excited closer to the center, and [FeVIII] and [OIII] excited at larger radii. Generally, [FeXIV] is the faintest iron line, and should be expected to be absent in some observations of TDE sources where [FeX] and [FeVII] are present. Note the different radial scales on the upper and lower panels. }
    \label{fig:coronal_lines}
\end{figure}

We further analyze this behavior in Figure \ref{fig:coronal_lines}, where we plot the reprocessed line luminosity of the three iron coronal lines, and [OIII], as a function of cloud radius from the galactic center, for a range of cloud parameters $n_H, f_V$. Again, this is a naive $L \propto r^3 j$ scaling and does not represent the actual observed line luminosity for a distant observer, which we shall properly compute shortly. 

We see that both high densities and filling factors are required to power observable [FeXIV] emission, which is generally a faint line. Cloud densities $n_H \lesssim 10^4\, {\rm cm}^{-3}$ are unlikely to ever result in observable coronal lines. A natural corollary of this is that coronal lines will only ever be observed in galactic centers, as the low densities required to have the correct ionization potentials at large radii will not result in observable levels of coronal line emission. 

We can go further than this however, and compute approximate ``light curves'' of the evolving coronal lines which switch on shortly after a TDE. Unlike the evolving EELR flux however, the ECLE emission evolves on timescales comparable to the evolution of the disc itself (or even shorter). This will complicate the calculation in two ways, firstly it will result in a fundamentally time dependent emissivity profile $j(r, t)$, as the spectrum of the disc changes on the same timescale as the emission. Secondly, the iso-delay contour will extend right from $r = 0$ (as emission is ongoing) to $\ell(t)$, modifying the precise form of the integrals required. In this first work we shall neglect the time dependence of $j(r, t)$, assuming it can be approximately described by an average profile like that computed above (while noting that this is an approximation of convenience). We begin with no approximations, in which case the line luminosity is given by 
\begin{equation}
    L_{\rm obs, \, line} =  \int_{0}^{x(t)} \int_{y_{\rm min}}^{y_{\rm max}} 2\pi \, x' \, f_V \, j_{\rm line}(\ell, t) \, {\rm d} y \, {\rm d}x' .
\end{equation}
This luminosity can be expressed in a more natural form after performing a variable transformation from $(x, y)$ to $(\phi, t)$, i.e., by using ${\rm d}x\, {\rm d}y = J \, {\rm d}\phi \,{\rm d}t$ where the Jacobian $J = \partial(x, y)/\partial(\phi, t) = |(\partial x/\partial \phi)(\partial y /\partial t) - (\partial x /\partial t)(\partial y/\partial \phi)|$. Explicitly then this line luminosity becomes
\begin{equation}
    L_{\rm obs, \, line}(t) = 2\pi c^3 \int_{0}^{t} \int_{\pi}^{\phi_\star(t')} \, t^{'2} f_V \, j_{\rm line}\left({ct'\over 1-\cos\phi}, t'\right){\left[\sin^3\phi - \sin\phi\cos\phi(1-\cos\phi) \right] \over (1-\cos\phi)^4} \,   \, {\rm d} \phi \, {\rm d}t'  .
\end{equation}
As noted above, this form of the luminosity integral, however, is complicated by the fact that  the emissivity of the surrounding gas is changing with time. If we were to take the above integral and simply replace $j_{\rm line}$ with an average profile, this integral would be simple to compute. However, while this will be a reasonable approximation for say H$_\alpha$ or [OIII] emission (which are relatively easy to excite), it is less appropriate for coronal lines which are likely only to be excited when the disk is at its most luminous (and spectrally hard -- as they require $E\gtrsim 100$ eV photons). We can therefore treat (for coronal lines) the ingoing disk luminosity field as an impulse, with spectrum similar to that at peak light. Under this simplification we can express 
\begin{align}
    L_{\rm obs, \, line} &=  \int_{0}^{x(t)} \int_{y_{\rm min}}^{y_{\rm max}} 2\pi \, x' \, f_V \, j_{\rm line}(\ell, t) \, {\rm d} y \, {\rm d}x' , \\ 
    &\approx \pi  \int_{y_{\rm min}}^{y_{\rm max}} \, x^2 \, f_V \, j_{\rm line}(\ell) \, {\rm d} y  , 
\end{align}
or,  after the variable transformation introduced above, as 
\begin{equation}
    L_{\rm obs, \,line}(t) \approx \pi c^3 t^3 \int^{\phi_\star(t)}_\pi {\sin^3 \phi \over (1 - \cos\phi)^4} \, f_V \, j\left({ct \over 1-\cos\phi}\right) \, {\rm d}\phi .
\end{equation}
Again, under the assumption of a tight spatially localized emissivity $j \approx j_0 \delta(\ell/\hat R - 1)$, and spherical symmetry of the emitting region, this integral can be evaluated explicitly 
\begin{equation}
    L_{\rm obs,\, line}(t) \approx \pi f_V \hat R^3 j_0 \left({2ct\over \hat R}\right) \left[1 - {ct \over 2 \hat R}\right], 
\end{equation}
which is a linear rise, rather than the $\sim t^{1/2}$ found for larger radius regions. 



\begin{figure*}
    \centering
    \includegraphics[width=0.48\linewidth]{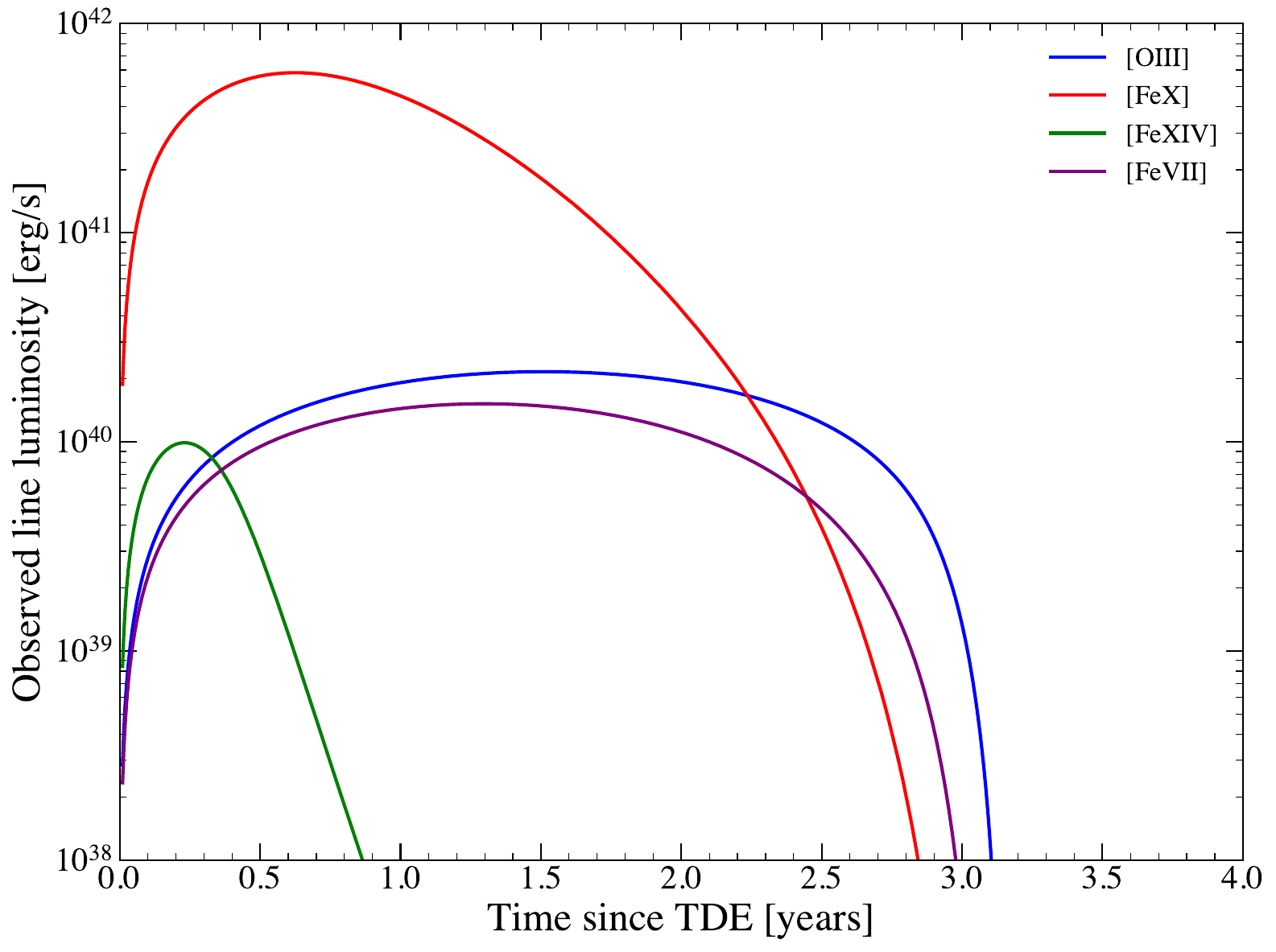}
    \includegraphics[width=0.48\linewidth]{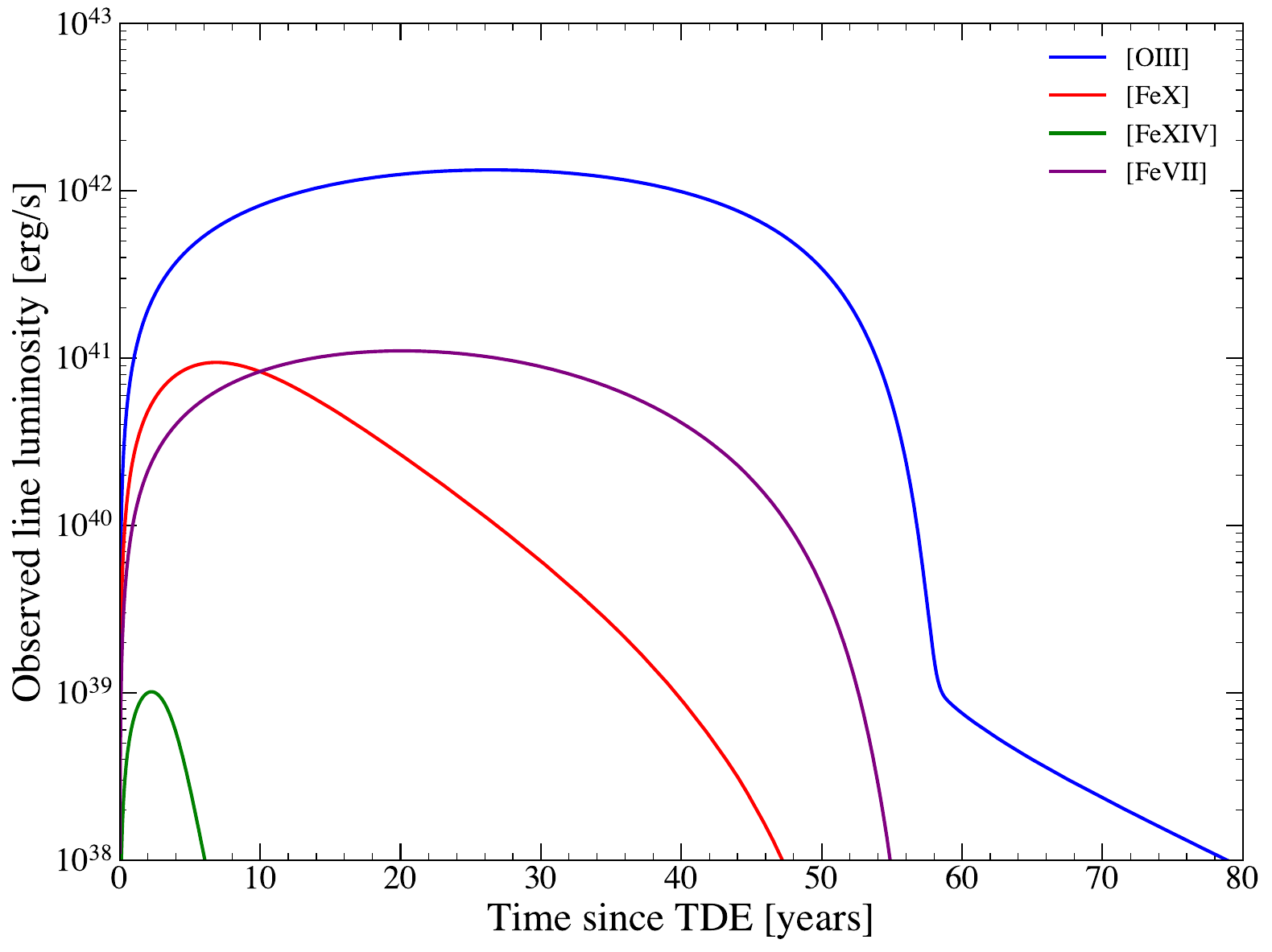}
    \caption{Extreme Coronal Line ``Light curves'' for the reprocessing of TDE disk emission through nuclear gas clouds of two different properties. {\bf Left:} $n_H = 10^8\, {\rm cm}^{-3}$ and $f_V = 10^{-4}$. {\bf Right:}  $n_H = 10^6\, {\rm cm}^{-3}$ and $f_V = 10^{-4}$. Owing to the higher densities $n_H \sim 10^6 - 10^8 {\rm cm}^{-3}$ required to produce coronal lines, these light curves evolve quickly (on $\sim$ year-to-decade timescales). Note the different horizontal and vertical axes on each plot.   }
    \label{fig:ecle_lc}
\end{figure*}

In Figure \ref{fig:ecle_lc} we display two examples of extreme coronal line ``light curves'' resulting from the reprocessing of TDE disk emission. The two scenarios differ in their assumed cloud densities ($n_H = 10^8\, {\rm cm}^{-3}$ for the left panel, and $n_H = 10^6\, {\rm cm}^{-3}$ for the right panel), and both assume $f_V = 10^{-4}$. (Note the different horizontal and vertical axes on each plot.) We see that, depending on the properties of the gas clouds surrounding the event, the relative luminosity of [OIII] and [FeX] can swap, and lines can evolve on $\sim$ year or $\sim$ decade timescales. The left (higher density) evolution is broadly in keeping with the evolution observed in AT2022upj (brighter [FeX] than [OIII], which both evolve on $\sim$ year timescales) although more detailed comparisons are necessary to be definitive. We remind the reader that the calculation performed when producing the left hand panel in Figure \ref{fig:ecle_lc} assumes that the TDE disk quickly cools below coronal line excitation energies (i.e., roughly $E\sim 100$ eV), something which may not be true for many TDEs. We stress that a more careful analysis must be performed when comparing to actual TDE systems. 

We note that (e.g.) [OIII] lines switching on rapidly ($\sim$ years post TDE) and then evolving relatively quickly ($\sim$ years-decades) as a result of the  interaction between high density gas and TDE radiation on galactic centre scales does not of course preclude much longer lived [OIII] EELR being produced in the same host galaxies on longer timescales from the lower density gas at large radial scales from the galactic nucleus (i.e., the analysis of section \ref{cloudy1}). In a realistic galaxy it is likely that both effects will happen one after each other. On a technical note, {\tt CLOUDY} does not have the capability to carry out such a simulation (and can only be used with one fixed density at a time), but we foresee minimal complications here\footnote{In principle the fact that photons that have ionized galactic center gas (i.e., to produce a coronal line photon) will not be able to ionize gas further out in the galaxy complicates the picture here (i.e., a sufficiently high $f_V$ near the nucleus could suppress EELR formation). The EELR structures of TDE host galaxies look sufficiency patchy (i.e., they visually look like they have a low $f_V$, e.g., Figures \ref{fig:tde_hosts}, \ref{fig:qpe_hosts}, \ref{fig:gsn_ne}) that we do not believe this effect will be important for the majority of systems, but this will be important for high $f_V$ systems.  }, and it is likely that real systems will undergo both behaviors. 

Finally, we note that there is an association between the detection of extreme coronal lines in TDE systems, and the presence of bright ``dust echoes'' \citep[e.g.,][and others]{Komossa09,Dou16,Lu2016,vanVelzen21b,Short2023,Hinkle2023,Clark2025}, as seen in mid infrared continuum observations taken by (e.g.) NEOWISE. We believe that this association is physically natural, and rather simple. In effect, what we believe is occurring in these systems is that both the light and dust echoes (i.e., both the coronal lines and the infrared continuum) are simply reflecting back the same transient disk emission to the observer. The modeling of this IR continuum echo is not the focus of this work, as it has already been thoroughly performed by, e.g., \citet{Lu2016}. However, our ECLEs-disk modeling allows us to note that the association between ECLEs and dust echoes arises naturally from the high gas densities required to produce observable coronal lines. This follows from the simple premise that regions with abundant gas also contain dust, given that the gas-to-dust ratio—approximately 100 by mass in the Milky Way—is not expected to vary by more than a factor of a few in external, non-active galaxies \citep{Draine2007}. The same ionizing photons which would cause coronal transitions to be excited would, if they encountered dust, cause dust to be heated which will eventually result in an infrared echo being detected. Those systems which have insufficient gas densities to result in observable coronal lines then likely have lower dust densities, and likely result in weaker (and therefore likely undetectable) dust echoes. In the following section we examine other possible observable features in the infrared region of the spectrum. 

\section{Infrared light echoes and prospects for JWST}\label{jwst}
While the analysis of the proceeding two sections focused on known observational features of TDE systems, we move in this section to some predictions of what may well be detected by future observations with new instruments, and in particular observations with the infrared instruments onboard the James Webb Space Telescope (JWST), more specifically the line features to be observed simultaneously to the IR-continuum dust  echos, months-years after the TDE, using e.g., the Mid-Infrared Instrument \citep[MIRI,][]{Rigby2023}. This analysis draws on the results of the proceeding section, and considers higher-density near-nuclear gas distributions.

The MIRI instrument observes across the mid-infrared (5-28$\mu$m) and in this section we repeat very similar analyses to before, but turn our focus to those lines which could in principle be detected throughout the mid-infrared part of the electromagnetic spectrum. In particular, we focus our attention on four Neon lines, namely [NeVI], [NeII], [NeV] (specifically the NeV line at 14.32 $\mu$m) and [NeIII], as they span interesting infrared regions from $\lambda \approx 7 \mu {\rm m}$ [NeVI] to $\lambda \approx 16 \mu {\rm m}$ [NeIII], and require high energy photons (at least $E \gtrsim 100$ eV for [NeVI], for example) to become excited, which would be the signature of disk-feedback induced transitions. These lines are of further interest, as their ratios are also used as accretion-driven ionization diagnostics \citep[e.g.,][]{Inami13, Feuillet24} in much the same way as optical emission lines are used in BPT diagrams. 

\begin{figure*}
    \centering
    \includegraphics[width=0.49\linewidth]{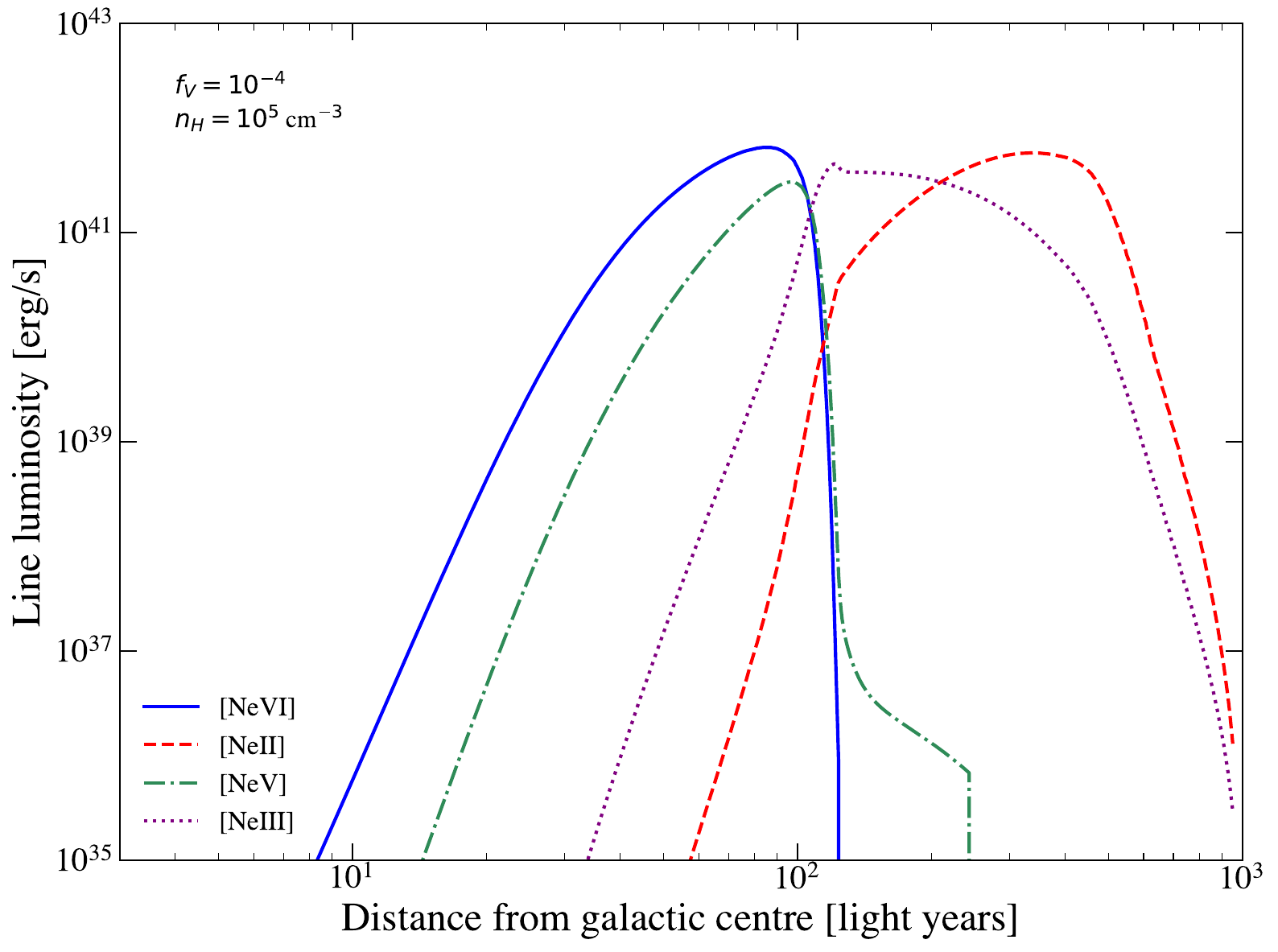}
    \includegraphics[width=0.49\linewidth]{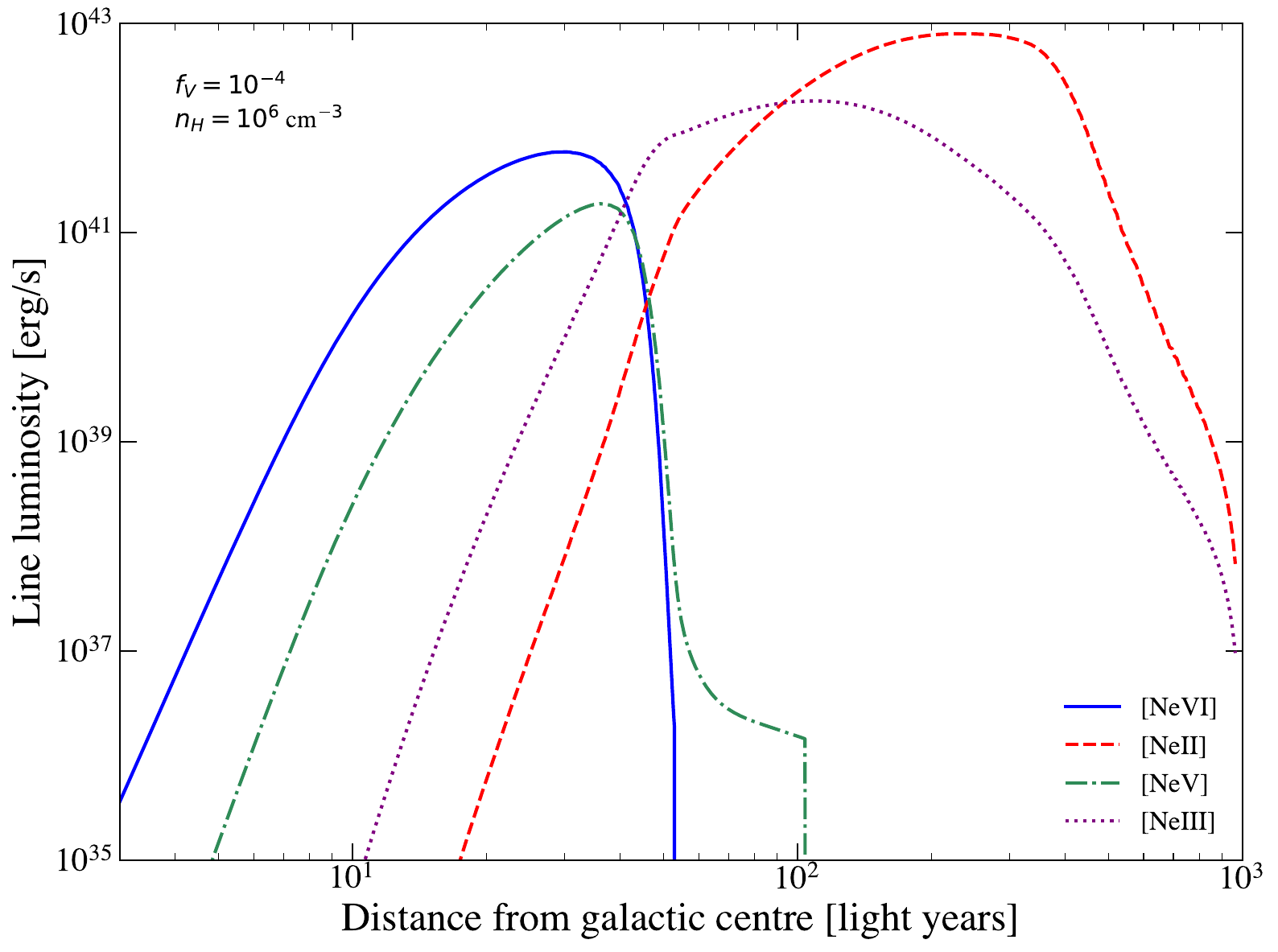}
    \includegraphics[width=0.49\linewidth]{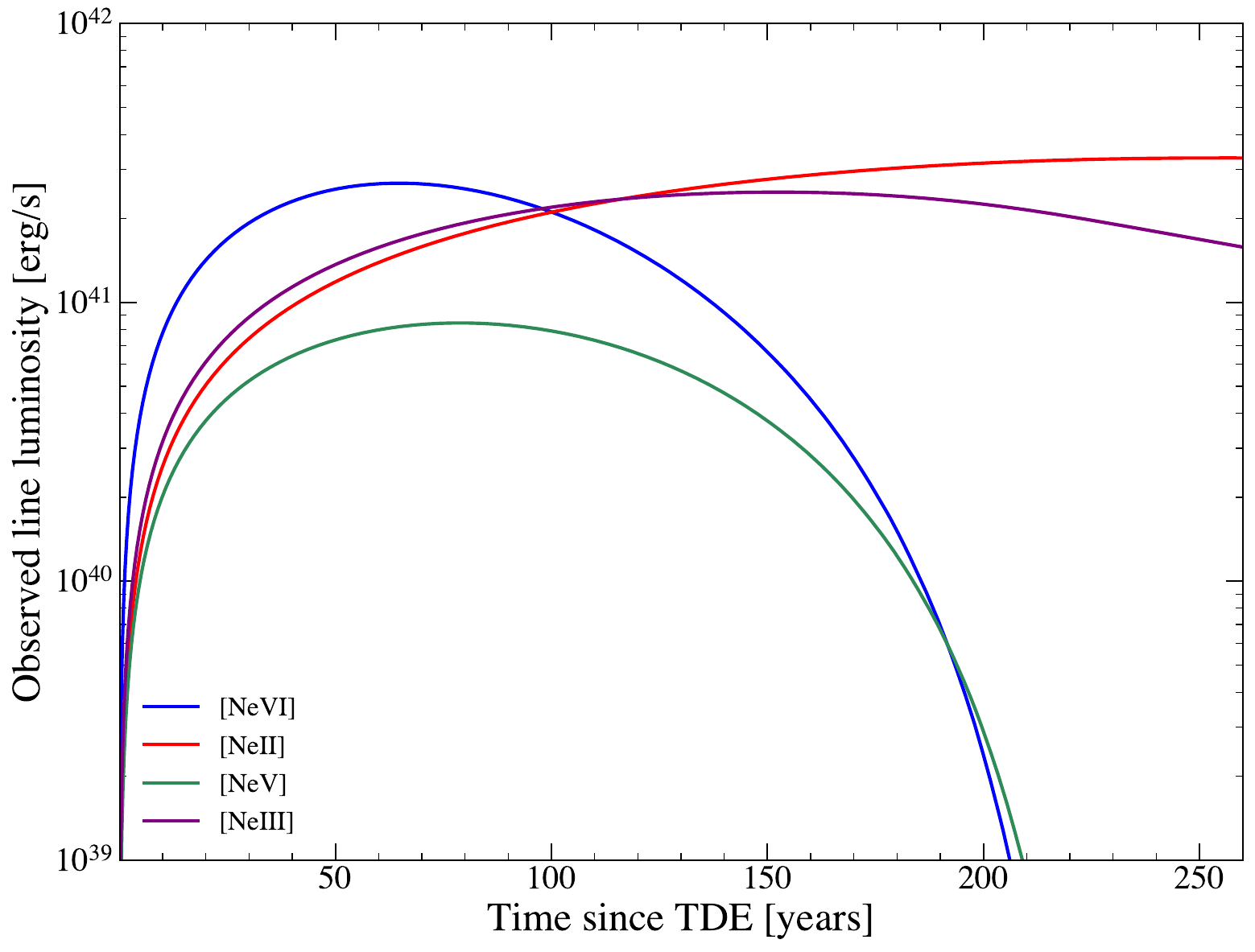}
    \includegraphics[width=0.49\linewidth]{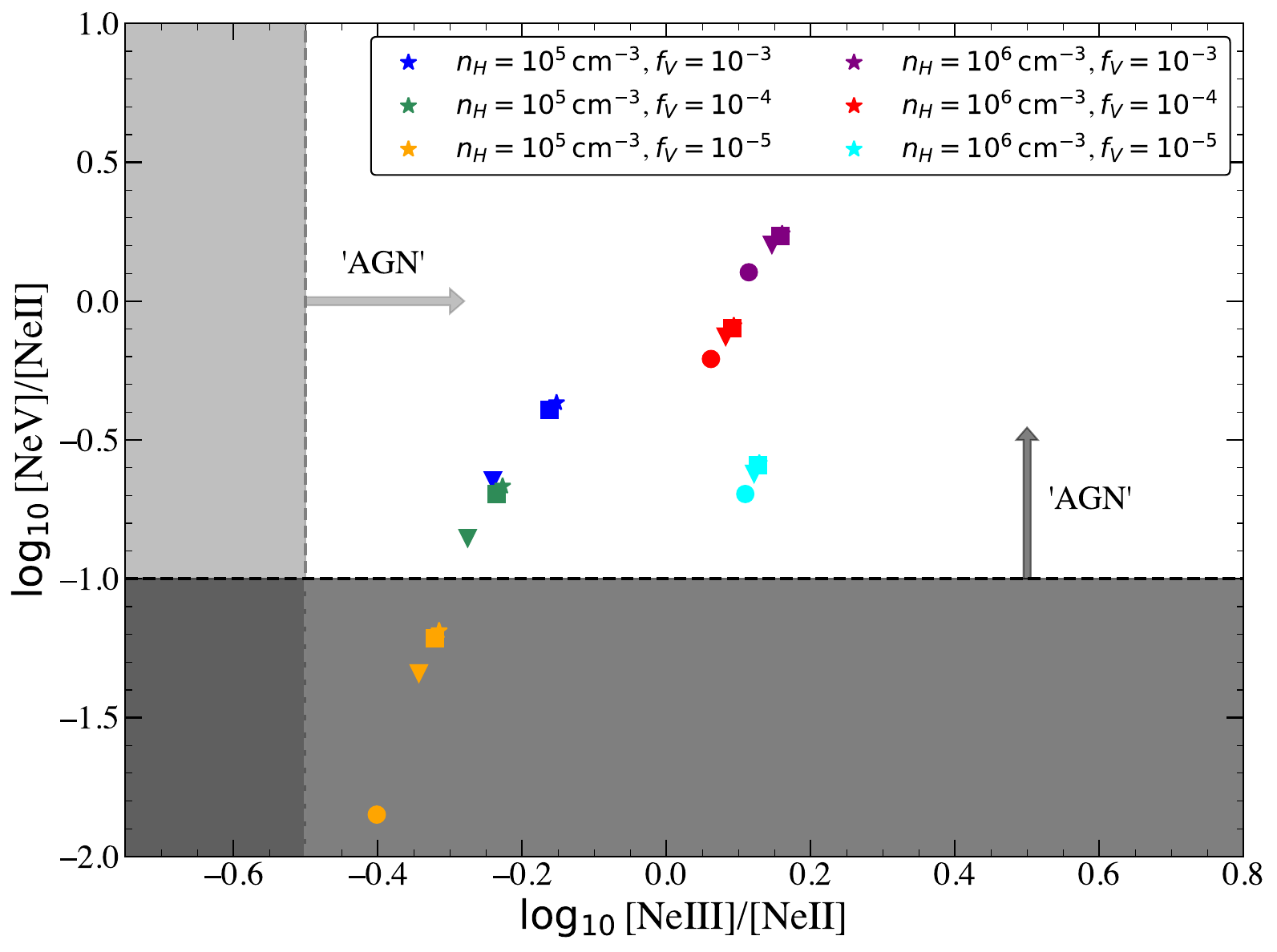}
    \caption{Various properties of infrared lines produced by excited Neon transitions, namely [NeVI], [NeII], [NeV] and [NeIII]. {\bf Upper right and left:} line emission profiles for the four neon lines, for different cloud densities (left $n_H = 10^5\, {\rm cm}^{-3}$, right  $n_H = 10^6\, {\rm cm}^{-3}$, both have $f_V = 10^{-4}$) as a function of distance from the galactic center. We reiterate the earlier point that this is now what a distant observer would see (see text). {\bf Lower left:} the ``light curves'' of the reprocessed infrared lines for the cloud properties $n_H = 10^5\, {\rm cm}^{-3}$ and $f_V = 10^{-4}$, showing that these Neon lines switch on rapidly ($\sim$ years post TDE) and remain bright for $\sim$ decades-centuries. {\bf Lower right:} these TDE reprocessed emission lines on a Neon diagnostic diagram, for a range of cloud parameters (see figure legend) and times since TDE (denoted by marker style: star = +1 year, square = +5 years, triangle = +20 years, circle = +50 years). We note that for a range of cloud parameters the Neon lines would indicate the presence of an ``AGN'' (see text for more details).   }
    \label{fig:neon}
\end{figure*}

We repeat the analysis of the preceding sections, and take the spectrum of a TDE disk at peak luminosity, and process it through {\tt CLOUDY} to compute the emissivities of the four infrared neon lines [NeVI], [NeII], [NeV] and [NeIII], for a range of different cloud properties. We found that gas densities of $n_H \sim 10^4-10^7\, {\rm cm}^{-3}$ produced bright Neon line features, which typically peak in emissivity at distances $\sim 10$'s $-100$'s light years from the galactic center for [NeV] and [NeVI], and at larger radii $(\sim 100$'s $-1000$'s of light years) for [NeII] and [NeIII].   This can be clearly seen in the naive line luminosities $(L = 4\pi r^3 j/3)$ plotted in the upper two panels of Figure \ref{fig:neon}. 

Moving to the observed luminosity as recorded by a distant observer, we compute TDE infrared echo light curves in the lower left panel of Figure \ref{fig:neon} (by solving the integrals introduced above). This was for the particular cloud parameters $n_H = 10^5\, {\rm cm}^{-3}$ and $f_V = 10^{-4}$, showing that these Neon lines switch on rapidly ($\sim$ years post TDE) and remain bright for $\sim$ decades-centuries. These should, therefore, be detectable with JWST observations of known TDE host galaxies, as sufficient time will have passed for these lines to have switched on. We note that, just as in previous sections, the precise amplitude and timescales of the Neon line evolution are set by the properties of the clouds, and the disk parameters. Generically, however, we find that Neon lines turn on quickly ($\sim$ months-years) for those systems which become observable. 

Finally, we turn to the question of whether TDE induced Neon line emission would place TDE host galaxies (or at least their nuclear regions) onto ``AGN''-like parts of diagnostic diagrams (we stress again, that this just mean that they are ionized by sources with spectra harder than star-forming regions and their {\it O}-stars). For this we turn to the line ratios $\log_{10}$[NeV]/[NeII] and $\log_{10}$[NeIII]/[NeII]. Specifically, \cite{Inami13} suggest that $\log_{10}$[NeV]/[NeII] $> -1$ is indicative of accretion power, and \cite{Feuillet24} similarly define $\log_{10}$[NeIII]/[NeII] $> -0.5$ as an ``AGN'' characteristic. These two ratios are simple to compute from our coupled TDE disk-{\tt CLOUDY} analysis, and we show the location of six different disk-cloud systems in the lower right panel of Figure \ref{fig:neon}. 

In this panel we display the TDE reprocessed emission lines on a Neon diagnostic diagram, for a range of cloud parameters (see figure legend) and times since TDE (denoted by marker style, where a star = 1 year post TDE, a square = 5 years, a triangle = 20 years and a circle = 50 years). We note that for a range of cloud parameters the Neon lines would indicate the presence of an ``AGN'', according to both the \cite{Inami13} and \cite{Feuillet24} diagnostics.

It appears that JWST observations of known TDE hosts would be an interesting probe of the physics of TDE accretion disks, and the circumnuclear gas of their hosts. Of course, there a huge range of atomic lines detectable in the infrared, and we have here only focused on Neon. Our analysis can trivially be extended to a much wider range of lines, should future observations with JWST find it necessary. 

While the focus of this section has thus far been concerned with mid-infrared JWST observations of known TDE host galaxies, it is worth noting that TDEs likely represent a contaminant for AGN studies which make use of infrared lines detected by JWST (much like they represent a containment of EELR studies discussed above). An important possible example of this is those galaxies which are observed at high redshifts, which make use of redshifted  optical emission lines with the aim of characterizing AGN in the early universe. It is beyond the scope of this work to analyze the combined effects of a changing TDE rate and different stellar populations with redshift, and their implications for the ionization signatures of TDEs in the early universe, but we believe it worth bearing in mind as more and more of these sources are discovered. 

Finally, \cite{Masterson24} have reported on a new population of heavily obscured TDEs candidates, where only a dust echo is detected without any optical/X-ray flare  (presumably a result of extremely high obscuration in the galactic nuclei $f_V \sim {\cal O}(1)$). While we have restricted our analysis to at most moderately $f_V \sim 10^{-3}$ obscured TDEs in this work, we can extrapolate the results derived here to the limit of higher obscuration. We see in Figure \ref{fig:neon} that larger obscuration increases the ratio of both [NeIII]/[NeII] and [NeV]/[NeII], meaning that these obscured TDEs may represent the best avenue for searching for Neon lines in TDEs. Further, increasing $f_V$ quite generally results in brighter emission lines with emissivities which peak closer to the centers of their host galaxies (Figure \ref{fig:cloudy1}), and so these lines should ``switch on'' quickly and likely be easily detectable in a distant observers frame. 

\section{Transient narrow line regions and Voorwerpen }\label{cloudy3}

In this section we discuss a simple, and important, extension of the analysis of sections \ref{cloudy1} and \ref{cloudy2}. Namely the ``switching on'' of galactic center narrow line regions in TDE host galaxies, which is an inevitable, although not widely discussed \citep[with some exceptions, e.g.,][]{Patra24} consequence of the formation of accretion flows in these events. 

The results of sections \ref{cloudy1} and \ref{cloudy2} imply that, should there be circumnuclear gas with the right properties (i.e., densities in the range $n_H \sim 10^5-10^8\, {\rm cm}^{-3}$), that galactic center narrow line regions ([OIII], [NII], H$\alpha$, etc.) should switch on relatively rapidly (in the distant observers frame) after a large fraction of TDEs. The exceptions to this will be those TDEs in `retired'/`quenched' gas-less galaxies, or in those galaxies with so much circumnuclear gas and dust that very little emission will escape at all,  although these seem like they will represent a minority of {\it observed} TDEs found in optical surveys. 

\begin{figure}
    \centering
    \includegraphics[width=0.48\linewidth]{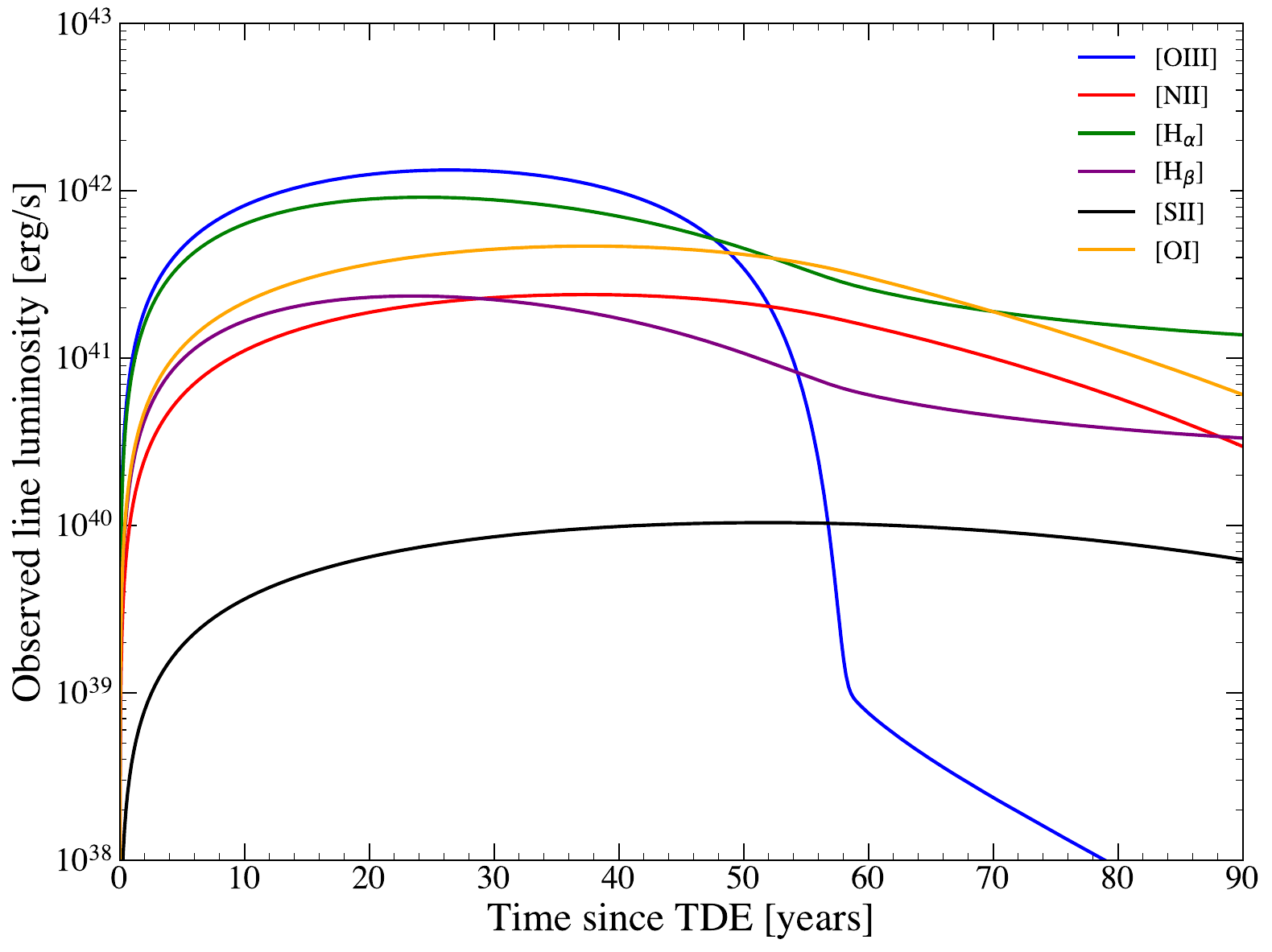}
    \includegraphics[width=0.48\linewidth]{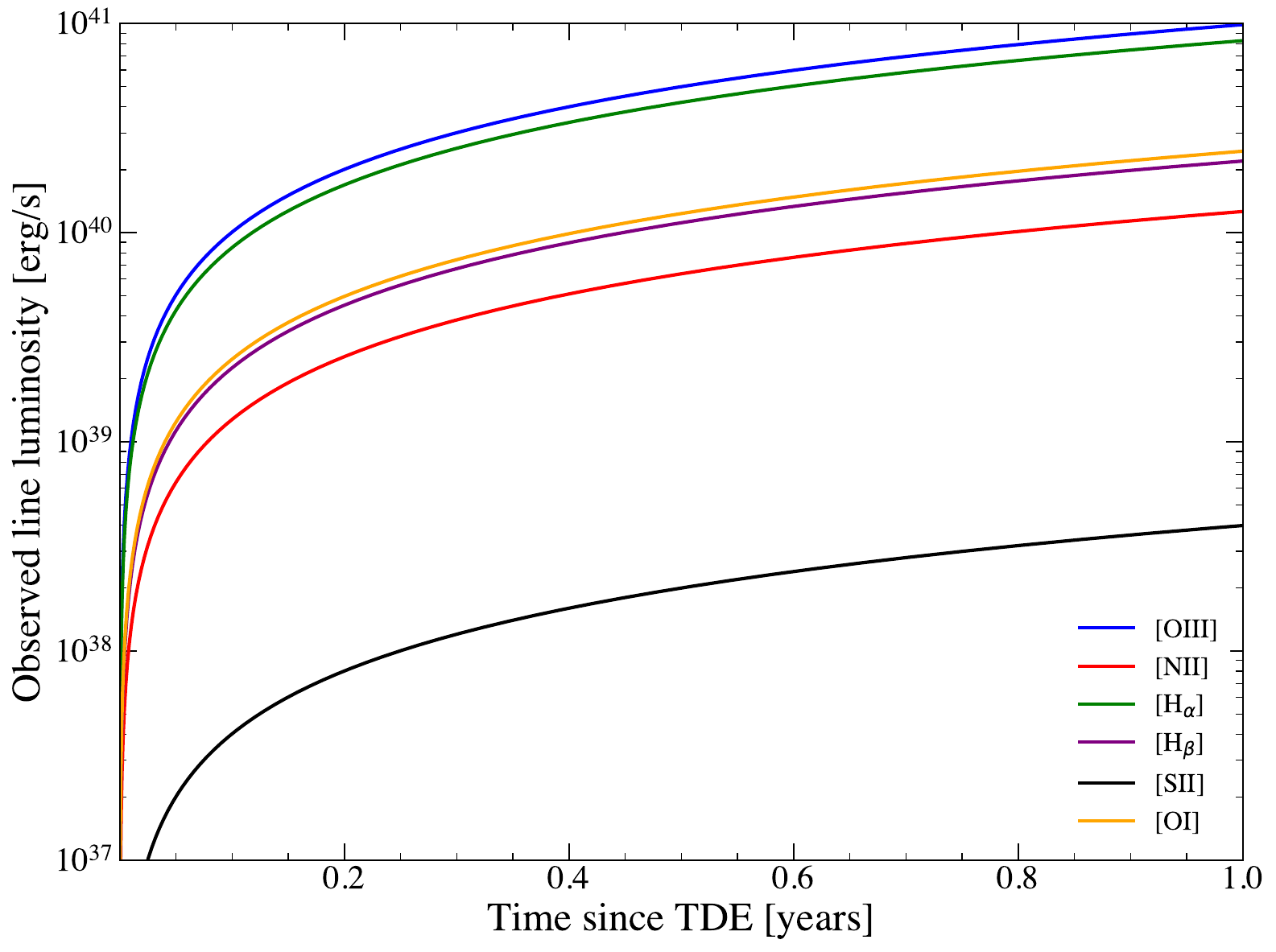}
    \caption{The evolving line luminosities of various ``typical'' narrow line region species, as a function of time since the onset of accretion for a cloud population of density $n_H = 10^6\, {\rm cm}^{-3}$ and $f_V = 10^{-4}$. On the left we show the long term evolution of the system, with line features evolving over $\sim$ decades. On the right we demonstrate just how quickly line features can switch on post disk-formation, by examining the system on even shorter timescales. Bright [OIII] detectable just $\sim$ months after the onset of accretion. This will inevitably complicate any analysis of nuclear emission line signatures of galaxies observed post-TDE.  }
    \label{fig:TNLR}
\end{figure}

To be explicit, we again turn to coupled TDE disk-{\tt CLOUDY} simulations. We now examine the effects of the reprocessing of TDE disk emission through dense circumnuclear clouds on timescales $\sim$ years-decades post TDE. Bearing in mind the caveats of probing timescales on which both the TDE disk spectrum and the reprocessed emission vary, we can examine both the observer-plane geometry and luminosity of various classical narrow line features. In Figure \ref{fig:TNLR} we display a series of evolving narrow line region species, as a function of time since the onset of accretion for a cloud population of density $n_H = 10^6\, {\rm cm}^{-3}$ and $f_V = 10^{-4}$. On the left we show the long term evolution of the system, with line features evolving over $\sim$ decades (this will likely be extended somewhat by a long lived accretion flow). On the right we demonstrate just how quickly line features can switch on post disk-formation, by examining the system on even shorter timescales. Bright [OIII] detectable just $\sim$ months after the onset of accretion. (We reiterate the point made in section \ref{cloudy2} that this of course does not preclude the same lines switching on at much later times from lower density gas in the outskirts of the same galaxy.) 

While not taken into account in the above plot (as the density was high), perhaps the most interesting post-TDE behavior of transient narrow line regions will be in those systems with galactic center gas clouds with densities in the $n_H \sim 10^4-10^5\, {\rm cm}^{-3}$ range. These systems (which may well be relatively common) have Hydrogen recombination times $[\tau_{{\rm rec},{\rm  H}} \sim 20 (n_H/10^4\, {\rm cm}^{-3})$ years$]$ which are long (in an observationally relevant sense), compared to the corresponding [OIII] and [NII] recombination timescales $\tau_{{\rm rec}, {\rm OIII}} \sim 0.1 \tau_{{\rm rec}, {\rm H}}$, but still have sufficient gas density for near-galactic center narrow lines to be excited. This means these systems will become (e.g.) [OIII] bright on $\sim$ year timescales which is significantly different from their Hydrogen emission switching on (on $\sim$ decade timescales). This will likely lead to dynamic evolution in the BPT plane, with this evolution taking place over timescales comparable to the lifetime of optical surveys. While the authors are not aware of any such system in the current (small) TDE sample, it seems likely that future systems will be detected, and will represent a particularly interesting class of objects. 

The above results has an obvious, but important, implication. Namely that much of the analysis on the properties of the galactic centers of TDE hosts takes place, naturally, after (sometimes by a few years) the TDE has been first detected.  We have demonstrated in this work that the presence of a TDE can switch on narrow line features, with reasonably large projected distances from the galactic center, rapidly following TDE disk formation. It is essential therefore that these features are not attributed to previous (recent) AGN activity without due care and attention, as they may simply be further observations of the current TDE (via a light echo). This requirement of extra care also applies to the modeling of stellar populations of TDE host galaxies which do not account for the bright and blue accretion flow which persists long after the initial flare. Not accounting for this disk emission will result in erroneously young stellar populations being inferred for the galactic centers of TDE hosts (the only stellar populations which can mimic the blue disk spectrum). Such a disk contribution can be accounted for carefully \citep[e.g.,][]{Guolo25}, for example by using a flexible fitting code such as {\tt Cigale} \citep{Boquien19}. 

Similarly, as TDEs happen frequently (once every $\sim 10^5$ years) in low mass galaxies, a reasonable fraction of the total galaxy population will have ionization signatures which can be traced to a previous TDE, but will have much fainter current nuclear activity (we anticipate that these will be galaxies which had a TDE $\sim 10–1000$ years ago). Many of these will be naively interpreted as AGN, owing to their location on various diagnostic diagrams, but will not be observed to have sufficient (current) galactic center emission (from other AGN diagnostics, e.g. infrared emission, broad line regions and hard X-ray corona) to explain the features. These may therefore be interpreted as ``fading AGN'', ``LINERs'' or ``Voorwerpen'', but will in reality be TDE light echoes. This may represent one route through which various discrepancies between AGN catalogues \citep[e.g.,][]{Ellison16} at different wavelengths may be resolved,  as these sources will only be designated an ``AGN'' if observed by an optical emission line survey, but will lack all other classical AGN features (i.e., an X-ray bright corona, radio emission, etc.). 

This is a potentially  interesting shift in our thinking about fading AGN candidates. As an explicit example \cite{Keel17} discuss a number of fading AGN candidates, all of which they inferred to have shown a significant fading event in the last $\sim 10^4$ years. We suggest that rather than a canonical AGN fading $\sim 10^4$ years ago, the host galaxies identified by \cite{Keel17} may have hosted a TDE on timescales $\sim 10^4$ years ago, which briefly injected a large ionizing flux into the galaxy, which is being echoed to us now. When observing the galactic center today, no remanent of a TDE $10^4$ years ago would be detectable (as the disk would have fully accreted on this timescale). 

One interesting, and testable,  observational prediction for this ``some Voorwerpen are TDE light echoes'' paradigm is related to a simple fact regarding tidal forces and black holes: if TDEs dominate (or even provide a moderate fraction of) the so-called ``fading-AGN'' population, then there should be a strong black hole mass (and therefore galaxy mass) suppression of the presence of ``fading AGN'' above $M_\bullet \sim 10^8 M_\odot$, for the following simple reason. For a tidal disruption event to ionize gas clouds in its host galaxy, the photons emitted in the event must be emitted from radial scales larger than the black holes event horizon, something which is not guaranteed for stellar tidal disruptions around all black hole masses. The (Newtonian) tidal acceleration is given by $a_T \sim GM_\bullet R_\star/r^3$, where $R_\star$ is the radius of the incoming star. To overcome the stars self gravity $a_\star \sim GM_\star/R_\star^2$, the star must enter a radial scale $r_T \sim R_\star (M_\bullet/M_\star)^{1/3}$. This radius, in units of the event horizon $r_+ \sim GM_\bullet/c^2$, is given by $r_T/r_+ \sim c^2 R_\star / G M_\bullet^{2/3} M_\star^{1/3}$. If this ratio drops below unity, or equivalently if the black hole mass exceeds $M_\bullet \gtrsim c^3 R_\star^{3/2} / G^{3/2} M_\star^{1/2}$ which for solar properties is $M_\bullet \sim 10^8 M_\odot$ \citep[known as the][mass]{Hills75}, then the star will be first disrupted inside the event horizon, and no ionizing radiation can be released into the galaxy. A fully relativistic expression for this mass scale can be derived in closed form \citep{Mummery24}, and for rapidly rotating black holes this Hills mass can be increased by up to an order of magnitude, but the qualitative effect of this simple Newtonian point is unchanged, and any observational feature which is produced by tidal disruption events should show a sharp (super-exponential) suppression above this characteristic mass scale. Studying the black hole (or more plausibly the galaxy) mass dependence of the presence (or lack thereof) of ``fading AGN'' would be a good test of the paradigm put forward here, which we actively encourage. 

\section{ The Milky Way }\label{MW}

\begin{figure}
    \centering
    \includegraphics[width=0.65\linewidth]{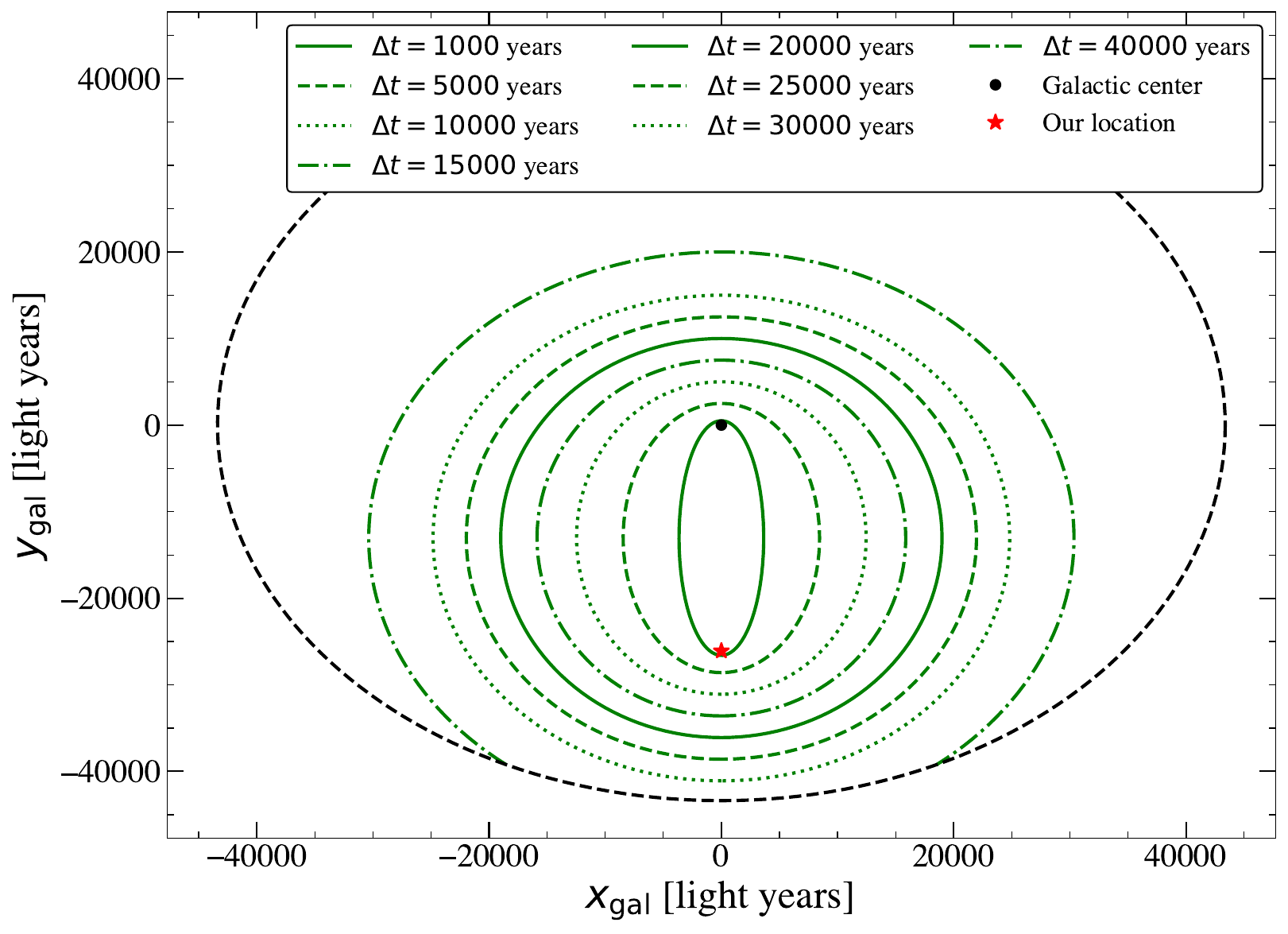}
    \caption{The iso-delay contours of reprocessed emission which will be observable in our frame from a TDE in the Milky Way which we ``missed'' by a time $\Delta t$ (i.e., the direct emission from the TDE passed through the Earths location a time $\Delta t$ ago). The size of each iso-delay contour grows with increasing $\Delta t$. As in Figure \ref{sizes} these contours can be formally rotated by $2\pi$ about the line joining the galactic center to the Earth (i.e., $x_{\rm gal} = 0$). Ionized emission detected on one (or more) of these contours would be indicative of past TDE activity in the Milky Way. 
    }
    \label{fig:milkyway}
\end{figure}

In this final section we speculate about the possibility that a previous TDE in the Milky Way (which is itself a green-valley galaxy \citep{Licquia15} with central black hole mass below that of the Hills mass, and therefore likely has an elevated TDE rate), could be inferred from ionization signatures in the Galactic outskirts. 

The actual line emissivity profiles that might be observable from reprocessed TDE emission in the Milky Way are in effect identical to those computed earlier (e.g., Figures \ref{fig:cloudy1}, \ref{fig:light_echo}, \ref{fig:BPTNII}, etc.), the major difference is of course the observed geometry of the light echo. 

In Figure \ref{MW} we display the isodelay contours of an observer within the same galaxy (we take a distance $D = 8$ kpc) as a TDE which they ``missed'' by a variety of times $\Delta t$. These isodelay contours would be relatively tight in the Milky Way, as they would presumably be limited in extent to only $\Delta r \sim c \Delta t_{\rm ION} \sim 100$ light years. Ionized emission detected along one (or more) of these contours would be indicative of past TDE (or at least nuclear) activity in the Milky Way.

We note, speculatively, that the [O III]-bright filament recently discovered in the direction of M31 by amateur astronomers \citep{Drechsler2023} appears to be Galactic in origin \citep{Lumbreras24}. Its properties suggest the need for an source of relatively high-ionizing radiation, particularly since it is brighter in [O III] than in any other optical emission line.
While other ionizing sources are possible, past nuclear activity—potentially a TDE—is worth considering, especially given that \citet{Lumbreras24} find no obvious alternative ionization sources near the filament. Although, \citet{Lumbreras24} argued that the line ratios e.g., [OIII]/H$_{\beta}$ are not high enough, as compared to typical EELR (powered by long-lived AGN), we have shown in section \ref{cloudy1} that TDE echoes can, effectively, populate any region of the BPT, depending of the properties of the TDE, the gas cloud, and the time since the event. A systematic search for similar highly ionized ISM structures with a Galactic narrow-band [O III] survey—akin to existing H$\alpha$ surveys \citep[e.g.,][]{Haffner2003,Barentsen2014}—could reveal new structures and further support the hypothesis that they trace echoes of past Galactic TDEs.

We note that the possibility of previous TDE activity in the Milky Way has been discussed in the community seeking to understand the Fermi/eROSITA bubbles \citep[see e.g.,][for a review]{Sarkar24}. In particular, \cite{Ko20, Scheffler25} suggest that this feature can be explained in its entirety by hydrodynamic feedback (i.e., mechanical feedback from the interaction between outflows and the surrounding medium) from regular tidal disruption events in the last $\sim$ million years.  If this paradigm is indeed correct, then accompanying these features from supposed mechanical feedback must be ionization signatures from the radiative feedback from a succession of TDE disks which have formed over the last few million years.  We suggest that searching for these ionization features may well provide important constraints on such a model for the the eROSITA bubbles, as well act as a possible probe of the gas distribution in the Milky Way. 




\section{ Discussion and conclusions }\label{conc}
In this paper we have sought to present a first analysis of the potential of those disks formed in tidal disruption events to act as drivers of radiative feedback in their host galaxies, seeking first to explain two interesting observational properties of TDE host galaxies: (i) the host galaxies of (current) tidal disruption events show a propensity for extended emission line regions with no recent (or current) classical active galactic nuclei present; (ii) many tidal disruption events are rapidly followed by the detection of extreme coronal lines, which are extremely rare in other galaxies. 

The argument put forward to explain these observations in this paper is at its heart very simple: (i) TDEs are more likely to be discovered in galaxies with a high TDE rate, and therefore in galaxies which have more recently (than usual) had a previous TDE; (ii) TDEs result in disks which produce a bright $\sim 10^{44}$ erg/s, long-lived (a minimum of one decade), ionizing source of radiation; (iii) this emission is sufficiently powerful to result in ionization features observed on galactic scales, provided that molecular gas is present with broadly the right characteristics ($n_H \sim 10^{1}-10^{3}\, {\rm cm}^{-3}$, $f_V \sim 10^{-4}-10^{-5}$), or produce coronal lines (and ``classical'' narrow lines) on galactic center scales for more dense $n_H \sim 10^5–10^8\, {\rm cm}^{-3}$ clouds. The first two points discussed above are observational facts, while the properties of the molecular gas required to produce galaxy-scale recombination features are similar to those of normal molecular clouds, and should therefore be relatively common. Indeed, such a density scale is exactly what is inferred from detailed observations of one (with the highest S/N) EELR in a TDE/QPE host galaxy (Figure \ref{fig:gsn_ne}). We expect that dense circumnuclear gas clouds will be present in a large fraction of TDE hosts, helping explain their propensity for producing coronal lines. 

We have shown that in such situations the typical line ratios that will be observed in such TDE systems (when superimposed on top of typical star forming galaxies), will appear above the \citep{Kewley01} line in the BPT diagrams, a.k.a in the ``AGN region'', (e.g., Figures  \ref{fig:obs_maps}, \ref{fig:BPTNII}, \ref{fig:fagn}, \ref{fig:BPTSII}, \ref{fig:BPTOI}), despite of course not being produced by a {\it classical} AGN.
We believe this is a natural resolution of the observations of an overabundance of  EELR in TDE host galaxies, and that what one is actually observing in these systems is in fact (at least) two TDEs at once, one unfolding in real time and another echoing off molecular clouds in the outskirts of its galaxy. 

Further, we have demonstrated that TDE disks are capable of producing coronal lines, such as [FeX] and [FeXIV], provided that the radiation is reprocessed by high density clouds near to the galactic center. The different ionization potentials $U$ of these coronal lines suggest that [FeXIV] should generally be fainter than [FeX] emission, and should fade quicker. These findings are entirely in-keeping with observations \citep[e.g.,][among others]{Newsome2024_22upj}. 

One of the important points we have discussed in this paper is the concept of an isodelay contour, and apparent superluminal motion in TDE light echoes (section \ref{geom}, Figure \ref{sizes}). While such apparent superluminal effects are well known in the radio community \citep[e.g.,][]{Rees66, Mirabel94, Davis91}, they do not appear to be so commonly applied in the TDE optical reverberation community. One of the important implications of the finite speed of light is that a wide range of host galaxy radii can be observed to be reprocessing a single shell of ionizing radiation simultaneously, with only the {\it lower} limit of the radii probed by this reprocessing set by the time between flares (not the {\it upper} limit which is often assumed). This means that: (i) a single shell of ionizing radiation can be observed to ``fill in'' the entire observed plane of the host galaxy (Figure \ref{fig:obs_maps}); (ii) it is an extremely non-trivial problem (in effect impossible) to infer the time between flares from spatially resolved emission line features, or visa versa the radial scale of reprocessing clouds from the time lag between the onset of lines and the flare origin. 

We now discuss other implications of this work. While we initially focused on large (galaxy) scale EELR features in this work, everything we  discussed on large scales applies equally well to small radial scales, provided that there are higher density clouds $n_H \sim 10^{4}–10^6\, {\rm cm}^{-3}$ on these smaller scales. 
Indeed, a similar analysis to our work, but focusing on narrow line regions observed on small radial scales in GSN 069 \citep{Miniutti2019}, has recently been performed by \cite{Patra24}, and we are in broad agreement with their findings. 

One result we have discussed here which was not discussed in \cite{Patra24} are the effects of apparent superluminal propagation of these ionization fronts on observations which probe small radial scales. Indeed, this superluminal motion of the light echo could result (provided there are clouds of gas between the observer and disk) in emission line structures becoming observable with projected radial offsets from the nucleus of $r \sim {\cal O}(10'{\rm s})$ of parsec scales on extremely short timescales ($\sim$ months) post TDE.  Indeed, such nuclear narrow line regions should be common in TDE systems. It is important therefore that care is taken when interpreting observations of galactic center emission line features from observations taken post TDE, even if the naive light travel time between the clouds is longer than the time post TDE flare (section \ref{geom}). 

Secondly, we have made a series of predictions of which infrared lines may be detected in future observations of TDE host galaxies by the JWST, particularly those with detectable infrared dust echoes. We focused on a series of Neon lines namely [NeVI], [NeII], [NeV] and [NeIII], as they span interesting infrared regions from $\lambda \approx 7 \mu {\rm m}$ [NeVI] to $\lambda \approx 16 \mu {\rm m}$ [NeIII], and require high energy photons (at least $E \gtrsim 100$ eV for [NeVI], for example) to become excited, which would be the signature of disk-feedback induced transitions. These lines are of further interest, as their ratios are also used for AGN diagnostics in much the same way as optical emission lines are used in BPT diagrams. We demonstrated that all four of these lines should be (i) bright and detectable, (ii) should switch on rapidly post-TDE ($\sim$ months to years), and (iii) their relative line ratios should place them in ``AGN'' like regions of characteristic diagnostic diagrams \citep[e.g.,][]{Inami13, Feuillet24}. This should be especially important for highly obscured systems \citep[like those discovered in][]{Masterson24}. 

Thirdly, tidal disruption events happen routinely (once every $\sim 10^4–10^5$ years), in a wide range of galaxies \citep[indeed, the typical SMBH likely accretes $\sim 10^6 M_\odot$ in tidally disrupted stars over its lifetime][]{Magorrian99}. The features we have discussed here should be represented at different relative rates in a broad population of all galaxies therefore (provided, as we discussed above, that they host a black hole with mass $M_\bullet \lesssim 10^8 M_\odot$), including those which have not just been discovered to host a (second) TDE. We can roughly estimate the fraction of galaxies which should host a given line feature (we will denote $f_l$) from the TDE rate (${\cal R}_{\rm TDE}$) and the timescale over which the features should be observable ($\Delta t_l$). We can estimate this fraction for both EELR and coronal lines which have very different values of $\Delta t_l$. 

Very roughly, the fraction of background galaxies with these features should be 
\begin{equation}
    f_{l} \sim f_{\rm cloud} \times {\cal R}_{\rm TDE} \times \Delta t_l \sim f_{\rm cloud} \times { {\cal R}_{\rm TDE}\times  R_{\rm cloud} \over c } ,
\end{equation}
where $R_{\rm cloud}$ is the rough size out to which the clouds with the correct structure (i.e., high density near-nuclear clouds for coronal lines, and low density galaxy-scale clouds for EELR) which can be ionized by a TDE (which sets the length of time post TDE at which EELR could be observed),  ${\cal R}_{\rm TDE} \sim 10^{-5}\, {\rm year}^{-1}$ is the TDE rate of a typical galaxy \citep[a number which can be derived on physical grounds, e.g.,][and constrained from observations \citealt{Yao2023}]{Rees88, Magorrian99}. Note that this intrinsic TDE rate will be lower than the TDE rate of those galaxies found in a survey to host TDEs \citep[e.g.,][]{Yao2023}, Finally, $f_{\rm cloud}$ is a simple parameter which describes the fraction of all galaxies which have gas clouds of broadly the right properties ($n_H, f_V$ etc.) which could possibly produce a given feature following a TDE. 

Taking ${\cal R}_{\rm TDE} = 10^{-5}\, {\rm year}^{-1}$, and EELR to have radial scales $R_{\rm cloud} \sim 10^4$ light years (e.g., Figures \ref{fig:tde_hosts}, \ref{fig:qpe_hosts}, \ref{fig:obs_maps}), then 
\begin{equation}
    f_{\rm EELR} \sim 0.1 f_{\rm cloud, \, EELR},
\end{equation}
while coronal lines are dominated by dense clouds on $\sim 1$ light year scales (e.g., Figures \ref{fig:22upj_cont}, \ref{fig:coronal_lines}), and so 
\begin{equation}
    f_{\rm coronal} \sim 10^{-5} f_{\rm cloud, \, coronal}.
\end{equation}

The relevant quantity to contrast these fractions with are the observational results of \cite{French23} who found that 6 out of 93 Post-Starburst Galaxies (which are known rather generally to have an elevated TDE rate) showed extended emission line regions, of which five out of six did not have current levels of nuclear emission which could have powered these features. \cite{French23} interpret this as evidence for a fading AGN in each of these galaxies. We suggest that it is much more natural to suppose that these post-starburst galaxies are showing signs of a (relatively) recent previous TDE. 

While roughly $\sim 5-10\%$ of Post-Starburst Galaxies show EELR (in keeping with our back of the envelope estimate),  coronal lines are much rarer, particularly in any random sample of galaxies (once obvious AGN are removed from the sample). Only $5$ non-AGN systems were found in the entire SDSS sample \citep{Komossa2008, Wang2011, Wang2012, Callow2024}, corresponding to a ballpark fraction of $f \sim 10^{-5}-10^{-6}$, again in keeping with our analysis.

We reiterate that our work raises an important point relevant for the broader studied of AGN duty cycles, namely that our work suggests that recent TDEs will act as a contaminant in populations studies of both fading AGN and also the broader optically selected AGN population. One way in which to examine if this is the case would be to study the galaxy mass (or in reality black hole mass) dependence of the presence of ``faded AGN''. Tidal disruption events can only occur below a maximum black hole mass (known as the Hills mass), where the tidal force required to disrupt a star is just reached at the horizon on the hole. For a Schwarzschild black hole this is a mass scale of roughly $\sim 10^8 M_\odot$ \citep[although this can be elevated by an order of magnitude for a maximally spinning black hole mass][]{Kesden12, Mummery24}, and the rate of TDEs is super-exponentially suppressed above this mass. A galaxy mass dependence of the prevalence of ``faded AGN'' could be taken as indirect evidence for the mechanism we lay out in this paper. 

Finally, we discuss the main shortcoming of our analysis of TDE-induced EELR, namely the assumption of statistical steady-state between the ionizing radiation field and the neutral gas, which is unlikely to be satisfied for the typical molecular clouds which we invoke here. The reason for this is entirely physical, low density clouds have recombination times of order the lifetime of a TDE disk system.  While we hope to rectify this in future studies, we discuss some possible implications of this behavior here. For example, it is likely to lead to interesting signatures on (e.g.) BPT diagrams, which may be a signature of a highly time varying (relative to the recombination timescale) ionizing source. Indeed, as the recombination timescales of [OIII] and [NII] are typically $\sim 0.1$ times the recombination timescale of hydrogen, such clouds may become (e.g.) [OIII] bright prior to hydrogen recombination, and lie significantly off the typical BPT location of similarly hard ionizing sources.   

Further, if there are spatial gradients in the gas clouds in the host galaxy (which there will of course be), then the fact these two timescales are different means that the (e.g.) Hydrogen photons probe a different region to the (e.g.) [OIII] photons, which may also lead to interesting signatures in observations.  

\begin{figure}
    \centering
    \includegraphics[width=0.6\linewidth]{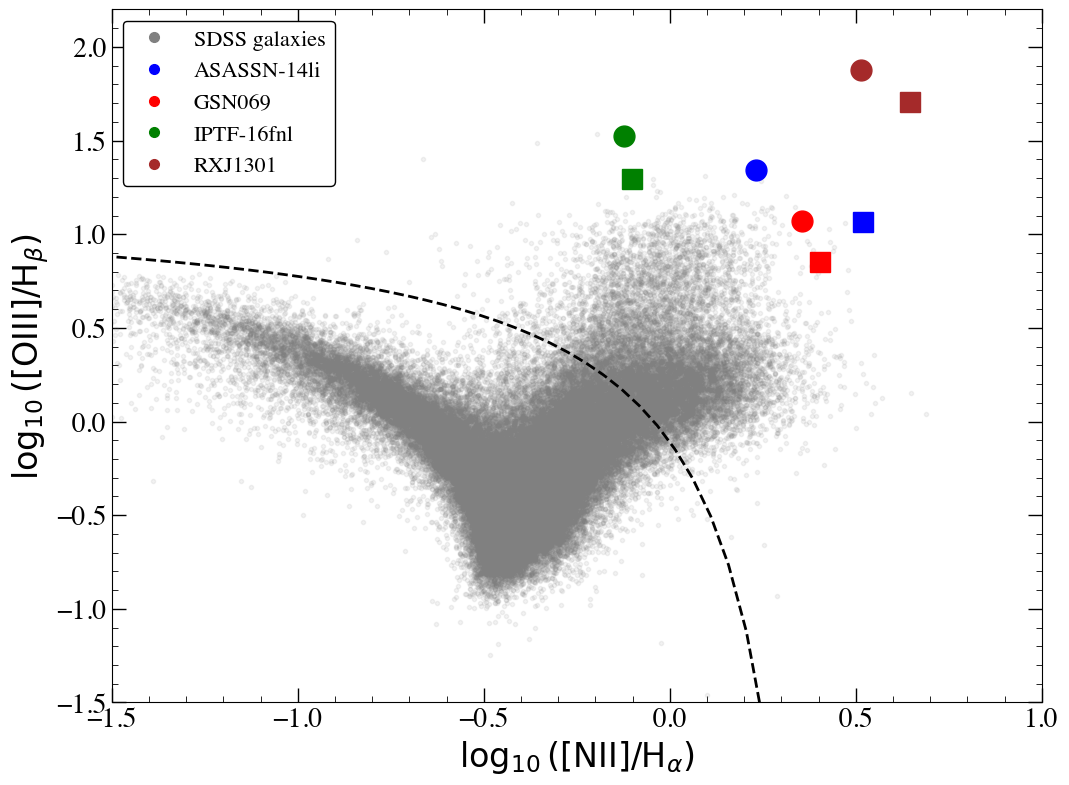}
    \caption{The observed location of different regions of the TDE and QPE host galaxies placed on the BPT diagram (circles = zone A, squares = zone B, see Figures \ref{fig:tde_hosts}, \ref{fig:qpe_hosts}).  We note that, despite representing an extremely small number of sources, TDE host galaxies represent large outliers in both [OIII]/H$_\beta$ and [NII]/H$_\alpha$, when compared to a significantly larger sample (200,000) of SDSS galaxies. We speculate that this may be related to the fact that TDE ionization fronts are likely to be out of recombination equilibrium, whereas most AGN-powered emission line regions are in equilibrium. As the recombination timescales of [OIII] and [NII] are typically $\sim 0.1$ times the recombination timescale of hydrogen, such clouds may become (e.g.) [OIII] \& [NII] bright prior to hydrogen recombination, and lie significantly off the typical BPT location of similarly hard ionizing sources.  }
    \label{fig:weird_tdes}
\end{figure}

In Figure \ref{fig:weird_tdes}, we plot the observed location of different regions of the TDE host galaxies placed on the BPT diagram (e.g., the host galaxies shown in Figures \ref{fig:tde_hosts}, \ref{fig:qpe_hosts}). We note that, despite representing an extremely small number of sources, TDE host galaxies represent large outliers in both  [OIII]/H$_\beta$ and [NII]/H$_\alpha$, when compared to a significantly larger sample of all SDSS galaxies. We speculate that this may be related to the fact that TDE ionization fronts are likely to be out of recombination equilibrium, whereas most AGN-powered emission line regions are in equilibrium. We believe that this observational fact is of real interest to future studies. 

\subsection{Conclusions}
To conclude, in this paper we have presented a broad analysis of the ionization signatures of those disks formed in tidal disruption events. We have demonstrated that these disks act as strong drivers of radiative feedback in their host galaxies, which has the potential to explain a number of observational properties of the host galaxies of (current) tidal disruption events. In particular, we focused initially on the fact that TDE host galaxies show a propensity for extended emission line regions with no recent (or current) classical active galactic nuclei present, and regularly produce (otherwise rare) extreme coronal lines. 

To summarize, our main results are:
\begin{itemize}
    \item Tidal disruption events produce long-lived, bright and blue accretion flows, which produce substantial ionizing fluxes (energy inputs of order $E_{\rm ION} \sim {\cal O}(10\%) M_\star c^2 \sim 10^{52}$ erg/s.) 
    \item TDEs can therefore, for the purposes of radiative feedback, be considered to produce transient ($\sim$ decades to centuries long) phases of an active galactic nucleus,  and will power ionization signatures in their host galaxies. 
    \item Reprocessed ``light echoes'' from previous tidal disruption events result in line ratios that will be observed (when superimposed on top of typical star forming galaxies), to appear in the ``AGN'' part of the BPT diagram (Figures \ref{fig:BPTNII}, \ref{fig:fagn}), despite of course not being produced by a ``classical'' AGN. 
    \item This may offer a natural resolution of the observations of EELR in TDE host galaxies:  what is actually being  observed in these systems is in fact (at least) two TDEs at once, one unfolding in real time and another echoing off molecular clouds in the outskirts of its galaxy. 
    \item If ionizing TDE disk radiation intercepts dense $n_H \sim 10^6-10^8\, {\rm cm}^{-3}$ circumnuclear gas clouds, the emission is capable of exciting coronal lines (such as [FeX], etc.), which will then evolve on much shorter ($\sim$ years) timescales. 
    \item Future observations of TDE host galaxies throughout the infrared with (for example) the JWST are likely to detect strong emission line features. In this paper we have focused on a series of Neon lines (namely [NeVI], [NeII], [NeV] and [NeIII]) although a wide range of infrared lines can be excited. 
    \item If these Neon lines are detected, we predict they will place TDE host galaxies in ``AGN'' like regions of characteristic diagnostic diagrams, just as optical emission lines place TDE hosts in the ``AGN'' regions of the classical BPT diagrams 
    \item TDE disks are capable of ``switching on'' nuclear narrow line regions in their hosts, which should be relatively common (gas cloud distribution dependent) in TDE hosts, and care must be taken when observations of TDE hosts are taken post-TDE, as these features can switch on rapidly. 
    \item TDEs can therefore act as contaminants of studies of AGN and ``fading'' AGN, and we have estimated that a moderate fraction of ``fading'' AGN candidates may in fact be TDE light echoes
    \item It is plausible that all ``anomalous'' (i.e., not obviously AGN powered) ECLEs are in reality tidal disruption disks at some stage of their evolution 
\end{itemize}

This work suggests a number of possible future observational tests/avenues for further study, some of which we discuss below. These questions represent a mixture of theoretical and observational avenues for future exploration 
\begin{enumerate}
    \item Do tidal disruption event disks power bright Neon lines? (Or other infrared lines?) If detected, do these Neon line ratios place them in the ``AGN'' part of diagnostic diagrams? These questions will presumably be answered soon when JWST spectra are taken of TDE hosts. 
    \item Can time dependent studies of coronal, optical and infrared lines be exploited to map out the circumnuclear gas density profiles of TDE hosts? This seems plausible given that direct observations of the TDE will allow constraints to be placed on the ionizing disk flux itself. These systems may in fact represent a unique opportunity for this science, as TDEs are in some sense ``cleaner'' than classical AGN. 
    \item What fraction of known TDE host galaxies result in the switching on of narrow line emission features? 
    \item What fraction of (e.g.,) SDSS [OIII] signatures show time evolution? These may well represent TDE powered, rather than AGN powered, line features. 
    \item By how much do the assumptions made by {\tt CLOUDY} (i.e., statistical steady state) effect the predictions of recombination emission from low-density galactic-scale molecular clouds? Can out-of-steady state effects explain the unusual placement of some TDE host galaxies on BPT diagrams (Figure \ref{fig:weird_tdes})? 
    \item Are relativistic disk effects important? Namely, does the anisotropy of the disk emission once Doppler and gravitational energy shifting effects are taken into account have any observational significance? It could plausibly be that (e.g.,) coronal lines are more likely to be excited at large inclinations to the disk axis, as the photons emitted in these directions are harder owing to Doppler boosting by the motion of the disk fluid. 
\end{enumerate}


\section*{Acknowledgments} 
This work was supported by a Leverhulme Trust International Professorship grant [number LIP-202-014]. MG is supported in part by NASA XMM-Newton grant 80NSSC24K1885. For the purpose of Open Access, AM has applied a CC BY public copyright license to any Author Accepted Manuscript version arising from this submission. 
 
\section*{Data accessibility statement}
All observational data used in support of this manuscript is publicly available. The MUSE data-cubes of the TDE/QPE host galaxies (Figures \ref{fig:tde_hosts} and \ref{fig:qpe_hosts}) are publicly available in the European Southern Observatory archive (\href{http://archive.eso.org/cms.html}{http://archive.eso.org/cms.html}). UV and X-ray data of ASASSN-14li (Figure \ref{fig:14li}) is publicly available through the High Energy Astrophysics Science Archive Research Center (HEASARC, \href{https://heasarc.gsfc.nasa.gov}{https://heasarc.gsfc.nasa.gov}). Optical-UV light curves of all current ZTF TDEs are publicly available at \href{https://github.com/sjoertvv/manyTDE}{https://github.com/sjoertvv/manyTDE}. The SDSS galaxy archive used in this work was accessed via \href{https://wwwmpa.mpa-garching.mpg.de/SDSS}{https://wwwmpa.mpa-garching.mpg.de/SDSS}. 

Numerical {\tt FitTeD}-{\tt CLOUDY} scripts will be made public upon acceptance. The base {\tt FitTeD} code package used in this work is available to download at the following repository:  
\href{https://bitbucket.org/fittingtransientswithdiscs/fitted_public/src}{https://bitbucket.org/fittingtransientswithdiscs/fitted\_public/src}, while {\tt CLOUDY} may be downloaded from 
\href{https://gitlab.nublado.org/cloudy/cloudy}{https://gitlab.nublado.org/cloudy/cloudy}.

\bibliographystyle{mnras}
\bibliography{andy}

\appendix

\section{Other diagnostic diagrams}\label{other}
\begin{figure}
    \centering
    \includegraphics[width=0.6\linewidth]{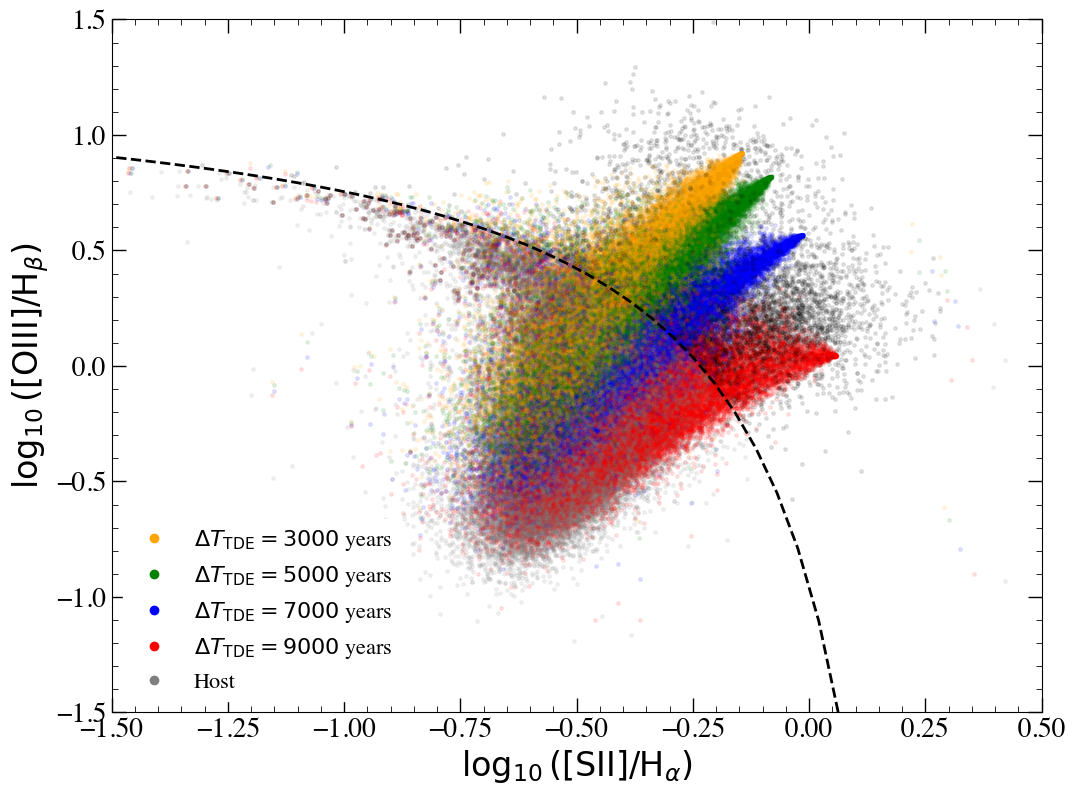}
    \caption{Same as for Figure \ref{fig:BPTNII}, except this figure displays the [OIII]-[SII] BPT diagram, a different diagnostic with the same broad features, namely that points above the dashed black curve cannot be produced by stellar populations.  }
    \label{fig:BPTSII}
\end{figure}

\begin{figure}
    \centering
    \includegraphics[width=0.6\linewidth]{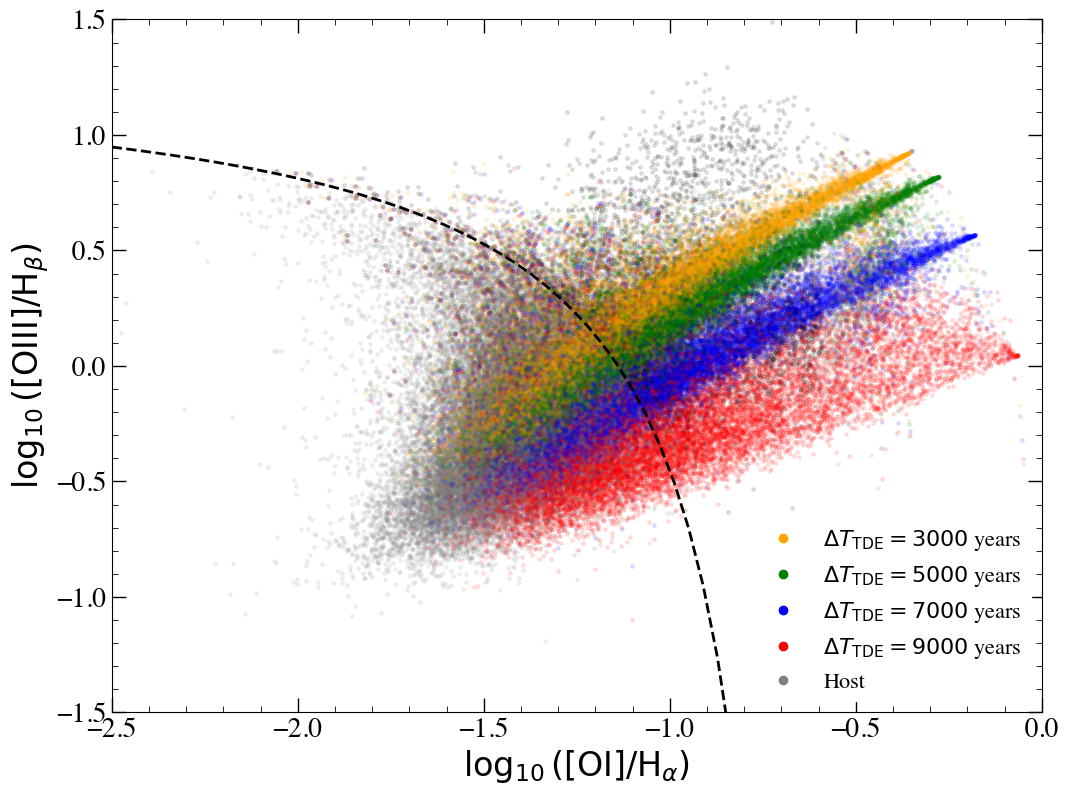}
    \caption{Same as for Figure \ref{fig:BPTNII}, except this figure displays the [OIII]-[OI] BPT diagram, a different diagnostic with the same broad features, namely that points above the dashed black curve cannot be produced by stellar populations.  }
    \label{fig:BPTOI}
\end{figure}

In this section we present the results of identical calculations presented in Figure \ref{fig:BPTNII}, but for the other classical BPT diagrams, namely [SII] (Figure \ref{fig:BPTSII}) and [OI] (Figure \ref{fig:BPTOI}). The results are very much in keeping with that of the main body of the text, with a substantial fraction (see Figure \ref{fig:fagn}) of SDSS host galaxies moving above \cite{Kewley01} lines, normally taken to be indicative of AGN activity.

\section{Data Analysis}\label{appB}
Actual observational data are presented in this paper only in Figures \ref{fig:tde_hosts}, \ref{fig:qpe_hosts}, \ref{fig:14li}, \ref{fig:gsn_ne}, \ref{fig:22upj}, and \ref{fig:weird_tdes}. All these data have already been published elsewhere in the same or similar form \citep[e.g.,][]{Prieto2016_MUSE_14li,Wevers24EELR, WeversFrench24, Newsome2024_22upj, GuoloMum24}, therefore here we provide only a brief description of the basic analysis performed, and we refer the reader to those papers for further details on the data.

To produce the continuum and line images shown in Figures \ref{fig:tde_hosts} and \ref{fig:qpe_hosts}, we downloaded the publicly available reduced Multi Unit Spectroscopic Explorer (MUSE) data cubes from the European Southern Observatory (ESO) for the TDE and QPE host galaxies. The images were generated by applying a synthetic narrowband filter in software, centered on the selected continuum and emission-line wavelength ranges. To measure the electron density of the host galaxy of GSN 069  (Fig.~\ref{fig:gsn_ne}), we measure the [SII] lines at $\lambda6716$\AA and $\lambda6732$\AA, the ratio of which ratios is know to correlate well with $n_e$ \citep[][at least in the $10-10^5 \ {\rm cm}^{-3}$ range]{Osterbrock06}. We employ the $n_e$ correlation of \citet{Proxauf14}, assuming an electron temperature of $T_e = 20,000K$. Emission-line measurements for regions A and B of each source were obtained using a circular 10\arcsec aperture. Stellar continuum modeling and subtraction, as well as emission-line flux measurements, were performed using the Penalized Pixel-Fitting (pPXF) method \citep{Cappellari2017}.

The UV light curve (and SED) and X-ray spectral reduction for ASASSN-14li, shown in Figure \ref{fig:14li} (upper panel), follow standard data reduction procedures in TDE literature. The full SED fitting and model description (bottom panel) are detailed in \citet{GuoloMum24} and are based on a novel implementation of the standard thin disk framework, which allows for simultaneous and self-consistent modeling of X-ray spectra and UV/optical/IR photometry.

The emission lines of AT 2022upj in Figure \ref{fig:22upj} were observed with the Goodman High Throughput Spectrograph on the Southern Astrophysical Research (SOAR) 4.1 m telescope, each using a 1\arcsec\ slit and the 400 mm$^{-1}$ line grating. The fluxes were calibrated to standard stars taken the same night of each observation with the same instrumental setup, and then photometrically calibrated to the subtracted UV-optical TDE light curve. See \cite{Newsome2024_22upj} for more details.

\label{lastpage}

\end{document}